%% file: main.tex
\documentclass[reprint,amsmath,amssymb,aps,prd,nofootinbib]{revtex4-2}

\usepackage{graphicx}
\usepackage{dcolumn}
\usepackage{bm}
\usepackage{verbatim}
\usepackage{rotating}
\usepackage{xcolor}
\usepackage[pdftex]{hyperref}
\usepackage{cleveref}

\begin{document}

\preprint{}

\title{Kinetic Sunyaev Zel'dovich velocity reconstruction from Planck and unWISE}

\author{Richard Bloch}
 \email{rbloch@my.yorku.ca}
\author{Matthew C. Johnson}
 \email{mjohnson@perimeterinstitute.ca}
\affiliation{Department of Physics and Astronomy, York University, Toronto, Ontario}
\affiliation{Perimeter Institute for Theoretical Physics, 31 Caroline St N, Waterloo, ON N2L 2Y5, Canada}

\date{\today}

\begin{abstract}
 The kinetic Sunyaev Zel'dovich (kSZ) effect is a blackbody cosmic microwave background (CMB) temperature anisotropy induced by Thomson scattering off free electrons in bulk motion with respect to the CMB rest frame. The statistically anisotropic cross-correlation between the CMB and galaxy surveys induced by the kSZ effect encodes the radial bulk velocity (more generally, the remote dipole field), which can be efficiently reconstructed using a quadratic estimator. Here, we develop a quadratic estimator for the remote dipole field for use with data from the Planck satellite and the unWISE galaxy redshift catalog. With this data combination, we forecast a signal-to-noise of order unity within $\Lambda$CDM assuming a simple model for the distribution of free electrons. Using reconstructions based on individual frequency temperature maps and a variety of component separated CMB maps, we characterize the impact of foregrounds and systematics. The dominant contaminant is a coupling between  the cosmic infrared background and large-scale galaxy survey systematics. We develop a method to minimize this effect, and demonstrate that after doing so the reconstructions are consistent with the expected level and properties of reconstruction noise. We use this reconstruction to constrain the multiplicative optical depth bias characterizing the amplitude of the remote dipole field to $b_v < 1.04$ at $68 \%$ confidence. Our fiducial signal model with $b_v =1$ is consistent with this measurement. Our results support an optimistic future for kSZ velocity reconstruction with near-term datasets.
\begin{description}
    \item[Keywords]
    CMB, Planck, LSS, kinetic Sunyaev-Zeldovich effect
\end{description}
\end{abstract}

\maketitle

\input{sections/introduction.tex}

\input{sections/methodology.tex}

\input{sections/data.tex}

\input{sections/discussion_results.tex}

\input{sections/posterior.tex}

\input{sections/conclusions.tex}

\appendix
\input{sections/bias_appendix.tex}

\input{sections/simulation_appendix.tex}

\bibliographystyle{unsrt}
\bibliography{planck_unwise_refs}

\end{document}

%% file: sections/introduction.tex
\section{Introduction}

Over the past three decades ever-more sensitive measurements of the Cosmic Microwave Background (CMB) have allowed us to gain unparalleled insight into the physics of the early universe. The Planck satellite \cite{PlanckCollaboration2020} has measured temperature anisotropies on the largest angular scales, the `primary' CMB, nearly to their cosmic variance limit. These measurements firmly established the standard cosmological model - $\Lambda$CDM. The new frontier in CMB science lies in the high-resolution, low-noise regime targeted by ground-based CMB experiments such as Atacama Cosmology Telescope (ACT)~\cite{Louis_2017}, South Pole Telescope (SPT)~\cite{Carlstrom_2011}, and Simons Observatory (SO)~\cite{Ade2019}. This regime is dominated by `secondary' CMB temperature anisotropies, which arise from interactions between CMB photons and large-scale structure (LSS) along the line of sight. The dominant blackbody component below arcminute angular scales is the kinetic Sunyaev Zel'dovich (kSZ) effect \cite{Sunyaev1980} - Thomson scattering of CMB photons from electrons in bulk motion. With existing CMB datasets, the kSZ effect has been detected with high significance using a variety of techniques, e.g.~\cite{Hand2012,Bernardis2017,Chen2022,Kusiak:2021hai}. With future datasets from e.g. SO, the kSZ effect will be detected with far higher significance. 

The kSZ effect is proportional to a line-of-sight integral over the product of the number density of free electrons and the locally observed CMB dipole projected along the line-of-sight - the remote dipole field.  The kSZ effect is  both a probe of  astrophysics through the (inhomogeneous) number density of electrons and cosmology through the remote dipole field. The astrophysical component of the kSZ effect is an important probe of the non-luminous `missing' baryons in the Universe (e.g.~\cite{PhysRevD.103.063513,Kusiak:2021hai,Hill:2016dta,Ferraro:2016ymw}), however the focus here will be on the cosmological information contained in the remote dipole field.

The CMB dipole seen by an observer like us is primarily sourced by peculiar velocities in non-linear structure and has a magnitude of a few mK, corresponding to velocities of order $10^2-10^3$ km/s. This component has a correlation length of order the size of galaxy groups and clusters, extending to distances of $\sim 50$ Mpc. Coarse-graining on $\sim100$ Mpc to Gpc scales, well into the linear regime, local velocities average down to the level of tens of km/s. On ultra-large scales, of order the cosmological horizon, the only contributions to the dipole are from last-scattering and the late-time Integrated Sachs-Wolfe effect, which are expected to be a few km/s in magnitude within $\Lambda$CDM. We  refer to this component as the `primordial' dipole~\footnote{This component is also known as the `intrinsic' or 'fundamencal' dipole. It is the component of the CMB dipole that would be observed in the rest frame of the CMB, e.g. as defined by the frame with zero aberration of the temperature anisotropies.}. The dipole field on these large scales has a correspondingly large correlation length of order Gpc. The remote dipole field sourcing the kSZ effect encodes both the local peculiar velocities as well as the primordial dipole, although the dominant component is from peculiar velocities. The remote dipole field can be an exquisite probe of the homogeneity of the Universe on the largest physical scales. New measurements on these scales can be used to probe e.g. large voids~\cite{Zhang:2010fc,Zhang:2010fa}, pre-inflationary relics~\cite{Zhang:2015uta}, anomalies in the primary CMB anisotropies~\cite{Cayuso:2019hen}, primordial non-Gaussianity~\cite{Munchmeyer2018,Contreras:2019bxy,AnilKumar:2022flx}, dark energy~\cite{Pen:2012hn}, modified gravity~\cite{Pan2019}, and  isocurvature~\cite{Hotinli2019,Kumar:2022bly}, among other scenarios.

In this paper we present a detailed analysis of a promising technique for extracting the remote dipole field: `kSZ tomography' or `kSZ velocity reconstruction'~ \cite{Shao:2010md,Terrana2017,Deutsch2018a,Smith2018,Cayuso2018,Cayuso2023}. This technique utilizes  non-Gaussian information in the small angular-scale cross-correlation between the kSZ component of CMB temperature anisotropies and a tracer of LSS to reconstruct the remote dipole field on large angular scales. Given tracers at a variety of redshifts, a tomographic reconstruction of the remote dipole field along the past light cone is possible. The simplest implementation of kSZ velocity reconstruction, which we employ here, is the quadratic estimator first introduced in Ref.~\cite{Deutsch2018a}~\footnote{At high signal-to-noise, a maximum likelihood estimator is superior~\cite{Kvasiuk:2023nje,Contreras:2022zdz}.}. This quadratic estimator was validated using N-body simulations~\cite{Cayuso2018,Giri:2020pkk}, and the potential impact of foregrounds and systematics assessed in Ref.~\cite{Cayuso2023}. This prior work forecasted a high signal to noise detection with near-term ground-based CMB experiments such as SO~\cite{Ade2019} in combination with photometric or spectroscopic galaxy redshift surveys such as the Vera C. Rubin Observatory LSST~\cite{LSSTScienceCollaboration2009} or DESI~\cite{DESI:2016fyo}. 

In preparation for this imminent flood of data, here we implement kSZ velocity reconstruction using existing data from the Planck CMB mission and galaxies from the Wide-Field Survey Infrared Explorer~\cite{2010AJ....140.1868W,Mainzer2014} (WISE) assembled into the unWISE catalogue~\cite{Lang_2014,Meisner_2017,Meisner_2017a,Schlafly_2019}. The Planck CMB temperature maps are well-characterized and provide a variety of ancillary datasets to assess the impact of foregrounds and instrumental systematics. The unWISE catalogue has $\sim 10^8$ objects split into three samples of increasing median redshift. Here, we focus on the `blue' sample of Ref.~\cite{Krolewski2020} which has $\sim$ 80 million objects over nearly the full sky. This dataset was chosen because large number densities and sky coverage are important for detecting the remote dipole field on large scales. This data combination has been used to detect the kSZ effect in Refs.~\cite{Hill:2016dta,Kusiak:2021hai}. Previous work~\cite{Cayuso2023} forecasted a total signal to noise of order unity for kSZ velocity reconstruction within $\Lambda$CDM with this data combination. In the absence of a statistically significant detection, our focus here is on demonstrating that systematics and foregrounds can be controlled at the level of statistical reconstruction noise - we find that they can! This result demonstrates that the future is promising for kSZ velocity reconstruction~\footnote{Indeed, the future is now here. Since the first version of this paper was posted on the arXiv preprint server, there have been a number of high-significance detections of the velocity field using kSZ velocity reconstruction including Refs.~\cite{Lague:2024czc,McCarthy:2024nik,Lai:2025qdw,Hotinli:2025tul}. }. 

We begin by tailoring the quadratic estimator introduced in Ref.~\cite{Deutsch2018a} to surveys with wide photometric redshift bins. The unWISE blue sample has a broad redshift distribution that spans $0.2 \alt z \alt 1$. The quadratic estimator yields a single two-dimensional map that is a weighted average of the dipole field over the unWISE survey volume. A key element of the quadratic estimator is the theoretical modeling of the galaxy-optical depth correlation function. If inaccurate, the estimator acquires a multiplicative 'optical depth' bias $b_v$. We demonstrate that photometric redshift uncertainties in the unWISE sample, uncertainties in the mean number density of electrons, and the degree of suppression of small-scale inhomogeneities in the electron distribution can in principle make significant contributions to the optical depth bias. We estimate that the optical depth bias can plausibly vary over the range $.5 \lesssim b_v \lesssim 1.1$.  We include $b_v$ as the only free parameter when comparing the measurements with the expected remote dipole signal, fixing other cosmological parameters. 

We then apply our estimator to individual frequency maps from Planck PR3 at 100, 143, 217, and 353 GHz to explore the impact of CMB foregrounds. We find that strong localized residuals in the reconstruction can be removed by masking. Correlations between unWISE and individual frequency maps in unmasked regions contribute primarily to the estimator monopole. This is consistent with the theoretical expectation~\cite{Cayuso2023} for a statistically isotropic cross-correlation. After masking and removing the monopole, the reconstructions at 100 and 143 GHz are consistent with the expected level of reconstruction noise from the primary CMB and instrumental noise only, with little evidence for foreground contamination. We find that the reconstruction at 217 GHz receives significant contributions from foregrounds at $\ell < 3$. The reconstruction at 353 GHz receives significant contributions from foregrounds on all scales, through a boost in power on large angular scales, and a large foreground contribution to the estimator variance. This result indicates that the cosmic infrared background (CIB), which increases in strength with frequency and correlates well with the unWISE galaxies, is the dominant foreground component to consider. 

Component-separated CMB maps in principle offer the highest signal-to-noise reconstruction available from Planck data, and can mitigate foreground contamination. We apply our estimator to CMB maps produced using a variety of component-separation techniques. In each case, strong residuals are mostly confined to regions that fall within the unWISE mask. Additionally, the estimator monopole has a magnitude far smaller than the 217 and 353 GHz maps, indicating that statistically isotropic correlations from CMB foregrounds are greatly reduced, as expected. We perform a set of tests for foreground residuals in unmasked regions of the sky, finding that reconstructions based on the SMICA map has the lowest level of foreground residuals on large angular scales. For our bottom-line CMB-unWISE reconstruction power spectrum we  therefore use the SMICA-based reconstruction.

Because they are sourced by the same underlying gravitational potentials, the signal component of the reconstruction is correlated with the unWISE galaxy density on large angular scales (for our model of unWISE, the correlation is significant for multipoles $\ell < 5$). Measuring this cross-power spectrum yields new information beyond the galaxy and estimator autospectra. The remote dipole (radial velocity) field on large angular scales is estimated from the small-scale CMB and unWISE maps, and therefore comes from an independent data combination. At a deeper level, the cross-correlation of the estimator with galaxy density is the squeezed limit of the galaxy-galaxy-temperature bispectrum. This data combination is particularly interesting since it can be used to measure a scale-dependent bias induced by primordial non-Gaussianity~\cite{Munchmeyer2018,Contreras:2019bxy,AnilKumar:2022flx} or isocurvature~\cite{Hotinli2019,Kumar:2022bly}. We find that the measured cross-spectrum is consistent with the expected sample variance. This is significant in light of the fact that the unWISE autospectrum is dominated by large angular-scale systematic effects at multipoles $\ell \lesssim 20$, and provides strong evidence for the future success of kSZ velocity reconstruction as a probe of non-Gaussianity and isocurvature~\footnote{In the months since the original version of this manuscript was posted on the arXiv preprint server several such analyses have been presented including Refs.~\cite{Krywonos:2024mpb,Lague:2024czc,Hotinli:2025tul}}. 

As a summary of the implications of our analysis for the remote dipole field, we compute the posterior over the optical depth bias $b_v$ given the reconstruction. We find an upper limit of $b_v < 1.4$ at $68 \%$ confidence. This is consistent with our expectation that the total signal-to-noise of the reconstruction is order one for this data combination. In a companion paper~\cite{Krywonos:2024mpb} we use the reconstructed remote dipole field presented here to constrain a variety of cosmological models.

The paper is laid out as follows. In Sec.~\ref{sec:methods} we review kSZ velocity reconstruction and develop a quadratic estimator for large photometric redshift bins characterizing the unWISE sample. We outline the expected statistics for this estimator and possible sources of systematic error. In Sec.~\ref{sec:data} we describe the properties of the datasets used as input for our reconstruction, various modeling assumptions implicit in the quadratic estimator, and predictions for the estimator response with these datasets. In Sec.~\ref{sec:results} we describe our analysis pipeline, and analyze reconstructions based on individual frequency and component-separated CMB maps. We characterize foregrounds and constrain the cross-correlation of the reconstruction with unWISE galaxy density. In Sec.~\ref{sec:posterior} we find the posterior probability distribution over the velocity bias $b_v$. We conclude and discuss the implications of our results in Sec.~\ref{sec:conclude}. We present a detailed assessment of the optical depth bias in Appendix~\ref{appendix:optical depth}. We perform a number of validation tests of the reconstruction pipeline in Appendix~\ref{appendix:pipeline_validation}.

%% file: sections/methodology.tex
\section{kSZ velocity reconstruction}
\label{sec:methods}

We begin by reviewing how kSZ velocity reconstruction can be used to reconstruct the remote dipole field - the locally observed CMB dipole projected on our past light cone. For a more detailed discussion of kSZ velocity reconstruction/kSZ tomography, we refer the reader to Refs.~\cite{Terrana2017,Deutsch2018a,Smith2018,Cayuso2023}.
 
The kSZ contribution to the CMB temperature is a line-of-sight integral:
\begin{equation}
    \Theta^{\mathrm{kSZ}}\left(\hat{n}\right) = - \int d\chi\ \dot{\tau}\left(\hat{n},\chi\right)v\left(\hat{n},\chi\right) \ ,
    \label{eq:ksz_temp}
\end{equation}
where $\dot{\tau}$ is the differential optical depth in direction $\hat{n}$ at comoving distance $\chi$ along the past light cone, and $v$ is the remote dipole field defined by
\begin{eqnarray}\label{eq:velocities}
v\left(\hat{n},\chi\right) &=& \hat{n} \cdot \vec{v} (\hat{n}, \chi)  + \sum_{m=-1}^1 v^m \left(\hat{n},\chi\right) Y_{1m} (\hat{n}) \ , \\
v^m \left(\hat{n},\chi\right) &\equiv& \int d^2 \hat{n}' \ \Theta(\hat{n},\chi,\hat{n}') Y^*_{1m}(\hat{n}') \ . 
\end{eqnarray}
The first contribution is from the peculiar velocity $\vec{v} (\hat{n}, \chi)$ projected along the line of sight $\hat{n}$. This term is sourced by local density perturbations. The second contribution is the primordial dipole, the local CMB dipole observed at rest and determined by the CMB radiation field $\Theta(\hat{n},\chi,\hat{n}')$. It receives contributions from the Sachs-Wolfe and integrated Sachs-Wolfe effects as well as Doppler shifts due to the velocity of plasma at last scattering. The primordial dipole directly probes the homogeneity of the Universe since it depends on the entirety of the surface of last scattering, not just the two-dimensional slice encoded in the primary CMB; for a detailed discussion see~\cite{Terrana2017,Deutsch2018a}.

Our focus here will be on cross-correlations with a photometric galaxy redshift survey, where the observed overdensity field is: 
\begin{equation}
    \delta^\mathrm{g}\left(\hat{n}\right) = \int d\chi\ W_\mathrm{g}\left(\chi\right)\delta^\mathrm{g} \left(\hat{n},\chi\right) \ ,
    \label{eq:LSS_moments}
\end{equation}
$\delta^\mathrm{g} \left(\hat{n},\chi\right)$ is the three-dimensional galaxy overdensity field and $W_\mathrm{g}$ is the galaxy window function defining the photometric sample. 

The basis of kSZ velocity reconstruction is the statistically anisotropic cross-correlation between kSZ temperature anisotropies and a tracer of LSS. Schematically, the kSZ effect is a product of density and velocity, and therefore $\langle \delta^g \Theta^{\mathrm{kSZ}}  \rangle \sim \int d\chi \langle \delta^g \dot{\tau} v \rangle$. This three-point function is dominated by `squeezed' configurations where velocity modes are far larger-scale than density modes~\cite{Deutsch2018a,Smith2018}. Therefore, the velocity field modulates small-scale power as $\langle \delta^g \Theta^{\mathrm{kSZ}}  \rangle \sim \int  d\chi \langle \delta^g \dot{\tau} \rangle \ v$, and we can estimate the velocity from a quadratic estimator given by $\hat{v} \sim \delta^g \Theta / \int d\chi \langle \delta^g \dot{\tau} \rangle$. Given multiple photometric redshift bins, one can break-up the kSZ line-of-sight integral to perform a tomographic reconstruction of $v$.

Since we have no direct measurement of the optical depth, the quadratic estimator relies on a model for the (statistical)  correlation between the LSS tracer and the optical depth as a function of redshift through $\int d\chi \langle \delta^g \dot{\tau} \rangle$. Mis-modeling this correlation leads to the `optical depth' or `velocity' bias (see e.g.~\cite{Battaglia:2016xbi,Smith2018} for further discussion), which can be due to incorrect assumptions about the galaxy-halo connection and gas-halo connection, as well as poor characterization of the redshift distribution of the LSS tracer. Fortunately, previous work has demonstrated that the optical depth bias is scale-independent on large scales~\cite{Giri:2020pkk,Cayuso2023,Kvasiuk:2023nje}, and can therefore be described by a manageable number of nuisance parameters. We discuss the optical depth bias in detail in the analysis below.

\subsection{Harmonic-space quadratic estimator}

The quadratic estimator used in our analysis is based on off-diagonal correlations between CMB temperature and galaxy density in harmonic space:
\begin{align}\label{eq:exactcross}
    \left\langle \Theta_{\ell m}\delta_{\ell^{\prime}m^{\prime}}^{g}\right\rangle 
     = - \sum_{LM} & w_{mm^{\prime}-M}^{\ell\ell^{\prime}L}  \\ & \times \int d\chi d\chi^{\prime}\,[C_{\ell^{\prime}}^{\dot{\tau} g}\left(\chi,\chi^{\prime}\right)]^{\rm t} v_{LM}\left(\chi\right) \ , \nonumber
\end{align}
where $[C_{\ell^{\prime}}^{\dot{\tau} g}\left(\chi,\chi^{\prime}\right)]^{\rm t}$ is the true cross-power between the differential optical depth and galaxies on the past light cone and 
\begin{align}
    w_{m m' -M}^{\ell \ell' L}&=\left(-1\right)^{M}\sqrt{\frac{\left(2\ell+1\right)\left(2\ell'+1\right)\left(2L+1\right)}{4\pi}} \nonumber \\
    & \times\begin{pmatrix}
        \ell & \ell' & L \\
        m & m' & -M
    \end{pmatrix}
    \begin{pmatrix}
        \ell & \ell' & L\\
        0 & 0 & 0
    \end{pmatrix} \ .
    \label{eq:w_3j}
\end{align}
To make progress, we make two crucial assumptions. First, since most of the signal-to-noise in the reconstruction comes from density modes on small angular scales, in the Limber approximation we can set:
\begin{eqnarray}
    [C_{\ell}^{\dot{\tau} g}\left(\chi,\chi^{\prime}\right)]^{\rm t} &\simeq& [C_{\ell}^{\dot{\tau} g}\left(\chi \right)]^{\rm t} \delta(\chi^\prime - \chi)
\end{eqnarray}
Next, we expand the galaxy-optical depth cross-power spectrum about some reference redshift $\chi = \bar{\chi}$ (typically the median redshift of the bin) and a reference scale $\ell = \bar{\ell}$ (typically $\bar{\ell} = 2\times 10^3$; see discussion below). Defining $\bar{C}_{\ell^{\prime}}^{\tau\mathrm{g}} \equiv C_{\ell^{\prime}}^{\dot{\tau}\mathrm{g}}\left(\chi=\bar{\chi}\right)\Delta\chi$ where $\Delta \chi$ is a normalization factor representative of the width of the redshift bin (defined more precisely below), we have
\begin{eqnarray}\label{eq:taugapprox}
    [C_{\ell}^{\dot{\tau} g}\left(\chi,\chi^{\prime}\right)]^{\rm t} 
    &\simeq&  [\bar{C}_{\ell}^{\tau g}]^{\rm t}  \frac{[C_{\ell=\bar{\ell}}^{\dot{\tau} g}\left(\chi\right)]^{\rm t}}{[\bar{C}_{\ell=\bar{\ell}}^{\tau g}]^{\rm t}} \delta(\chi^\prime - \chi) \ .
\end{eqnarray}
With these assumptions,
\begin{align}\label{eq:map_crosspower}
    \langle\Theta_{\ell m}\delta^\mathrm{g}_{\ell^\prime m^\prime} \rangle \simeq - \sum_{LM} &  w^{\ell \ell^\prime L}_{m m^\prime -M} [\bar{C}_{\ell^\prime}^{\tau\mathrm{g}}]^{\rm t} \\
    & \times \int d\chi \ \frac{[C_{\ell=\bar{\ell}}^{\dot{\tau} g}\left(\chi\right)]^{\rm t}}{[\bar{C}_{\ell=\bar{\ell}}^{\tau g}]^{\rm t}}  v_{LM}\left(\chi\right)  \ . \nonumber
\end{align}
When this approximation is valid, the scale- and redshift-dependent factors in the cross-correlation can be separated.

Given Eq.~\eqref{eq:map_crosspower} we can write down a simple quadratic estimator in analogy with those first presented in Refs.~\cite{Deutsch2018a,Smith2018,Cayuso2023}:
\begin{align}    \label{eq:estimator}
    \hat{v}_{\ell m} = - N_\ell \sum_{\ell_1 m_1;\ell_2 m_2} &\left(-1\right)^m
    \begin{pmatrix}
        \ell_1 & \ell_2 & \ell \\
        m_1 & m_2 & -m
    \end{pmatrix} \\
    & \times G_{\ell_1 \ell_2 \ell}
    \Theta_{\ell_1 m_1} \delta_{\ell_2 m_2} \ ,
    \nonumber
\end{align}
where
\begin{equation}
    N_\ell = \left( 2\ell +1\right) \left( \sum_{\ell_1 \ell_2} G_{\ell_1 \ell_2 \ell}\ f_{\ell_1 \ell_2 \ell} \right)^{-1} \ ,
    \label{eq:variance_general}
\end{equation}
and (neglecting cross-correlations between the non-kSZ components of the CMB and the galaxy survey)
\begin{equation}
    G_{\ell_1 \ell_2 \ell} \equiv \frac{f_{\ell_1 \ell_2 \ell}}{C_{\ell_1}^{\mathrm{TT}}C_{\ell_2}^{\mathrm{gg}}} \ .
    \label{eq:estimator_filter}
\end{equation}
 $C_\ell^{\mathrm{TT}}$ includes the primary CMB, instrumental noise, kSZ, as well as galactic and extragalactic foregrounds. $C_\ell^{\mathrm{gg}}$ is the galaxy power spectrum including the clustering signal as well as shot noise and survey systematics. The CMB and galaxy power spectra are ideally based on self-consistent theoretical models. In practice it is acceptable to use the empirically measured power spectra of the input maps; see Appendix~\ref{appendix:pipeline_validation} for further discussion of the pitfalls in doing so. The function $f_{\ell_1 \ell_2 \ell}$ is  defined as:
\begin{equation}
    f_{\ell_1 \ell_2 \ell} \equiv \sqrt{\frac{\left(2\ell_1 +1\right)\left(2\ell_2 +1\right)\left(2\ell+1\right)}{4\pi}}
    \begin{pmatrix}
        \ell_1 & \ell_2 & \ell \\ 
        0 & 0 & 0
    \end{pmatrix}
    \bar{C}_{\ell_2}^{\tau\mathrm{g}} \ ,
    \label{eq:Gamma}
\end{equation}
where $\bar{C}_{\ell_2}^{\tau\mathrm{g}}$ is a {\em model} for the galaxy-optical depth power spectrum (denoted by the absence of the `t' superscript) at the reference redshift.  

Computing the estimator mean, we find:
\begin{equation}
    \langle \hat{v}_{\ell m}\rangle =
    \int d\chi\ W_v\left(\chi\right) v_{\ell m}\left(\chi\right) \ ,
    \label{eq:estimator_mean}
\end{equation}
where 
\begin{equation}\label{eq:vwindow}
 W_v \left(\chi\right) \equiv \frac{\sum_{\ell_1 \ell_2}  G_{\ell_1 \ell_2 \ell}\ f_{\ell_1 \ell_2 \ell} \ [C_{\ell_2}^{\dot{\tau} g}\left(\chi \right)]^{\rm t}  / \bar{C}_{\ell_2}^{\tau g}  }{\sum_{\ell_1' \ell_2'} G_{\ell_1' \ell_2' \ell}\ f_{\ell_1' \ell_2' \ell} } \ .
\end{equation}
The estimator variance is given by
\begin{eqnarray}\label{eq:est_harm_power}
    \langle \hat{v}^*_{\ell m} \hat{v}_{\ell' m'} \rangle &=& C_\ell^{\hat{v} \hat{v}} \delta_{\ell \ell'} \delta_{m m'} \\ 
    &=& \int d\chi d\chi' \ W_v\left(\chi\right) W_v\left(\chi' \right) C_\ell^{vv} \left(\chi, \chi' \right) \delta_{\ell \ell'} \delta_{m m'} \nonumber \\ 
    &+& N_\ell \ \delta_{\ell \ell'} \delta_{m m'} \ . \nonumber
\end{eqnarray}
Within $\Lambda$CDM the power spectrum for the remote dipole field is related to the primordial power spectrum $\mathcal{P}(k)$ by
\begin{equation}
C_\ell^{vv} \left(\chi, \chi' \right) = \frac{2}{\pi} \int \frac{dk}{k} \Delta^v_\ell (k,\chi) \Delta^v_\ell (k,\chi') \mathcal{P}(k) \ .
\end{equation}
The transfer function for the remote dipole field is
\begin{eqnarray}\label{eq:vsources}
\Delta^v_\ell (k,\chi) = \frac{i^\ell}{2 \ell + 1} \left[ S^{\rm LD} (k,\chi) + S^{\rm P} (k,\chi)  \right] \nonumber \\ \times \left[ \ell j_{\ell -1}(k\chi) - (\ell+1) j_{\ell+1} (k\chi) \right] \ ,
\end{eqnarray}
where $S^{\rm LD} (k,\chi)$ is the `local Doppler' source induced by the radial peculiar velocity field and $S^{\rm P} (k,\chi)$ is the source for the `primordial' dipole field induced by the Sachs-Wolfe, Integraged Sachs-Wolfe, and primordial Doppler components. The full form of the source functions can be found in Ref.~\cite{Deutsch2018a}.

Note that in general $W_v \left(\chi\right)$ is dependent on $\ell$ (see Eq.~\ref{eq:vwindow}). However, this scale dependence is weak at low-$\ell$ (the case of interest below), and therefore we have suppressed the $\ell$-dependence in the argument of this function. When the ratio $[C_{\ell_2}^{\dot{\tau} g} \left(\chi \right)]^{\rm t} / \bar{C}_{\ell_2}^{\tau g}$ is independent of $\ell_2$, the approximation in Eq.~\eqref{eq:taugapprox} is exact, and $W_v = [C_{\ell=\bar{\ell}}^{\dot{\tau} g}\left(\chi\right)]^{\rm t} / \bar{C}_{\ell=\bar{\ell}}^{\tau g}$. We find below that this is an excellent approximation within our model. The velocity window function and estimator weights depend on $\bar{C}_{\ell_2}^{\tau g}$, which comes from a {\em model} for the galaxy-optical depth cross-correlation $C_{\ell_2}^{\dot{\tau} g} \left(\chi \right) \neq [C_{\ell_2}^{\dot{\tau} g} \left(\chi \right)]^{\rm t}$. We must incorporate this model uncertainty when comparing a reconstruction against theoretical expectations. We outline how a mis-match between the true and fiducial optical depth-galaxy cross-power contribute to the optical depth bias in Appendix \ref{appendix:optical depth}. Finally, we fix the normalization parameter $\Delta \chi$ by
\begin{equation}
    \Delta \chi = \int d\chi \ C_{\ell=\bar{\ell}}^{\dot{\tau} g}\left(\chi\right)/ C_{\ell=\bar{\ell}}^{\dot{\tau} g}\left(\chi =\bar{\chi}\right) \ ,
\end{equation}
given a model $C_{\ell_2}^{\dot{\tau} g} \left(\chi \right)$.

\subsection{Pixel-space quadratic estimator}

In the presence of incomplete sky coverage and masking, it is preferable to use a pixel-space form of the estimator in Eq.~\eqref{eq:estimator}. This takes a particularly simple form when we neglect the scale dependence of $N_\ell$ defined in Eq.~\eqref{eq:variance_general}, which is an excellent approximation in the limit where $\ell_1, \ell_2 \gg \ell$ where :
\begin{equation}\label{eq:const_est_n}
N_\ell \simeq N \equiv \left[ \sum_{\ell_1} \frac{2 \ell_1 + 1}{4 \pi} \frac{(\bar{C}_{\ell_1}^{\tau g})^2  }{C_{\ell_1}^{TT} C_{\ell_1}^{gg}}\right]^{-1} \ .
\end{equation}
We first define filtered CMB and galaxy fields:
\begin{equation}
    \xi \left( \hat{n} \right) = \sum_{\ell m} \Theta_{\ell m} \frac{1}{C_\ell^{\mathrm{TT}}}Y_{\ell m} (\hat{n}), 
    \ \ \
    \zeta \left( \hat{n} \right) = \sum_{\ell m} \delta_{\ell m} \frac{\bar{C}_\ell^{\tau\mathrm{g}}}{C_\ell^{\mathrm{gg}}} Y_{\ell m} (\hat{n}) \ .
    \label{eq:map_filters}
\end{equation}
The filtering operation in $\xi \left( \hat{n} \right)$ acts as a high-pass filter for the CMB, suppressing large-scale correlations. The filtering operation in $\zeta \left( \hat{n} \right)$ will in general preserve fluctuations in the galaxy field on large scales where baryons and galaxies trace the underlying dark matter distribution, while suppressing power on scales ($\sim 1-10$ Mpc) affected by feedback processes. The pixel-space form of the quadratic estimator is
\begin{equation}
\hat{v}\left(\hat{n}\right) = - N \ \xi\left(\hat{n}\right) \zeta\left(\hat{n}\right) \ .
\label{eq:estimator_coeffs}
\end{equation}

Due to the nature of the filtering, the product $\xi \left( \hat{n} \right) \zeta \left( \hat{n} \right)$ is sensitive only to local correlations between the CMB and galaxy survey. This property is advantageous when dealing with masking and partial sky coverage. 

The mean of the pixel-space estimator is
\begin{equation}
    \langle \hat{v}\left(\hat{n}\right) \rangle = \int d\chi W_v(\chi)  v(\hat{n},\chi) \ ,
\end{equation}
which is consistent with Eq.~\eqref{eq:estimator_mean}. Turning to the pixel-space estimator variance, the contribution from the reconstruction noise deserves some discussion. Assuming that both $\xi(\hat{n})$ and $\zeta (\hat{n})$ are Gaussian random fields, the one-point function is a normal product distribution:
\begin{equation}\label{eq:normalproductdist}
    P(\hat{v}) = \frac{1}{\pi N \sigma_\xi \sigma_\zeta} K_0 \left[\frac{|\hat{v}|}{N \sigma_\xi \sigma_\zeta} \right] \ ,
\end{equation}
where $K_0(x)$ is the modified Bessel function of the second kind and 
\begin{eqnarray}
    \sigma_\xi^2  &=& \langle \xi(\hat{n})^2 \rangle \nonumber \\
    &=& \sum_{\ell_1} \frac{2 \ell_1 + 1}{4 \pi} \frac{1}{C_{\ell_1}^{TT}} \ ,
\end{eqnarray}
and
\begin{eqnarray}
    \sigma_\zeta^2  &=& \langle \zeta(\hat{n})^2 \rangle \nonumber \\
    &=& \sum_{\ell_2} \frac{2 \ell_2 + 1}{4 \pi} \frac{(\bar{C}_{\ell_2}^{\tau g})^2}{C_{\ell_2}^{gg}} \ .
\end{eqnarray}
The coincident two-point function is straightforward to compute from the one-point function:
\begin{equation}
    \langle v(\hat{n})^2 \rangle = N^2  \sigma_\xi^2 \sigma_\zeta^2 \ .
\end{equation}
Because the reconstruction noise is non-Gaussian, even under the approximation that $N_\ell$ is independent of scale, the reconstruction noise is correlated in pixel space. However, coarse-graining by convolving the map with a beam, by the Central Limit Theorem, the distribution approaches a Gaussian. We can therefore treat reconstruction noise on large angular scales, where our signal lies, as Gaussian. However, retaining all scales, we speculate that the non-Gaussian properties of the reconstruction noise can be used as an additional method to distinguish it from the underlying Gaussian dipole field signal. We explore some of these aspects below.

\subsection{Possible sources of systematics}\label{sec:possible_systematics}

There are a number of potential systematic effects that could lead to a biased reconstruction. Previous work~\cite{Cayuso2018,Cayuso2023} explored a variety of these effects using simulations, but the influence of systematics in an analysis of real data has not yet been performed - this is one of the primary goals of this paper. Here, we present the expected systematics at a qualitative level.

In general, we can classify potential systematics into the following categories:
\begin{itemize}
    \item {\bf Optical depth bias}: When $C_{\ell_2}^{\dot{\tau} g}\left(\chi \right) \neq [C_{\ell_2}^{\dot{\tau} g}\left(\chi \right)]^{\rm t}$, the estimator mean will be biased against the true remote dipole field - referred to as the `optical depth' or `velocity' bias. This modelling error can arise because of a poor understanding of the galaxy-halo and/or gas-halo connection, environmental/selection effects, and inaccurate/uncertain redshift distributions, among other factors arising from our limited knowledge of the distribution of baryons. Fortunately, so long as we focus on the reconstruction on large scales, this bias is scale independent~\cite{Cayuso2023,Giri:2020pkk}. 
    In Appendix~\ref{appendix:optical depth} we estimate the possible magnitude of this bias by computing the expected signal  over a range of model assumptions, finding that variations in the range $0.5 \lesssim b_v \lesssim 1.1$ are plausible. 
    \item {\bf Statistically isotropic CMB-galaxy cross-correlations}: An isotropic correlation between the CMB and galaxy survey, e.g. due to extragalactic CMB foregrounds such as the Cosmic Infrared Background (CIB) or thermal Sunyaev Zel'dovich effect (tSZ), contributes to the estimator weights. With a detailed model of the cross-correlation, this can be incorporated into $G_{\ell_1 \ell_2 \ell}$ defined in Eq.~\eqref{eq:estimator_filter}. Neglecting these contributions yields a slightly sub-optimal estimator (e.g. the variance is not as low as possible). Additionally, statistically isotropic correlations  contribute to the monopole of the reconstruction:
\begin{eqnarray}
\langle \hat{v}_{00} \rangle &=& - \frac{\sum_{\ell_1 \ell_2}  G_{\ell_1 \ell_2 \ell}\ f_{\ell_1 \ell_2 \ell} \ C_{\ell_2}^{T g}  / \bar{C}_{\ell_2}^{\tau g}  }{\sum_{\ell_1' \ell_2'} G_{\ell_1' \ell_2' \ell}\ f_{\ell_1' \ell_2' \ell} } \\
&\simeq& - N \left[ \sum_{\ell_1} \frac{2 \ell_1 + 1}{4 \pi} \frac{\bar{C}_{\ell_1}^{\tau g} C_{\ell_1}^{T g}  }{C_{\ell_1}^{TT} C_{\ell_1}^{gg}}\right]
\end{eqnarray}
    
    \item {\bf Statistically anisotropic CMB-galaxy cross-correlations}: The quadratic estimator in Eq.~\eqref{eq:estimator} is in principle sensitive to any effect that modulates the cross-correlation between the CMB and galaxies across the sky. A variety of potential foregrounds and systematics were considered in Ref.~\cite{Cayuso2023}. Such effects lead to an additive bias. For example, given a signal in the CMB temperature $\Theta^M = M(\hat{n}) \delta^M (\hat{n})$ where $\delta^M (\hat{n})$ is correlated with LSS, the mean estimator response is
    \begin{eqnarray}\label{eq:cmb_addbias}
        \langle \hat{v}^M_{\ell m} \rangle &=& -\frac{\sum_{\ell_1 \ell_2}  G_{\ell_1 \ell_2 \ell}\ f_{\ell_1 \ell_2 \ell} \ C_{\ell_2}^{M g}\left(\chi \right)  / \bar{C}_{\ell_2}^{\tau g}  }{\sum_{\ell_1' \ell_2'} G_{\ell_1' \ell_2' \ell}\ f_{\ell_1' \ell_2' \ell} } M_{\ell m} \nonumber \\
        &=& \langle \hat{v}_{00} \rangle M_{\ell m}
        \ .
    \end{eqnarray}
    In the second line we have assumed that $\Theta^{M}$ is the sole contribution to the correlation with $\delta^g$. 
    This type of systematic can arise from physical effects such as CMB or galaxy lensing as well as  relativistic effects modulating point source number counts (e.g.~\cite{Planck:2020qil}). It can also arise from instrumental systematics such as anisotropic beams or anisotropic levels of  foreground removal. Likewise, systematics and physical effects in the galaxy survey that modulate a component correlated with the CMB $\delta^P = P(\hat{n}) \Theta^P (\hat{n})$  lead to a mean estimator response:
    \begin{eqnarray}\label{eq:lss_addbias}
        \langle \hat{v}^P_{\ell m} \rangle &=& -\frac{\sum_{\ell_1 \ell_2}  G_{\ell_1 \ell_2 \ell}\ f_{\ell_1 \ell_2 \ell} \ C_{\ell_2}^{P T}\left(\chi \right)  / \bar{C}_{\ell_2}^{\tau g}  }{\sum_{\ell_1' \ell_2'} G_{\ell_1' \ell_2' \ell}\ f_{\ell_1' \ell_2' \ell} } P_{\ell m} \nonumber \\
        &=& \langle \hat{v}_{00} \rangle P_{\ell m}\ .
    \end{eqnarray}
    Again, in the second line we assume that $\delta^P$ is the sole source of correlation with $\Theta$. A physical effect leading to this is the relativistic modulation of number counts. Systematic effects include  extinction, redshift calibration errors, anisotropic depth, and background effects where e.g. the presence of nearby stars makes it difficult to isolate extragalactic sources. In any of the cases described above, the estimator response is proportional to the reconstruction monopole. We utilize this observation below. 
    \item {\bf Higher order noise bias:} There are additional contributions to the estimator variance presented in Eq.~\eqref{eq:est_harm_power} beyond those we have considered here~\cite{Giri:2020pkk}. These arise from  $\langle \Theta^{\rm kSZ} \delta^g \Theta^{\rm kSZ} \delta^g\rangle$. The disconnected components of this correlator are referred to, in analogy with CMB lensing, as the $N^{(1)}$ bias; the connected component due to non-linear clustering is referred to as the $N^{(3/2)}$ bias. These effects are only relevant in the high-signal-to-noise regime, and will not be discussed further here.
\end{itemize}

%% file: sections/data.tex
\section{Data and Theoretical Modelling}
\label{sec:data}

Before proceeding, we review the properties of ideal datasets for kSZ velocity reconstruction. The quadratic estimator relies on reconstructing the remote dipole field on large scales from anisotropic CMB-galaxy cross-power on small angular scales. The performance of the estimator therefore benefits from small-scale modes probed by a high resolution, low-noise CMB dataset and a galaxy survey that has a high number density of objects. Because the remote dipole field has power primarily on large angular scales, it is also desirable to have large sky coverage. In addition to these factors influencing the expected statistical error of the reconstruction, we must also worry about a variety of potential systematics, as described in the previous section. It is therefore desirable to utilize surveys with well-characterized foregrounds and instrumental systematics. Finally, it is desirable to have many redshift bins to perform a three-dimensional reconstruction of the dipole field along our past light cone.

In this paper, we use a data combination that has many of these desirable properties: Planck temperature anisotropies and galaxies from the unWISE `blue' sample. The main shortcoming of this combination is the limited sensitivity of Planck and the lack of redshift resolution in the unWISE sample. We describe in this section the data products we utilize and our modelling assumptions.

\subsection{Planck Temperature Maps}

Our analysis utilizes temperature maps from the the Planck Data Release 3 (PR3) \cite{PlanckCollaboration2020}. We analyze both individual frequency maps at 100, 143, 217, 353, 545, and 857 GHz as well as the Planck SMICA, NILC, SEVEM, and Commander component-separated CMB maps~\cite{Planck:2018yye}. In addition, we analyze NILC maps produced using the {\tt pyilc} code~\cite{McCarthy:2023hpa} where the contribution from specific components such as tSZ and CIB are explicitly nulled. In our assessment of various foregrounds and systematics, we additionally use a variety of CMB-subtracted maps and other ancillary Planck data products. Here, we describe the relevant properties of these data products.

For our main nalysis, we utilize PR3 individual frequency maps at 100, 143, 217, and 353 GHz. Each of these frequency maps have a corresponding CMB-subtracted map for each component separation technique, providing estimates of the sum of all foregrounds and noise at each frequency. We apply the estimator to each of these maps to determine the influence of foregrounds on the reconstruction at each frequency. Individual frequency maps are debeamed with a Gaussian beam of FWHM 9.68, 7.30, 5.02, and 4.94 arcminutes for 100, 143, 217, and 353 GHz frequencies respectively; we work at the Planck native resolution of $N_{\rm side} =2048$. We utilize simulated (galactic) foreground maps produced using the 10th Planck full focal plane simulation set (FFP10)~\cite{Planck:2015txa} including free-free, synchrotron, and thermal dust components. We also utilize instrumental noise realizations from the FFP10 simulations. Finally, we utilize PR3 individual frequency maps at 545 and 857 GHz to provide an empirical estimate of velocity reconstruction systematics, as described in greater detail below.


We analyze Planck blackbody maps produced using four different component separation techniques, SMICA, NILC, SEVEM, and Commander, described in Ref.~\cite{Planck:2018yye}. The CMB maps have an effective Gaussian beam with FWHM of 5 arcminutes, which we remove in our analysis; we work at the Planck native resolution of $N_{\rm side} =2048$. As our fiducial component separated CMB map we choose SMICA (Spectral Matching Independent Component Analysis). This technique is based on a linear weighting of each Planck frequency map in harmonic space~\cite{Tegmark2003}, such that the variance of a desired spectral component is minimized -- here the blackbody spectrum, containing the primary CMB and kSZ. On the small angular scales relevant to the analysis presented below, the harmonic weights are largest in magnitude at 217, 353, and 857 GHz. 

Finally, we analyze several maps where tSZ or CIB have been explicitly nulled ('deprojected') in the component separation algorithm. We use a version of the SMICA map ('SMICA no SZ') produced by the Planck collaboration~\cite{Planck:2018yye} that explicitly nulls anisotropies with the spectral dependence of the thermal Sunyaev Zel'dovich (tSZ) signal. We utilize various NILC maps produced using the {\tt pyilc} code~\cite{McCarthy:2023hpa} that explicitly null tSZ or CIB signals. These maps were created using data from the Planck Data Release 4 (PR4)~\cite{Planck:2020olo}. For maps that null the CIB, the CIB spectral energy distribution is modeled as a modified blackbody with two free parameters; see Ref.~\cite{McCarthy:2023hpa} for further details.

\subsection{unWISE Galaxy Map}

The unWISE~\cite{Schlafly_2019} catalogue contains over 500 million galaxies between $0\leq z \leq 2$ constructed from NEOWISE data \cite{2010AJ....140.1868W,Mainzer2014}. unWISE provides the largest currently available extragalactic catalogue with a measured redshift distribution, making it a desirable dataset for kSZ velocity reconstruction. Various catalogues derived from WISE data have already been successfully utilized in a wide variety of CMB cross-correlation studies including e.g.~\cite{Ferraro:2014msa,Hill:2016dta,Ferraro:2016ymw,Marques:2019aug,Krolewski2020,Krolewski2021,Kusiak:2021hai,Krolewski:2021znk,Kusiak:2022xkt,Farren:2023yna,2023arXiv230905659F,Yan:2023okq}.

Here, we use the unWISE catalogue described in Ref.~\cite{Krolewski2020}. Objects in the full unWISE catalogue were cross-checked against Gaia DR2~\cite{GaiaCollaboration2018}
sources to reduce stellar contamination and divided into three large color/redshift bins labeled `red’, `green’, and `blue’ in order of descending median redshift.  We focus on the blue sample for our analysis as it has the strongest confidence in redshift measurements and has the highest number density of galaxies. Future analyses could utilize all three samples to provide a true tomographic reconstruction of the remote dipole field. 

\begin{figure}
    \centering
    \includegraphics[width=1.\columnwidth]{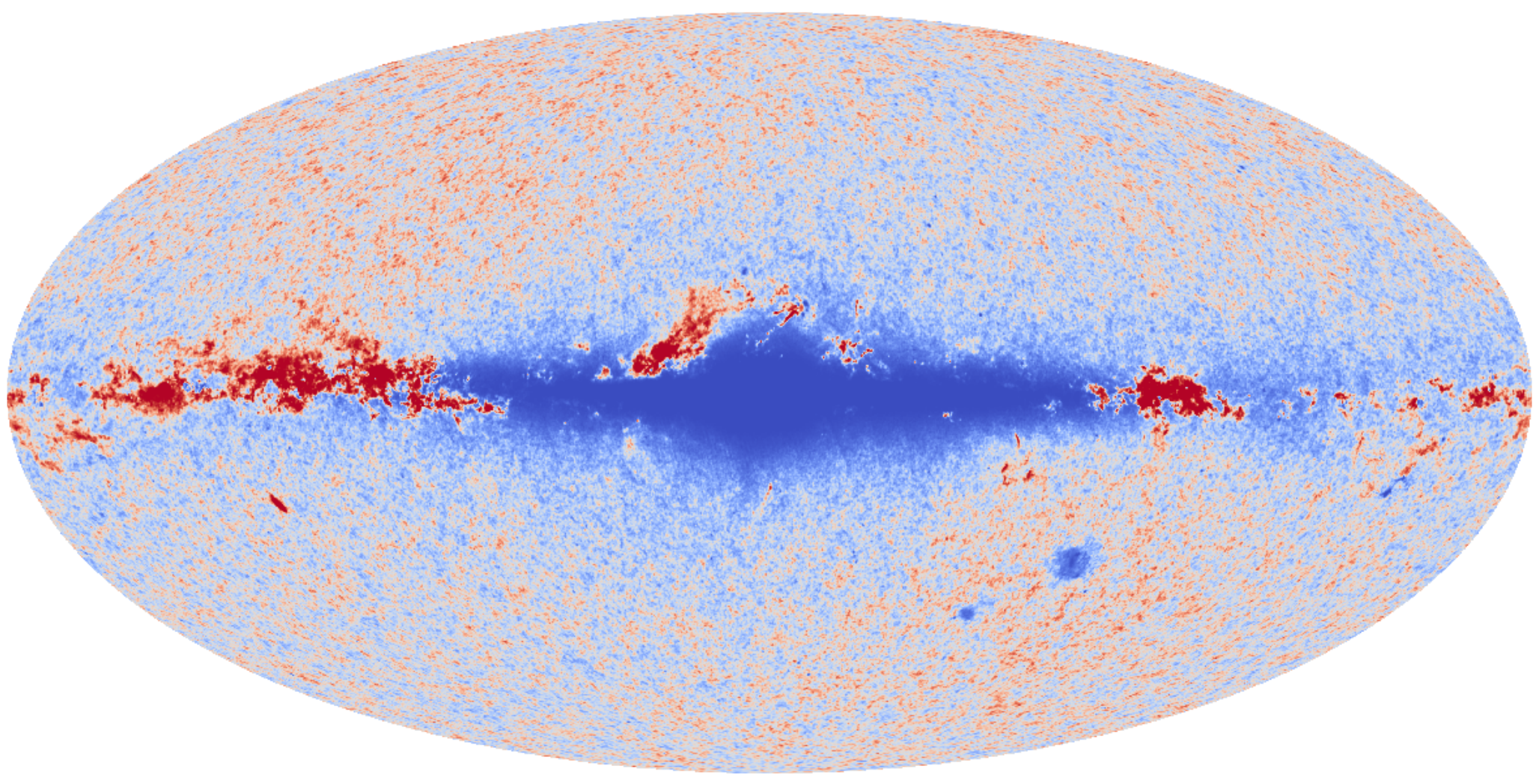}
    \caption{The unWISE blue sample galaxy density. Small density fluctuations are shown with a linear color scaling and large density fluctuations are logarithmically scaled to enhance the cosmological signal.}
    \label{fig:unwise_input}
\end{figure}

In Fig.~\ref{fig:unwise_input} we show the galaxy number density of the unWISE blue sample, defined from the number counts $N^{\mathrm{g}}(\hat{n})$ by
\begin{equation}
\delta^{\mathrm{g}} (\hat{n}) = (N^{\mathrm{g}}(\hat{n}) - \bar{N}^{\mathrm{g}})/\bar{N}^{\mathrm{g}} \ ,
\end{equation}
where $N^{\mathrm{g}}(\hat{n})$ is the number of objects per pixel and the mean is defined by $\bar{N}^{\mathrm{g}} \equiv N_{\rm tot}/N_{\rm pix}$ with $N_{\rm pix}$ the number of un-masked pixels at $N_{\rm side} = 2048$ resolution and $N_{\rm tot}$ the total number of objects in un-masked pixels. For the unWISE blue sample with the mask defined in the following subsection, this is $\bar{N}^{\mathrm{g}} = 2.8$, corresponding to a number density of $0.95 \ {\rm arcmin}^{-2}$. Visible in Fig.~\ref{fig:unwise_input} are large over- and under-densities concentrated along the galactic plane. The clustering signal is visible far from the galactic plane.

The unWISE blue sample is known to have significant correlations with foregrounds (e.g. Galactic dust, stellar density) and survey depth. A set of weights to correct for the most significant correlations were derived in Ref.~\cite{2023arXiv230905659F}; these weights are shown in Fig.~\ref{fig:unwise_mask}. The most significant correlations are with Gaia stellar density and the limiting magnitude in the W2 (4.6 $\mu$m) WISE band (determined by the survey strategy). In Sec.~\ref{sec:foregroundssystematics_template} we estimate systematics weights directly from the anisotropic cross-correlations between the PR3 857 GHz map and the unWISE blue sample. The two systematics weight maps are clearly highly correlated. Quantitatively, the correlation coefficient $r_\ell = C^{AB}_\ell/\sqrt{C^{AA}_\ell C^{BB}_\ell}$ is large at low-$\ell$, with $r_{\ell=1} = 0.98$, $r_{\ell=2} = 0.90$, $r_{\ell=3} = 0.64$, $r_{\ell=4} = 0.89$. We use these weights maps in our analysis below to explore and correct for the impact of survey systemtics on kSZ velocity reconstruction.

\subsection{Masking}\label{sec:masking}

\begin{figure}
    \centering
    \includegraphics[width=1.\columnwidth]{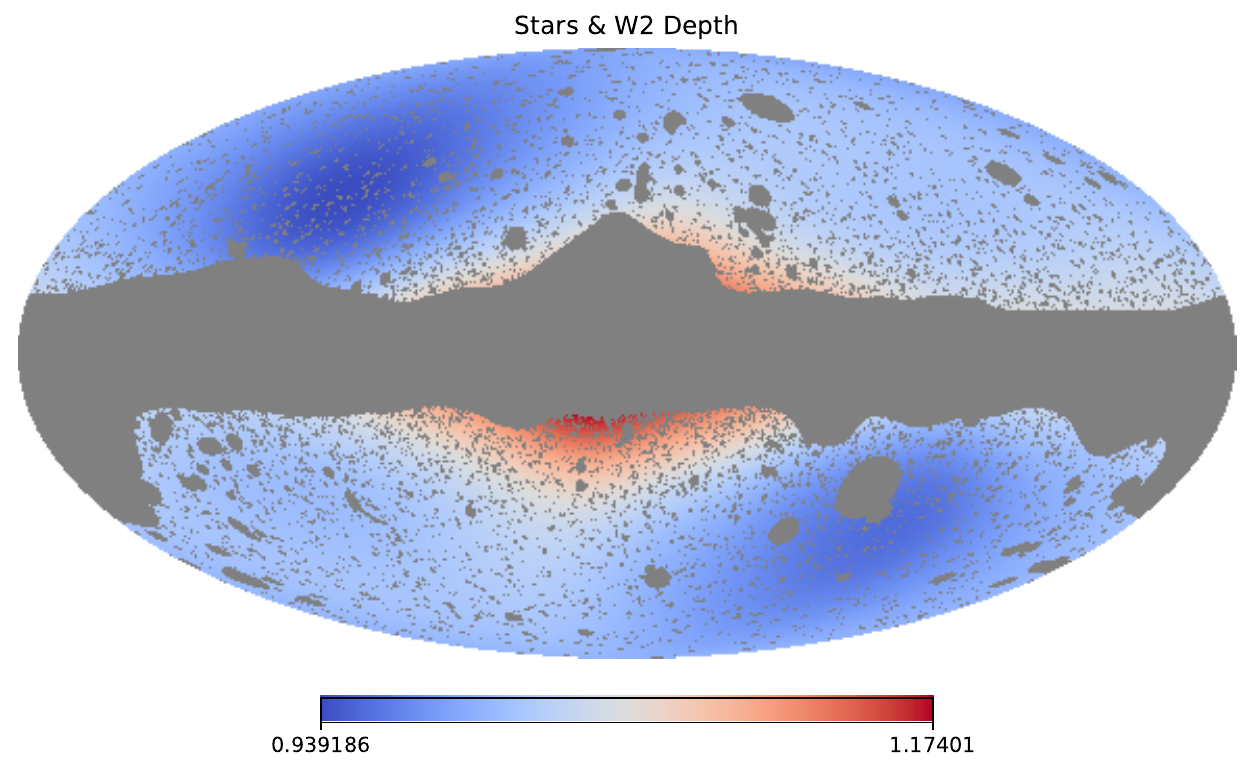}
    \includegraphics[width=1.\columnwidth]{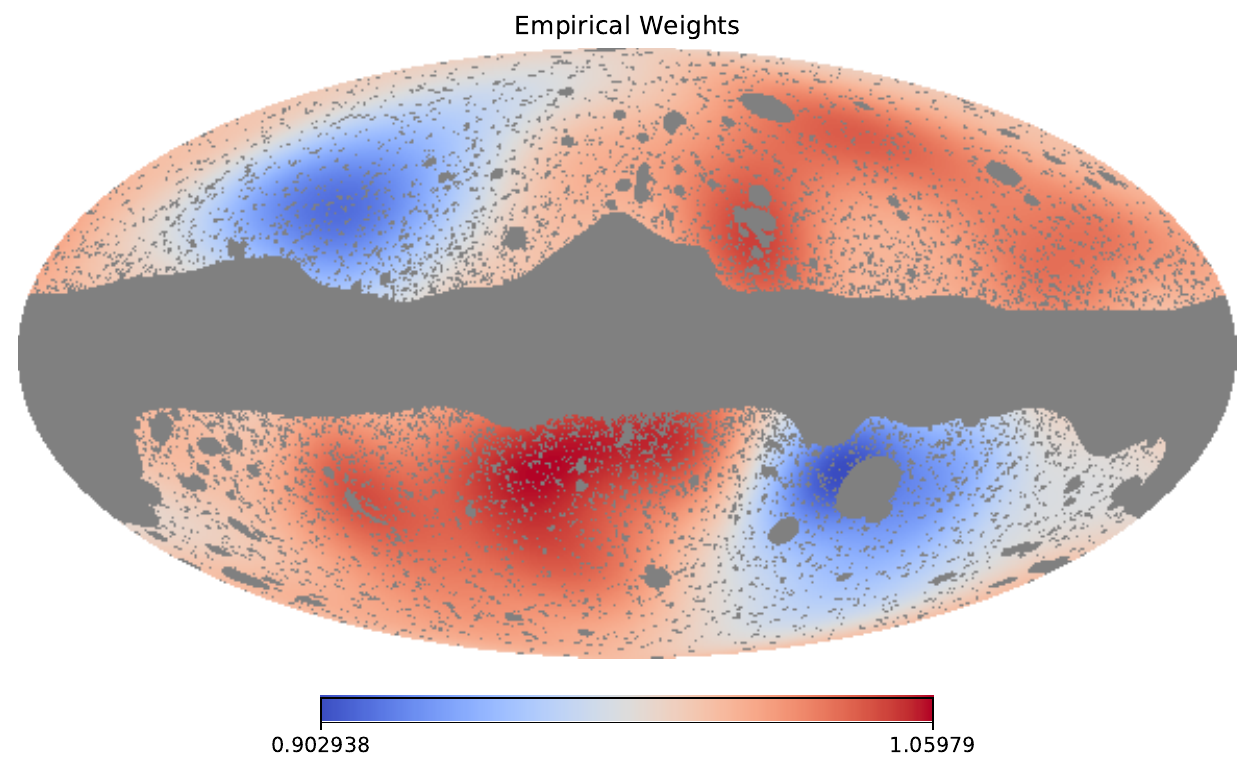}
    \caption{The reconstruction mask and two types of systematics weights applied to the unWISE map prior to reconstruction. The mask is the union of the binary unWISE mask from \cite{Krolewski2020} and the Planck CMB temperature confidence mask of Ref.~\cite{Akrami2020}. The final uncut sky fraction for this mask is $f_{\mathrm{sky}}=0.58$. On top, we show the weights derived in Ref.~\cite{2023arXiv230905659F} that correct for Gaia stellar density and the limiting magnitude in the W2 (4.6 $\mu$m) WISE band; see Ref.~\cite{2023arXiv230905659F} for futher details. On bottom, we show an estimate of systematic weights based on the statistically anisotropic cross-correlation of the PR3 857 GHz and unWISE blue maps; see Sec.~\ref{sec:foregroundssystematics_template} for details.}
    \label{fig:unwise_mask}
\end{figure}

To minimize the effect of a variety of galactic and extragalactic foregrounds in the Planck and unWISE maps, we employ a set of masks in our analysis. First, we must estimate the CMB power spectrum for use in the estimator weights. This computation is performed for individual frequency maps by applying the HFI point source mask and a galactic cut retaining $70 \%$ of the sky at 100 GHz and $60\%$ of the sky at other frequencies. For Planck component separated maps, we use the SMICA-based confidence mask~\cite{Akrami2020}. For the analysis of {\tt pyilc} component separarated maps, we use the union of the fiducual masks used in Ref.~\cite{McCarthy:2023hpa} and a $60\%$ galactic cut. Next, we estimate the galaxy power spectrum by applying the binary unWISE mask used in Ref.~\cite{Krolewski2020}, composed of a galactic plane cut retaining $70 \%$ of the sky as well as masking of stars, planetary nebulae, and bright sources. The resulting mask has an uncut sky fraction of $58\%$. For the analysis of reconstructions from the Planck component separated maps we create a mask which is the union of the unWISE mask and the SMICA-based confidence mask. This mask, shown in Fig.~\ref{fig:unwise_mask}, is applied to the final reconstructions to find power spectra. The sky coverage is nearly the same as the unWISE mask, also preserving $58\%$ of the sky. For the analysis of {\tt pyilc}-based reconstructions, we use the union of the fiducial {\tt pyilc} mask and the unWISE mask; this mask preserves $56\%$ of the sky.

\subsection{Modelling assumptions}
\label{ssec:spec_models}

A necessary input to the quadratic estimator is the optical depth-galaxy cross-power spectrum; see Eq.~\eqref{eq:exactcross}. This is not currently directly measured from existing datasets (though it may be in the future, e.g.~\cite{Madhavacheril:2019buy}), and so we must construct a model. Schematically, we connect unWISE galaxy number counts to electron density by modeling the relation of both to the underlying dark matter distribution.

A variety of previous works have attempted to constrain the relation between unWISE galaxies and dark matter, e.g.~\cite{Farren:2023yna,Krolewski:2021znk,Krolewski2020,Krolewski2021,Kusiak:2021hai,Kusiak:2022xkt,Yan:2023okq}. We adopt the simplest linear-bias model used to model unWISE galaxies described in Ref.~\cite{Krolewski2020}. The galaxy power spectrum is:
\begin{eqnarray}
C_\ell^{gg} &=& \int d\chi d\chi' \ W_\mathrm{g}(\chi) W_\mathrm{g}(\chi') C_{\ell}^{mm} (\chi, \chi') \nonumber \\
&+& N_{\mathrm{shot}} \ ,
\end{eqnarray}
where $C_{\ell}^{mm} (\chi, \chi')$ is the matter angular power spectrum
\begin{eqnarray}
    C_{\ell}^{mm} (\chi, \chi') = \frac{2}{\pi} \int \frac{dk}{k} \Delta_\ell^g (k,\chi) \Delta_\ell^g (k,\chi') \mathcal{P}(k) \ ,
\end{eqnarray}
with
\begin{equation}\label{eq:gsource}
    \Delta_\ell^g (k,\chi') = S^m (k,\chi) j_\ell (k \chi) \ ,
\end{equation}
where $S^m (k,\chi)$ is the source function for matter density. We neglect sub-dominant contributions from redshift space distortions and magnification that are relevant mostly on large angular scales. The shot noise for the unWISE blue sample is $N_\mathrm{shot}=9.2\times10^{-8}$ (steradians) and the galaxy window function $W_\mathrm{g}$ is defined by:
\begin{equation}
    W_\mathrm{g} \left(\chi\right) \equiv b_\mathrm{g}\left(z\right) \frac{dN}{dz} H\left(z\right),
    \qquad
    b_\mathrm{g}\left(z\right) \equiv 0.8 + 1.2z \ .
    \label{eq:galaxy_window}
\end{equation}
Here, $b_\mathrm{g}\left(z\right)$ is the galaxy bias and $dN/dz$ is the redshift distribution; both were  empirically determined for the unWISE blue sample in Ref.~\cite{Krolewski2020}. The redshift distribution $dN/dz$ was determined from matching sources with deep photometric redshifts in COSMOS~\cite{Laigle:2016jxn}; the galaxy bias $b(z)$ was determined through cross-correlation with SDSS spectroscopic galaxies in narrow bins. 
Within our model, the clustering signal dominates over shot noise for $\ell \lesssim 10^3$. 

\begin{figure}
    \centering
    \includegraphics[width=1.\columnwidth]{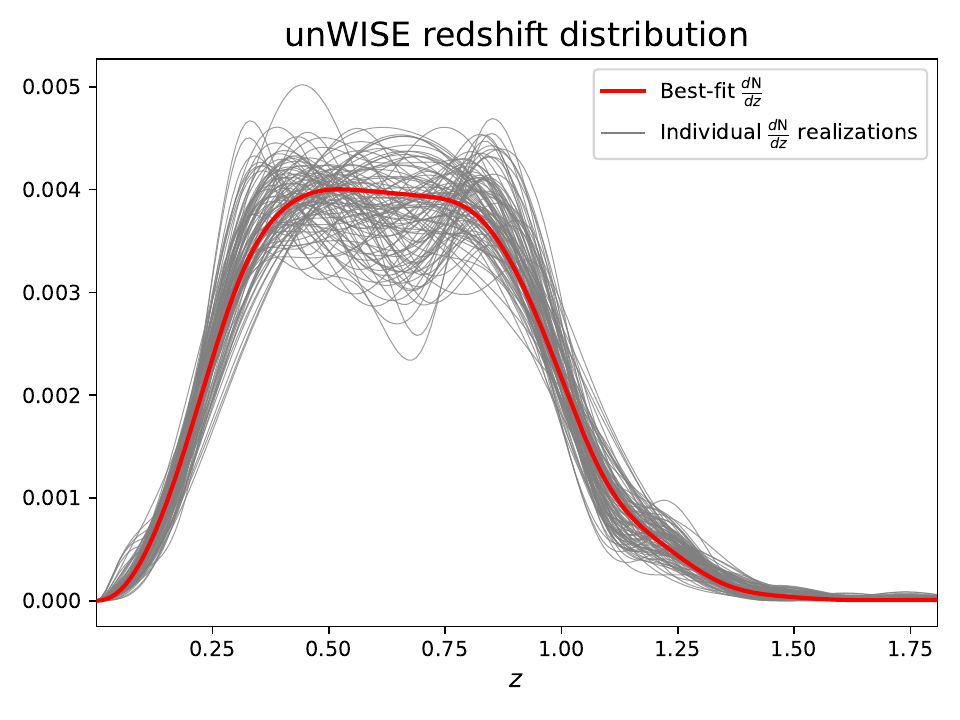}
    \caption{Normalized redshift distribution $\frac{d\mathrm{N}}{dz}$ for the unWISE blue sample. The thin grey lines are 100 individual samples of $\frac{d\mathrm{N}}{dz}$ consistent with the expected error. The thick red line indicates the measured  $\frac{d\mathrm{N}}{dz}$, taken to be the fiducial redshift distribution used in our analysis.}
    \label{fig:dndz}
\end{figure}

Uncertainty in the redshift distribution $\frac{d\mathrm{N}}{dz}$ for the unWISE blue sample is an important systematic in our analysis. In Ref.~\cite{Krolewski2020} this was quantified by determining the variance in the redshift distribution over 44 different patches observed by HSC, each with the same area as COSMOS, and then drawing 100 samples consistent with this expected error. The 100 $\frac{d\mathrm{N}}{dz}$ realizations used in Ref.~\cite{Krolewski2020} are shown in Fig.~\ref{fig:dndz} as thin grey lines, with the thick red line indicating the best-fit fiducial  $\frac{d\mathrm{N}}{dz}$. This spread illustrates the degree of uncertainty in the unWISE blue redshift distribution, and will be important in the discussion below.

The differential optical depth is proportional to the inhomogeneous distributions of electrons. We relate this to the dark matter distribution through a scale-dependent linear bias $\delta_e (\vec{k},\chi) = b_e (k, \chi) \delta_m (\vec{k},\chi)$, employing the model~\footnote{Various window functions relating the distribution of electrons to matter have been employed in previous work, e.g.~\cite{Shaw:2011sy} used a combined exponential and power law while ~\cite{Smith:2016lnt} used Eq.~\eqref{eq:matter_electron_bias} with $b_* = {\rm const}$, $\gamma = 1$, $k_* = \left(\alpha h / D(z)\right) \ {\rm Mpc}^{-1}$. All of these fitting functions share the qualitative feature that $b_e$ falls off beyond some wave-number, corresponding to the impact of baryonic feedback on the distribution of gas.}:
\begin{equation}
    b_e \left(z, k \right) = b_\star \left( z\right)\left[1 + \left(\frac{k}{k_\star\left(z\right)}\right)^{\gamma\left(z\right)}\right]^{-\frac{1}{2}} \ ,
    \label{eq:matter_electron_bias}
\end{equation}
where we use the fit to the IllustrisTNG simulations~\cite{Springel:2017tpz} from Ref.~\cite{Takahashi2020} as our fiducial model
\begin{align}
    b_\star \left(z\right) &= \sqrt{- 0.013z + 0.971} \ ,  \nonumber  \\
    \gamma\left(z\right) &= 0.10z^2 - 0.59z + 1.91 \ , \\
    k_\star \left( z\right) &= \left(-0.42z^3 +3.10z^2 -3.24z +4.36\right) {\rm Mpc}^{-1} \ . \nonumber 
\end{align}
Heuristically, $b_\star$ controls the redshift-dependence of the amplitude, $k_\star$ controls the scale on which electron inhomogeneities are suppressed as compared to dark matter, and $\gamma$ controls the abruptness of this transition. This model differs somewhat from the halo model-based approach introduced in Ref.~\cite{Smith2018}, which predicts (for the fiducial 'Battaglia profiles' for gas in halos) less power suppression compared to dark matter than our fiducal model. We further contrast these models in Appendix~\ref{appendix:optical depth}. 

In the Limber approximation, the cross-power is
\begin{eqnarray}\label{eq:ctaugmodel}
    C_\ell^{\dot{\tau}\mathrm{g}} (\chi) &=& \frac{1}{\chi^2} W_\mathrm{g} \left( \chi \right) W_\tau\left( \chi \right) b_e \left(\chi,k=\frac{\ell+\frac{1}{2}}{\chi}\right) \nonumber \\ &\times& P_{mm}\left(\chi,k=\frac{\ell+\frac{1}{2}}{\chi} \right)  \ .
\end{eqnarray}
$W_\tau$ is the optical depth window function:
\begin{equation}
    W_\tau \left(\chi\right) \equiv \sigma_\mathrm{T}  \bar{n}_{e,0} (1+z\left(\chi\right))^{2}\ , 
\end{equation}
where $\sigma_\mathrm{T}$ is the Thomson cross section, $a\left(\chi\right)$ is the scale factor, and $\bar{n}_{e,0}$ is the average number density of electrons today. We model $\bar{n}_{e,0}$ as
\begin{equation}\label{eq:ne0}
\bar{n}_{e,0} = \frac{f_{\rm gas} X \Omega_{b,0} \rho_{\rm crit,0}}{\mu_e m_p} \ ,
\end{equation}
where $f_{\rm gas}$ is the mass fraction of baryons in ionized gas, $X$ is the fraction of the total number of electrons that are ionized, $\mu_e m_p$ is the mean baryon mass per electron, $\Omega_{b,0}$ is the present-day baryon density parameter, and $\rho_{\rm crit,0}$ is the present-day critical density. Assuming a primordial helium abundance of $Y_p = 0.24$, and assuming that all helium is doubly ionized within the redshifts probed by unWISE (helium reionization is expected to have happened at $z \agt 2.5$ near the peak of quasar activity~\cite{Calura_2012,Viel_2013,Boera_2015,Upton_Sanderbeck_2016,La_Plante_2017,Plante_2018,Bolton_2016}), we have $X = 1$~\footnote{The fraction of ionized electrons is defined as
\begin{equation}
X \equiv \frac{1-Y_p(1-N_{\rm He}/4)}{1-Y_p/2} \ ,
\end{equation}
where $N_{\rm He} = 0, 1, 2$ for neutral, singly-ionized, and doubly-ionized helium. For a primordial helium abundance of $Y_p = 0.24$, this takes values $X = 0.86, 0.93, 1.0$ for these three ionization states, respectively.} and $\mu_e = 1.14$. We assume that $10 \%$ of baryonic matter by mass is cold (neutral) or bound up in stars, yielding $f_{\rm gas} = 0.9$. We do not consider redshift-evolution of the ionization state of Helium in the intergalactic medium or the ionized gas fraction over cosmological epochs probed by the  unWISE blue sample~\footnote{With future surveys covering a broader range of redshifts, it will be possible to measure Helium reionization using kSZ tomography~\cite{Hotinli:2022jna,Hotinli:2022jnt,Caliskan:2023yov}.}. With our fiducial cosmological parameters, we have $\sigma_T \bar{n}_{e,0} \simeq 4.08\times 10^{-7} \ {\rm Mpc}^{-1}$. 

Uncertainty in the distribution of electrons, here quantified by the scale-dependent bias $b_e(z,k)$, is an important systematic to consider. Within our assumed model, $b_e$ is close to unity on large physical/angular scales, where baryons trace the underlying distribution of dark matter. Baryonic feedback effects become relevant on physical scales $k > k_*$, washing out baryon fluctuations; this is modeled through a decrease in $b_e(z,k)$. At the median redshfit of the unWISE blue sample $\bar{z} \simeq 0.6$, this transition corresponds to physical scales $k \sim 1 \ {\rm Mpc}^{-1}$ and angular scales $\ell \agt k \chi \sim 2300$. At this angular scale there is still significant signal-to-noise in the Planck CMB and the clustering signal in unWISE, implying that there will be some sensitivity to variations about the fiducial model.

\begin{figure}
    \centering
    \includegraphics[width=1.\columnwidth]{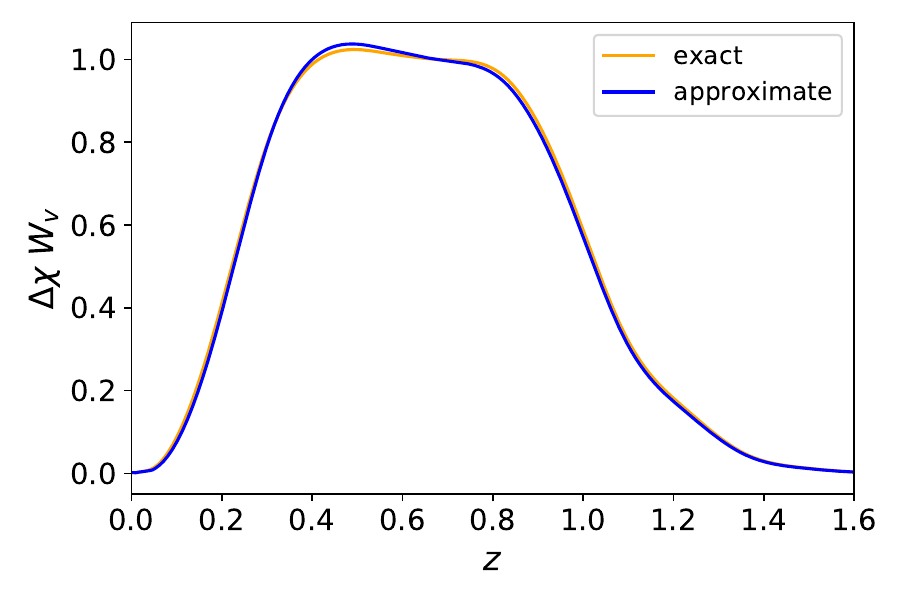}
    \caption{The velocity window function $W_v (z)$ Eq.~\eqref{eq:vwindow} (orange solid) relating the estimator mean to the underlying velocities compared with the approximate window function $W_v \simeq C_{\ell=\bar{\ell}}^{\dot{\tau} g}\left(\chi\right)/\bar{C}_{\ell=\bar{\ell}}^{\tau g}$ used in our analysis (blue).}
    \label{fig:W_v}
\end{figure}

As described in Sec.~\ref{sec:methods}, the estimator mean has the simplest interpretation when we can factorize the scale- and redshift-dependence of the optical depth-galaxy cross-power. This allows us to approximate $C_\ell^{\dot{\tau}\mathrm{g}} (\chi) \simeq \bar{C}_\ell^{\tau \mathrm{g}} \ (C_{\ell=\bar{\ell}}^{\dot{\tau}\mathrm{g}} (\chi) / \bar{C}_{\ell = \bar{\ell}}^{\tau \mathrm{g}} ) $ on small angular scales where the estimator receives the greatest contributions. 
In Fig.~\ref{fig:W_v} we plot the velocity window function computed using the approximate expression $W_v \simeq C_{\ell=\bar{\ell}}^{\dot{\tau} g}\left(\chi\right)/\bar{C}_{\ell=\bar{\ell}}^{\tau g}$ (blue) against the exact expression Eq.~\eqref{eq:vwindow} (orange). The agreement is excellent, demonstrating that the approximations used in the derivation of the estimator are accurate within this model. We comment on bias induced by mis-modeling in Sec.~\ref{sec:opticaldeptbias}. The distribution is relatively flat over the redshift range $0.3 \lesssim z \lesssim 0.9$, and is very nearly given by the normalized redshift distribution shown in Fig.~\ref{fig:dndz}.

\subsection{Predicted estimator variance}\label{sec:predictedvariance}

\begin{figure}
    \centering
    \includegraphics[width=1.\columnwidth]{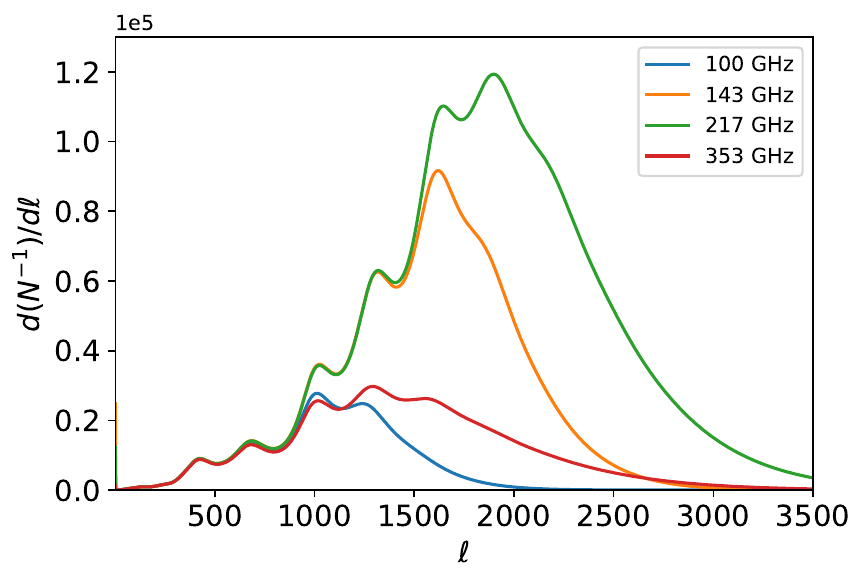}
    \caption{The predicted summand in Eq.~\eqref{eq:const_est_n} for reconstructions from individual frequency maps. The range in $\ell$ where the summand is largest determines the scales that make the largest contribution to the estimator variance.}
    \label{fig:Nsummand}
\end{figure}

We now have everything necessary to compute the estimator variance defined in Eq.~\eqref{eq:est_harm_power}. We can estimate the expected level of reconstruction noise from   Eq.~\eqref{eq:const_est_n} using the fiducial model for $\bar{C}_\ell^{\tau g}$ and $C_\ell^{gg}$ described above and estimating $C_\ell^{TT}$ as the the sum of the primary CMB and an effective white noise level of $77.4$, $33.0$, $46.8$, $153.6$ $\mu$K-arcmin for the $100$, $143$, $217$, and $353$ GHz channels respectively. In Fig.~\ref{fig:Nsummand} we show the resulting summand in Eq.~\eqref{eq:const_est_n}. The larger the summand, the smaller the reconstruction noise. The multipoles over which the summand is significant determines which scales contribute most to the estimator variance. From this plot, we see that the $217$ GHz map is expected to yield the lowest reconstruction noise (before incorporating foregrounds) and that scales $\ell \simeq 2000$ are most relevant to the reconstruction. This motivates our choice of $\bar{\ell} = 2000$ as the reference scale for our model of the galaxy-optical depth power spectrum. 

\begin{figure}
    \centering
    \includegraphics[width=1.\columnwidth]{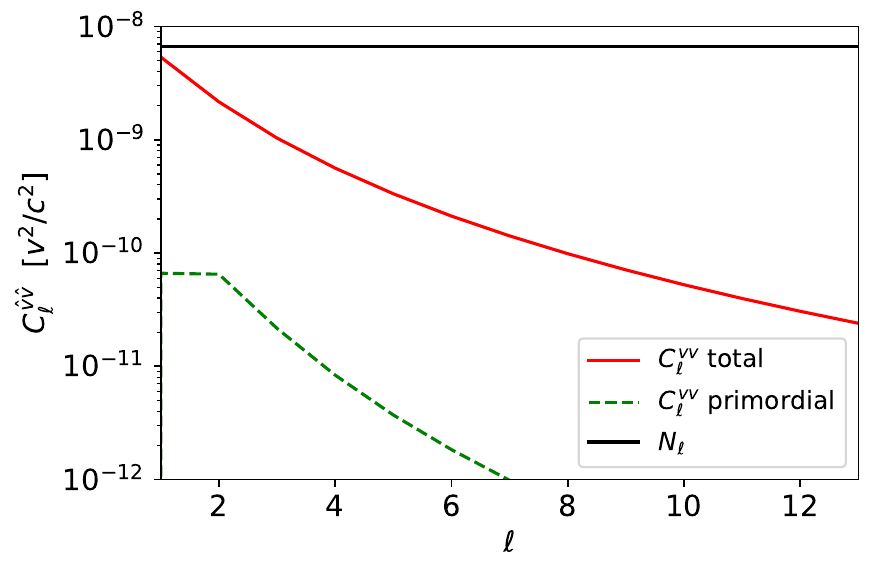}
    \caption{The predicted signal and noise power spectra for the reconstructed remote dipole field. The full remote dipole spectrum (red solid) is several orders of magnitude larger than the spectrum associated with the primordial dipole (green dashed); see Eq.~\eqref{eq:vsources}. The expected reconstruction noise for the 217 GHz map (black solid) is comparable to the expected signal on the largest angular scales.}
    \label{fig:SNR}
\end{figure}

We compare the reconstruction noise expected for the $217$ GHz Planck map against the expected signal contributions in Fig.~\ref{fig:SNR}. The reconstruction noise computed as described above (black solid) is comparable to the total predicted signal (red solid, computed using Eq.~\eqref{eq:est_harm_power} with the fiducial velocity window function shown in Fig.~\ref{fig:W_v}) at the very lowest $\ell$, falling steeply with $\ell$. The primordial component (green dashed; computed using the primordial dipole source term in Eq.~\eqref{eq:vsources}) does not significantly contribute to the predicted signal for this data combination~\footnote{The suppression of power at $\ell = 1$ is due to the cancellation of contributions to the locally observed CMB dipole that occurs for long-wavelength adiabatic modes. See Refs.~\cite{Erickcek:2008jp,Terrana2017} for a detailed discussion.}. The expected total signal-to-noise of the map-level reconstruction defined as:
\begin{equation}\label{eq:Sndefinition}
    {\rm SN}^2 = \sum_\ell \frac{2 \ell + 1}{2} f_{\rm sky} \left( \frac{C_{\ell}^{vv}}{N_\ell} \right)^2 \ ,
\end{equation}
is ${\rm SN} = 0.89$, with most of the contribution coming from $\ell < 5$ (and roughly half from $\ell=1$). Note that incorporating the mask as a factor of $f_{\rm sky}$ is likely inaccurate at such large angular scales (see e.g. Ref.~\cite{Alizadeh_2012} for a related discussion). Additionally, this forecast does not incorporate any degradation from foregrounds and systematics. Therefore, we do not anticipate a conclusive detection for this data combination within $\Lambda$CDM. Note however that the statistical reach of the quadratic estimator for bulk radial velocities on Gpc scales is an impressive $\sqrt{N} \sim 25 \ {\rm km/s}$! Future CMB experiments such as Simons Observatory have the statistical power to achieve ${\rm SN} >  5$ in cross-correlation with the unWISE blue sample (and ${\rm SN} >  100$ in combination with LSST)~\cite{Cayuso2023}; detection of the primordial dipole signal is possible in this high SN regime~\cite{Contreras:2019bxy}.  

\subsection{Optical depth bias}\label{sec:opticaldeptbias}

Uncertainties in the modelling choices used to construct $C_\ell^{\dot{\tau} g}$ appear as a bias on the estimator mean and variance, known as the optical depth bias (see Sec.~\ref{sec:possible_systematics}). We present a detailed computation and assessment of the optical depth bias in  Appendix~\ref{appendix:optical depth}, collecting the main results here. In general, we can define the optical depth bias by the estimator mean evaluated using the true temperature-galaxy cross-correlation, but with fiducial estimator weights:
\begin{equation}
    \langle \hat{v}_{\ell m}\rangle^{\rm t} = \int d\chi \ b_v (\chi) W_v (\chi) v_{\ell m} (\chi) \ ,
\end{equation}
where the `t' superscript indicates this is evaluated on the `truth' values for the temperature-galaxy cross-correlation. In Appendix~\ref{appendix:optical depth}, we  demonstrate that this can be approximated by
\begin{eqnarray}\label{eq:bv_definition}
    b_v (\chi) 
    &\simeq& \frac{ \sum_{\ell_1} \frac{2 \ell_1 + 1}{4 \pi} \frac{\bar{C}_{\ell_1}^{\tau g}  [\bar{C}_{\ell_1}^{\tau g}]^{\rm t}}{C_{\ell_1}^{TT} C_{\ell_1}^{gg}}  }{  \sum_{\ell_2} \frac{2 \ell_2 + 1}{4 \pi} \frac{(\bar{C}_{\ell_2}^{\tau g})^2  }{C_{\ell_2}^{TT} C_{\ell_2}^{gg}}}
    \frac{[C_{\ell=\bar{\ell}}^{\dot{\tau} g}\left(\chi\right)]^{\rm t}}{C_{\ell=\bar{\ell}}^{\dot{\tau} g}\left(\chi\right)} \frac{\bar{C}_{\ell=\bar{\ell}}^{\tau g}}{[\bar{C}_{\ell=\bar{\ell}}^{\tau g}]^{\rm t}} \ .
\end{eqnarray}
Given a range of possible models for the 'truth,' we can assess the range of $b_v$ we might plausibly expect. Note that in general the optical depth bias is a $\chi-$dependent function. However, on large angular scales it manifests as a multiplicative constant relating the reconstruction and the true velocity. We define $b_v$ without the explicit $\chi$ dependence as this constant multiplicative factor:
\begin{equation}
    \langle \hat{v}_{\ell m}\rangle^{\rm t} \simeq b_v \langle \hat{v}_{\ell m}\rangle, \ \ [C_{\ell}^{\hat{v} \hat{v}}]^{\rm t} \simeq b_v^2 C_{\ell}^{\hat{v} \hat{v}} \ .
\end{equation}
Note that the reconstruction noise $N$ is defined by the fiducial model in the estimator, so using the reconstruction noise to place limits on any hypothetical underlying signal requires an understanding of $b_v$. For example, the total expected signal-to-noise defined in Eq.~\eqref{eq:Sndefinition} scales as $[SN]^{\rm t} \simeq b_v^2 SN$. This implies that if we under-estimate the strength of feedback, corresponding to $b_v < 1$, we overestimate the SN. 

In Appendix~\ref{appendix:optical depth} we estimate the range of values that $b_v$ could plausibly take by varying the unWISE photometric redshift distribution (see Fig.~\ref{fig:dndz}), the model parameters determining $b_e$, and the mean number density of electrons. We find that all three of these uncertainties can individually contribute to the optical depth bias at the $\mathcal{O}(10\%)$-level. Under the variations we consider, we find that it is difficult to increase $b_v$ beyond $b_v \sim 1.1$. Under the largest variations we consider in the scale and abruptness of electron power suppression determining $b_e$, we can plausibly obtain values of $b_v$ as small as $b_v \sim 0.5$. Under the assumption that our model is flexible enough to encompass the true underlying spectrum, we expect that $b_v$ lies in the range $0.5 \lesssim b_v \lesssim 1.1$. Although we do not pursue it here, we note that it is in principle possible to derive a prior on $b_v$ using a variety of measurements and upper-limits on e.g. the kSZ power spectrum~\cite{2012ApJ...755...70R}, pairwise velocity~\cite{Hand2012} or projected field~\cite{Kusiak:2021hai,Hill:2016dta,Ferraro:2016ymw} kSZ estimators, numerical simulations including baryonic feedback, etc. Future analyses with more precisely calibrated photometric or spectroscopic redshifts will also mitigate contributions to the optical depth bias from uncertainty in the redshift distribution. 

Finally, we note that the numerical value of $b_v$ depends on the data sets used in an analysis. This is because $b_v$ as defined in Eq.~\eqref{eq:bv_definition} depends on $C_\ell^{TT}$ and $C_\ell^{gg}$. These spectra depend on the noise and beam of the CMB map as well as the bias and shot noise associated with the galaxy sample. Therefore, even if we fix $\bar{C}_{\ell}^{\tau g}$ and $[\bar{C}_{\ell}^{\tau g}]^{\rm t}$, different data combinations have different expected $b_v$. Care should be taken when comparing kSZ velocity reconstruction analyses based on different data combinations.

\subsection{Galaxy-velocity reconstruction cross-correlation}\label{eq:cross_recon_theory}

\begin{figure}
    \centering
    \includegraphics[width=1.\columnwidth]{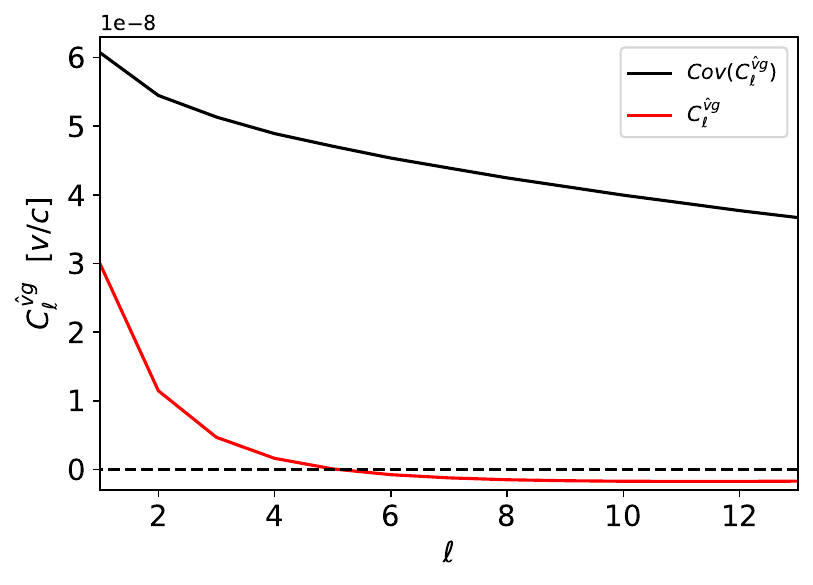}
    \caption{The predicted cross-correlation between the galaxy density and estimator mean (red solid; see Eq.~\eqref{eq:vgtheory}) and the expected covariance based on the theoretical reconstruction and galaxy auto spectra (black solid; see Eq.~\ref{eq:vgnoise}).}
    \label{fig:gv}
\end{figure}

The reconstructed remote dipole field is correlated with the galaxy density on large angular scales. This cross-correlation is equivalent to the squeezed limit of the temperature-galaxy-galaxy bispectrum~\cite{Smith2018} (e.g. $\langle T_S \delta^g_S \delta^g_L \rangle $ where $S$ denotes small angular scales and $L$ denotes large angular scales). We can estimate the predicted signal from our theoretical models for the estimator mean and the galaxy density
\begin{equation}\label{eq:vgtheory}
C_\ell^{\hat{v} g} = \int d\chi \int d\chi' W_v (\chi) W_g (\chi') C_{\ell}^{v g} (\chi,\chi') \ .  
\end{equation}
The galaxy-velocity cross-correlation is
\begin{equation}
    C_{\ell}^{v g} (\chi,\chi') = \frac{2}{\pi} \int \frac{dk}{k} \Delta^v_\ell (k,\chi) \Delta^g_\ell (k,\chi') \mathcal{P}(k) \ ,
\end{equation}
where $\Delta^v_\ell$ and $\Delta^g_\ell$ are defined in Eq.~\eqref{eq:vsources} and~\eqref{eq:gsource} respectively. This is shown in Fig.~\ref{fig:gv}. There is a significant positive correlation for $\ell < 5$, and a slowly decreasing anti-correlation for $\ell > 10$. The expected variance on the cross-spectrum based on the theoretical model for the galaxy power spectrum and remote dipole estimator variance is 
\begin{eqnarray}\label{eq:vgnoise}
    \langle (C_\ell^{\hat{v} g})^2 \rangle = \frac{C_{\ell}^{gg} C_{\ell}^{\hat{v} \hat{v}}}{2 \ell + 1}  \ .
\end{eqnarray}
We discuss this in detail in the next subsection. The variance is roughly twice the expected signal at $\ell = 1$ and more than an order of magnitude larger at $\ell \sim 10$, with the ratio growing roughly linearly with $\ell$ thereafter. We therefore do not expect a detection of a signal in the cross-power for Planck and unWISE. 

\subsection{Likelihood and posterior}\label{sec:likelihoodandposterior}

Assuming that the reconstructed dipole field and unWISE galaxy density are Gaussian random fields (a good approximation on large angular scales), the likelihood for the observed spectra $\mathbf{\hat{C}}_\ell$ given theory spectra $\mathbf{C}_\ell$ is at each $\ell$ given by a Wishart distribution:
\begin{align}\label{eq:Wishart}
    p(\mathbf{\hat{C}}_\ell | \mathbf{C}_\ell ) \propto &\frac{\left[{\rm det}(\mathbf{\hat{C}}_\ell) \right]^{(\nu-3)/2}}{\left[{\rm det}(\mathbf{C}_\ell) \right]^{\nu/2}} \\
    & \times \exp\left[ - \frac{\nu}{2} {\rm Tr} \left(\mathbf{C}_\ell^{-1} \cdot \mathbf{\hat{C}}_\ell \right) \right] \ , \nonumber
\end{align}
where $\nu \equiv 2 \ell + 1$, the measured spectra are assembled into the matrix
\begin{equation}
    \mathbf{\hat{C}}_\ell = 
\begin{pmatrix}
\hat{C}_\ell^{\hat{v} \hat{v}} &  \hat{C}_\ell^{\hat{v} g} \\
\hat{C}_\ell^{\hat{v} g} & \hat{C}_\ell^{g g}
\end{pmatrix} \ ,
\end{equation}
and the theory spectra are assembled into 
\begin{equation}
    \mathbf{C}_\ell = 
\begin{pmatrix}
C_\ell^{\hat{v} \hat{v}} & C_\ell^{\hat{v} g} \\
C_\ell^{\hat{v} g} & C_\ell^{g g}
\end{pmatrix} \ .
\end{equation}
For large-$\ell$, the Wishart distribution approaches a multi-variate Gaussian distribution over the spectra. By marginalizing over $\hat{C}_\ell^{g g}$ and $\hat{C}_\ell^{\hat{v} g}$ (e.g. discluding these observables from our data vector) we obtain the likelihood for $\hat{C}_\ell^{\hat{v} \hat{v}}$ which is a Gamma function:
\begin{equation}\label{eq:cvvlikelihood}
    p(\hat{C}_\ell^{\hat{v} \hat{v}} | C_\ell^{\hat{v} \hat{v}}) = \frac{\left( \nu/2\right)^{\nu/2}}{C_\ell^{\hat{v} \hat{v}} \Gamma(\nu/2)}   \left(\frac{\hat{C}^{\hat{v} \hat{v}} }{C_\ell^{\hat{v} \hat{v}}}\right)^{\nu/2} \exp \left[ -\frac{\nu \hat{C}_\ell^{\hat{v} \hat{v}}}{2 C_\ell^{\hat{v} \hat{v}}}\right] \ .
\end{equation}
The mean is $C_\ell^{\hat{v} \hat{v}}$ and the variance $2 {C_\ell^{\hat{v} \hat{v}}}^2 / \nu$; at high $\ell$ this approaches a Gaussian with this mean and variance. We will also be interested below in the likelihood over $\hat{C}_\ell^{\hat{v} g}$ given a set of theory spectra. Marginalizing Eq.~\eqref{eq:Wishart} over $\hat{C}_\ell^{\hat{v} \hat{v}}$ and $\hat{C}_\ell^{g g}$ we have
\begin{eqnarray}\label{eq:cross_likelihood}
    p(\hat{C}_\ell^{\hat{v} g} | \mathbf{C}_\ell) &=& \frac{\nu 2^{(1-\nu)/2}}{\sqrt{\pi} \Gamma(\nu/2) [{\rm det}(\mathbf{C}_\ell)]^{1/2}} \left[\frac{(\nu \hat{C}_\ell^{\hat{v} g})^2}{C_\ell^{\hat{v} \hat{v}} C_\ell^{gg}}  \right]^{(\nu-1)/4} \\
    &\times& \exp\left[ \frac{\nu \hat{C}_\ell^{\hat{v} g} C_\ell^{\hat{v} g}}{{\rm det}(\mathbf{C}_\ell)}\right] K_{\frac{\nu-1}{2}} \left[ \frac{\nu |\hat{C}_\ell^{\hat{v} g}| \sqrt{C_\ell^{\hat{v} \hat{v}} C_\ell^{gg}} }{{\rm det}(\mathbf{C}_\ell)}\right] \ . \nonumber
\end{eqnarray}
The mean is $C_\ell^{\hat{v} g}$ and the variance for small $C_\ell^{\hat{v} g}$ is approximately equal to $ C_\ell^{\hat{v} \hat{v}} C_\ell^{g g} / \nu$; at high $\ell$ this approaches a Gaussian with this mean and variance. 

On the full sky, the spectra at different multipoles $\ell$ are independent, and the joint likelihood for the full spectrum can be constructed by simply multiplying the likelihood at each $\ell$:
\begin{equation}\label{eq:full_likelihood}
    p( \mathbf{\hat{C}} | \mathbf{C}) \propto \prod_{\ell = \ell_{\rm min}}^{\ell_{\rm max}} p(\mathbf{\hat{C}}_\ell|\mathbf{C}_\ell ) \ .
\end{equation}
Below, we must introduce sky cuts to mitigate foreground contamination in the reconstruction and unWISE galaxy density. This is a non-trivial component of the analysis because one must determine how to account for the change in variance due to missing information (changing $\mathbf{C}_\ell$) as well as the fact that spectra measured on the masked sky are no longer statistically independent (violating the assumption that led to  Eq.~\eqref{eq:full_likelihood}). In general, one employs pseudo-$C_\ell$ (PCL) or quadratic maximum likelihood (QML) estimates of the full-sky spectra to contend with this. This is required in a full analysis of the velocity reconstruction and the galaxy density, including their covariance. However, if we focus on the velocity reconstruction alone, this heavy machinery is not necessary. This is because from the analysis in the previous sections we expect the auto-spectrum of the reconstruction ($\hat{C}_{\ell}^{\hat{v} \hat{v}}$) to be dominated by reconstruction noise. For a constant noise power spectrum, the cut-sky spectra at different $\ell$ remain statistically independent and follow the same distribution as the full-sky spectra, 
but with $C_\ell^{\hat{v}\hat{v}} \rightarrow f_{\rm sky} C_\ell^{\hat{v}\hat{v}}$ in Eq.~\eqref{eq:cvvlikelihood}. 

We restrict the scope of our analysis here to finding an upper limit on the velocity bias $b_v$ from the velocity reconstruction auto-power spectrum. We derive this from measurements of the cut-sky power spectrum $\hat{C}^{\hat{v} \hat{v}; C}$ using
\begin{equation}\label{eq:bvposterior}
    p( b_v | \hat{C}^{\hat{v} \hat{v}; C} ) = A \prod_{\ell = \ell_{\rm min}}^{\ell_{\rm max}} p(\hat{C}_\ell^{\hat{v} \hat{v}; C} | C_\ell^{\hat{v} \hat{v}; C} (b_v) ) \ ,
\end{equation} 
We assume a flat prior over $b_v$, and fix the normalization constant $A$ by integrating over $b_v$. The likelihood function $p( b_v | \hat{C}^{\hat{v} \hat{v}; C} )$ is defined in Eq.~\eqref{eq:cvvlikelihood}. We model the full-sky spectrum as
\begin{equation}\label{eq:reconmodel}
    C_\ell^{\hat{v} \hat{v}; C} = b_v^2 C_\ell^{vv; C} + f_{\rm sky} N  \ .
\end{equation}
The reconstruction noise $N$ is fixed by Eq.~\eqref{eq:const_est_n} for each data combination. We compute $C_\ell^{vv; C}$ from an ensemble of masked Gaussian simulations produced using the fiducial set of cosmological parameters. For the datasets and masks considered here $C_\ell^{vv; C}\sim f_{\rm sky} C_\ell^{vv}$  for $\ell \agt 4$; at lower $\ell$ there is some additional suppression of the masked spectrum.

In Appendix~\ref{appendix:pdftest}, we validate this approach using simulations. Specifically, we demonstrate that the cut-sky spectra are very nearly  statistically independent for the data combinations we consider and that the distribution of cut-sky spectra is as assumed in Eq.~\eqref{eq:bvposterior}. These simulations indicate that (for a typical realization) Planck and unWISE can only set an upper limit on $b_v$, as expected from the signal-to-noise estimate Eq.~\eqref{eq:Sndefinition}. 

%% file: sections/discussion_results.tex
\section{Results}
\label{sec:results}

We now proceed to describe our analysis using Planck individual frequency and component-separated maps and the unWISE blue sample number counts. We begin by describing the analysis pipeline. We then present the dipole field reconstruction based on individual frequency maps, and investigate the impact of various foregrounds and systematics, followed by an analysis of reconstructions based on the component-separated CMB maps. Finally, we measure the cross-correlation of the reconstruction and unWISE galaxy density on large angular scales. 

\subsection{Anaysis pipeline}

The analysis pipeline proceeds as follows:
\begin{enumerate}
\item {\bf Compute input spectra:} We first pre-compute the various quantities necessary to construct the estimator. The galaxy-optical depth cross-power $\bar{C}_{\ell}^{\tau g}$ is computed from the model (Eq.~\eqref{eq:ctaugmodel}) evaluated at the reference redshift $\bar{z} = 0.68$, corresponding to $\bar{\chi} = 2505$ Mpc. Where spectra are computed at a reference multipole, we use $\bar{\ell} = 2000$. We estimate $C_\ell^{TT}$ by computing the power spectrum of the masked and de-beamed temperature map using the pymaster software~\cite{Alonso_2019}~\footnote{One could alternatively compute the power spectrum of the masked map directly, re-scaling by $f_{\rm sky}^{-1}$ and dividing by a Gaussian beam of the appropriate width: $C_\ell^{TT} = [C_\ell^{TT}]^{\rm masked \ map}/f_{\rm sky}/B_\ell(\theta_{\rm FWHM})^2$. However, this produces a sub-optimal estimator whose variance is not as expected. We explore this in detail in Appendix~\ref{appendix:pipeline_validation}.} The choices of mask, $f_{\rm sky}$, and beam width $\theta_{\rm FWHM}$ used in our analysis are recorded in Table~\ref{table:cmb_choices}. We apodize the mask by a $10$ arcmin beam before computing the power spectrum. This spectrum is used in the estimator normalization (Eq.~\eqref{eq:const_est_n}) and in the inverse-variance filtering operation (Eq.~\eqref{eq:map_filters}). We then estimate the unWISE blue galaxy power spectrum by computing the power spectrum of the masked unWISE blue number density map re-scaled by $f_{\rm sky}$: $C_\ell^{gg} = [C_\ell^{gg}]^{\rm masked \ map}/f_{\rm sky}$~\footnote{Identical results are obtained when deconvolving the mask with pymaster.}. We use the unWISE mask described in Sec.~\ref{sec:masking}, with $f_{\rm sky} = 0.58$. The power spectra described above are used to compute the estimator normalization in Eq.~\eqref{eq:const_est_n}.
\item {\bf Filter:} The inputs to the pixel-space quadratic estimator are the filtered CMB field $\xi(\hat{n})$ and galaxy field $\zeta(\hat{n})$ defined in Eq.~\eqref{eq:map_filters}. To construct $\xi(\hat{n})$ we perform a forward spherical harmonic transform of the unmasked maps at healpix resolution $N_{\rm side} = 2048$. We then apply a high- and low-pass filter that nulls all harmonic coefficients $\ell < 100$ and $\ell > 4000$. From Fig.~\ref{fig:Nsummand} this range of scales should include all significant contributions to the estimator variance, while mitigating the impact of foregrounds and systematics on very large and very small angular scales. We divide by the power spectrum $C_\ell^{TT}$ computed as described above, which is representative of the expected CMB power in un-masked regions of the sky. We then perform an inverse spherical harmonic transform to obtain $\xi(\hat{n})$. To construct $\zeta(\hat{n})$ we perform a forward spherical harmonic transform of the unmasked unWISE blue number density map at healpix resolution $N_{\rm side} = 2048$. For our fiducial analysis, we apply the empirically derived systematic weights shown in the bottom panel of Fig.~\ref{fig:unwise_mask} before performing the harmonic transform. We then apply a high- and low-pass filter that nulls all harmonic coefficients $\ell < 100$ and $\ell > 4000$. We filter in harmonic space by the ratio $\bar{C}_{\ell}^{\tau g}/C_{\ell}^{gg}$ where $\bar{C}_{\ell}^{\tau g}$ and $C_{\ell}^{gg}$ are computed as described above. We inverse spherical harmonic transform to obtain $\zeta(\hat{n})$.
\item {\bf Assemble and analyze the reconstruction:} The quadratic estimator for the dipole field Eq.~\eqref{eq:estimator_coeffs} is simply the product of the $\xi(\hat{n})$ and $\zeta(\hat{n})$ maps at full resolution of $N_{\rm side} = 2048$, rescaled by $N$ defined in Eq.~\eqref{eq:const_est_n}. For visualization purposes, below we filter maps with a Gaussian kernel of width  $\sigma_{\rm FWHM} = 5.75^\circ$. For analysis, we apply the reconstruction mask described in Sec.~\ref{sec:masking} with $f_{\rm sky} = 0.58$, and then estimate/remove the monopole from the unmasked pixels (using the healpix function \texttt{fit\_monopole} with healpy masked arrays).
\end{enumerate}

\begin{table}
\centering
\begin{tabular}{l|c|c|c}
    Temperature Map & Mask & $f_{\rm sky}$ & $\theta_{\rm FWHM}$  \\
   \hline 
   \hline
     100 GHz & Pt Src \& 70\% Gal & 0.696 & 9.68 \\
    \hline
     143 GHz & Pt Src \& 60\% Gal & 0.598 &  7.30 \\
    \hline
     217 GHz & Pt Src \& 60\% Gal & 0.598 & 5.02 \\
     \hline 
     353 GHz & Pt Src \& 60\% Gal &0.598 & 4.94 \\
     \hline 
     545 GHz & Pt Src \& 60\% Gal &0.594 & 4.80 \\
     \hline 
     857 GHz & Pt Src \& 60\% Gal &0.589 & 4.60 \\
     \hline 
     Planck CMB & Common CMB & 0.779 & 5.0 \\
     \hline 
     PyILC & PyILC \& 60\% Gal & 0.569 & 5.0 \\
\end{tabular}
\caption{Analysis choices for estimating the CMB temperature autospectrum through $C_\ell^{TT} = [C_\ell^{TT}]^{\rm masked \ map}/f_{\rm sky}/B_\ell(\theta_{\rm FWHM})^2$. This quantity is used in Eq.~\eqref{eq:const_est_n} and \eqref{eq:map_filters}. Pt Src refers to the PR3 HFI point source mask at the corresponding frequency, Gal refers to the PR3 HFI galactic mask, and Common CMB refers to the PR3 common CMB mask.}
\label{table:cmb_choices}
\end{table}

We perform a set of validation tests on the reconstruction using a suite of simulations described in Appendix~\ref{appendix:pipeline_validation}. We confirm that the velocity is reconstructed without bias when the fiducial spectra used in the estimator weights match the true spectra associated with an injected signal. We also confirm that the estimator pre-factor $N$ matches the estimator variance. The analysis pipeline is followed to produce reconstructions of the remote dipole field corresponding to each input Planck temperature map.

\subsection{Reconstruction from Individual Frequency Maps}

\begin{figure*}
\centering
    \includegraphics[width=\textwidth]{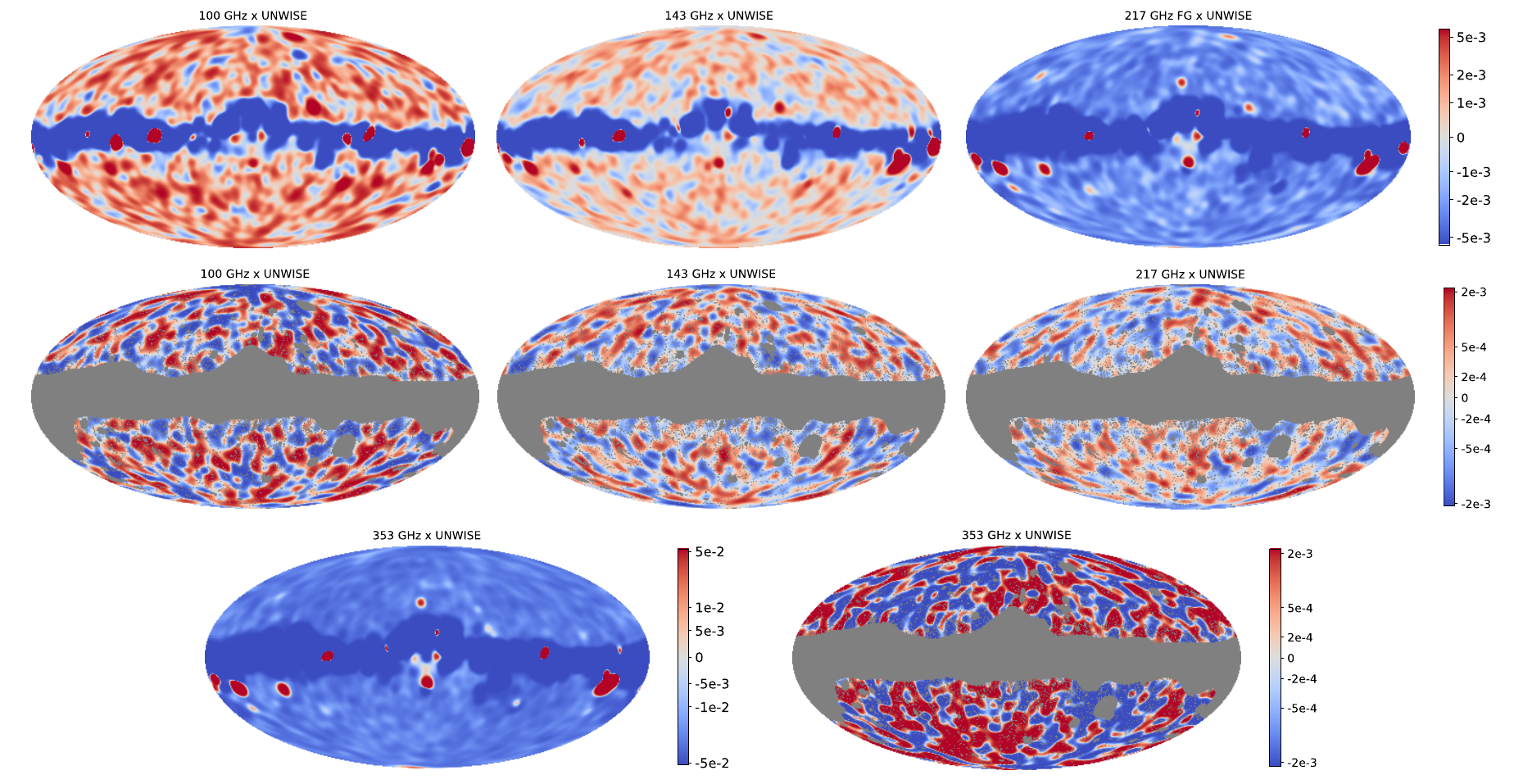}
    \caption{Dipole field reconstructions based on the Planck 100, 143, 217, and 353 GHz individual frequency maps and the unWISE blue sample in units of $v/c$ and smoothed with a $5.75^\circ$ Gaussian beam. Unmasked maps (top row and bottom left) are plotted on a linear scale at low amplitudes and a logarithmic scale at high amplitude. Masked maps (middle row and bottom right) are plotted on a linear scale. Large-amplitude features in the reconstruction confined to the galactic plane are effectively removed by masking. A large monopole (positive for low frequency, negative for high frequency) is visible in the unmasked maps. Masked maps are shown with the dipole and monopole removed. The amplitude of fluctuations is visibly lowest for the masked 217 GHz reconstruction - this is the cleanest channel for kSZ velocity reconstruction.}
    \label{fig:freq_maps}
\end{figure*}

The remote dipole field reconstructed from Planck individual frequency maps at 100, 143, 217, and 353 GHz and unWISE blue galaxies are shown in Fig.~\ref{fig:freq_maps}. All maps contain localized features in the galactic plane with an amplitude orders of magnitude larger than the typical fluctuations away from the galactic plane (note the non-linear color scaling in the figures - small amplitudes are linear while large amplitudes are logarithmic). One such feature is a negative amplitude band encompassing the galactic plane. The other are localized spots with a large positive amplitude, confined near the galactic plane. These features are correlated among the different frequencies - the negative amplitude band has nearly the same width/morphology and the positive amplitude features are located at the same positions. The presence of foreground artifacts concentrated near the galactic plane is not surprising given the strong galactic emission in the Planck maps as well as the visible galactic contamination and removal of stars in the unWISE galaxy density (see Fig.~\ref{fig:unwise_input}). The large amplitude features at any frequency are within the reconstruction mask, as seen in Fig.~\ref{fig:freq_maps} (middle row and bottom-right). The quadratic estimator is local (it relies on the small angular-scale cross-correlations), so there is little leakage between contaminated and uncontaminated regions; we conclude that galactic foregrounds can be effectively mitigated by masking. 

For further analysis, we apply the reconstruction mask described in Sec.~\ref{sec:masking}. Clearly visible in the individual unmasked frequency maps is a large monopole, which is positive at 100 GHz, almost null at 143 GHz, and increasingly negative at 217 and 353 GHz. We observe the same trend at 545 and 857 GHz. We use the healpix function \texttt{fit\_monopole} to estimate and remove the best-fit monopole from the masked maps. The magnitudes of the monopole (defined as the average over un-masked pixels) are recorded in Table~\ref{table:monodipo}. 

\begin{table*}
\centering
\begin{tabular}{l|c|c|c|c}
    Temperature Map & Monopole  $[v/c]$ & No-CMB Monopole $[v/c]$ & Noise Monopole $[v/c]$ & Reconstruction Noise $[(v/c)^2]$ \\
   \hline 
   \hline
     100 GHz & $1.38 \times 10^{-3}$ & $5.64 \times 10^{-4}$ & $8.52 \times 10^{-5}$ & $4.94 \times 10^{-8}$ \\
    \hline
     143 GHz & $3.81 \times 10^{-4}$ & $-2.69 \times 10^{-4}$ & $-1.06 \times 10^{-5}$ & $1.35 \times 10^{-8}$ \\
    \hline
     217 GHz & $-2.51 \times 10^{-3}$ & $-3.10 \times 10^{-3}$ & $3.24\times 10^{-7}$ & $8.37 \times 10^{-9}$ \\
     \hline 
     353 GHz & $-2.10 \times 10^{-2}$ & $-2.17 \times 10^{-2}$ & $-4.63\times10^{-5}$ & $1.13 \times 10^{-7}$ \\
     \hline 
     545 GHz & $-3.45 \times 10^{-1}$ & xx & xx & $7.72 \times 10^{-6}$\\
     \hline
     857 GHz & $-41.9$ & xx & xx & $4.87 \times 10^{-2}$ \\
     \hline 
     SMICA & $5.42 \times 10^{-4}$ & N/A & N/A & $5.88 \times 10^{-9}$ \\
     \hline
     SMICA no SZ & $-5.65 \times 10^{-4}$ & N/A & N/A & $6.13 \times 10^{-9}$ \\
     \hline 
     Commander & $-2.39 \times 10^{-3}$ & N/A & N/A & $5.91 \times 10^{-9}$ \\
     \hline
     SEVEM & $-3.99 \times 10^{-4}$ & N/A & N/A & $6.16 \times 10^{-9}$ \\
     \hline
     NILC & $3.83 \times 10^{-4}$ & N/A & N/A & $5.91 \times 10^{-9}$ \\
     \hline
     pyILC & $8.41 \times 10^{-4}$ & N/A & N/A & $5.83 \times 10^{-9}$ \\
     \hline
     pyILC no SZ& $-9.13 \times 10^{-6}$ & N/A & N/A & $6.04 \times 10^{-9}$\\
     \hline
     pyILC no CIB & $2.61 \times 10^{-4}$ & N/A & N/A & $5.89 \times 10^{-9}$ \\
     \hline
\end{tabular}
\caption{The best-fit monopole (the average of unmasked pixels) and reconstruction noise for the masked reconstructed dipole field based on various temperature maps correlated with unWISE galaxies. The monopole is defined as the average of the unmasked pixels.}
\label{table:monodipo}
\end{table*}

\begin{figure*}
\centering
    \includegraphics[width=.75 \textwidth]{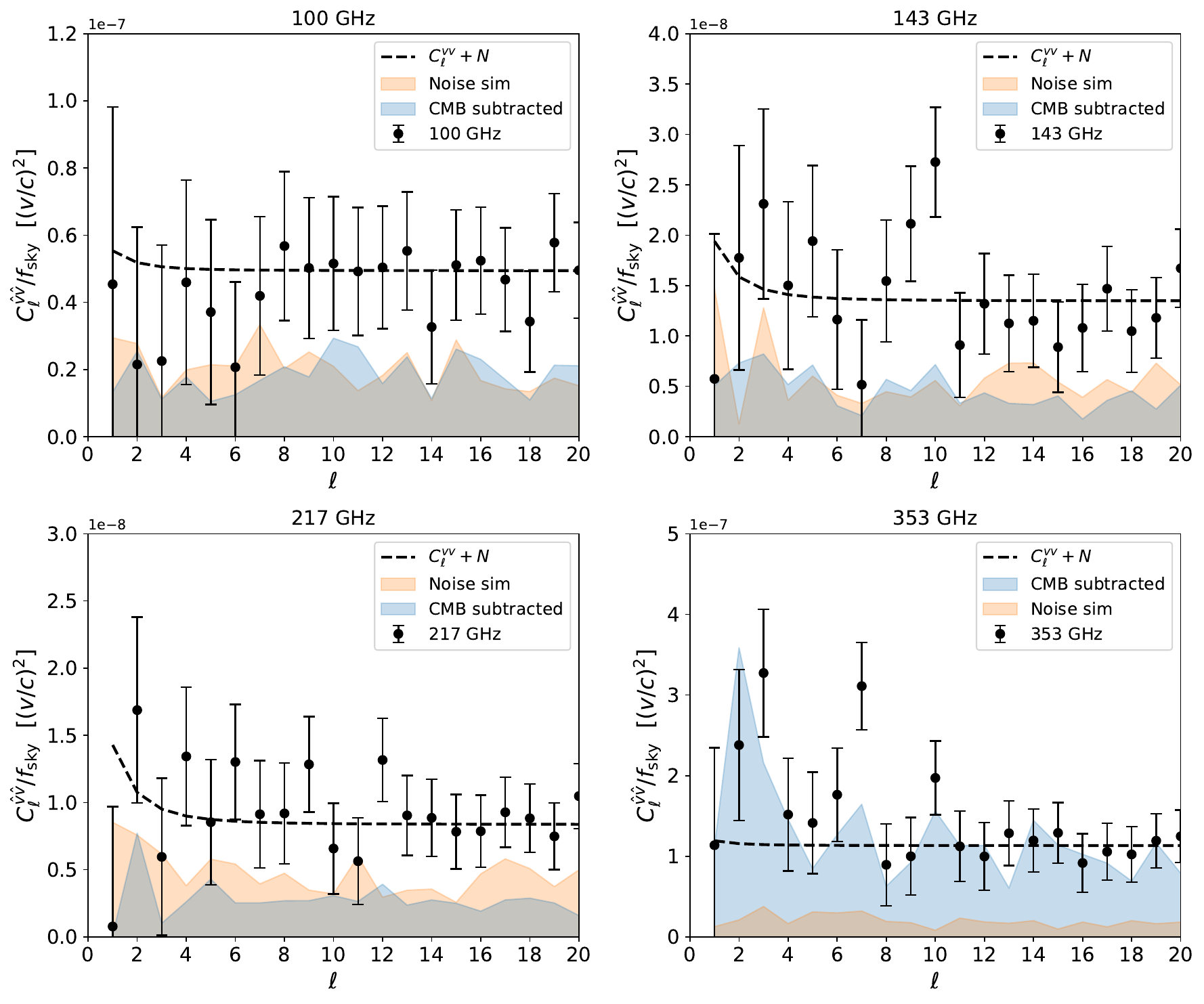}
    \caption{Power spectra of the masked reconstructed dipole field after the monopole has been subtracted. The reconstruction power produced using each individual frequency map (black circles) is compared to the predicted signal and noise power spectrum $C_\ell^{\hat{v} \hat{v}} + N$ (black dashed). The error bars represent the expected cosmic variance from reconstruction noise $\pm \sqrt{2/(2\ell+1)/f_{\rm sky}} N$. To determine how much the primary CMB, foregrounds, and CMB detector noise contribute to the reconstruction variance, we also plot reconstructions using CMB-subtracted maps (blue shading) and a noise realization from the Planck FFP10 simulation suite (orange shading).}
    \label{fig:freq_pspec}
\end{figure*}

After subtracting the monopole, we compute the angular power spectra of the masked maps and divide by the $f_{\rm sky}$ of the reconstruction mask (an estimate of the full-sky power spectrum for white noise). The results for each frequency are shown in Fig.~\ref{fig:freq_pspec}. We overplot the cosmic variance error bars expected from reconstruction noise. Comparing the reconstruction power spectrum (black data points) to the expected reconstruction signal plus noise on the full sky (black dashed) we find good general agreement. Comparing the level of reconstruction noise against the expected signal amplitude (see Fig.~\ref{fig:SNR}), our results are consistent with the expectation that our measurements are in the noise-dominated regime. 

\begin{figure}
\centering
    \includegraphics[width=.8\columnwidth]{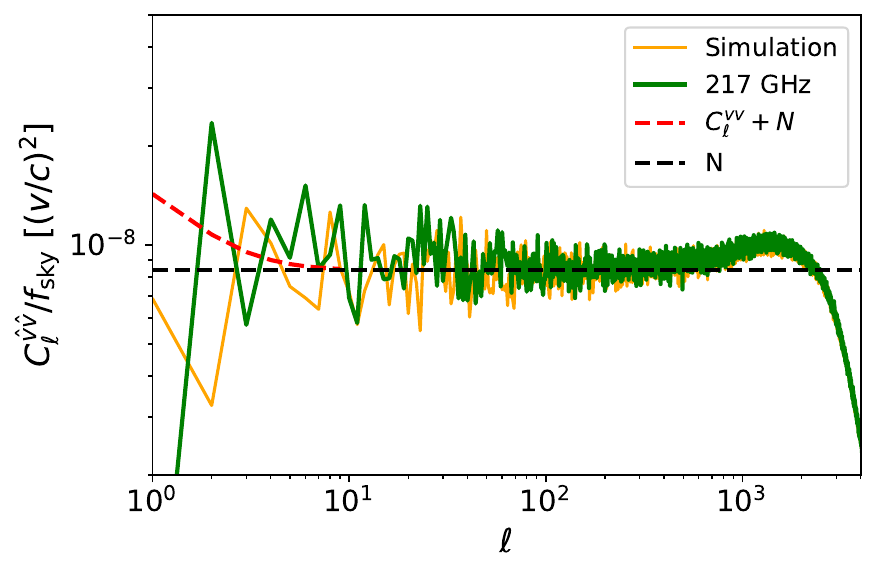}
    \caption{Power spectrum of the masked, monopole subtracted, and $f_{\rm sky}$ re-scaled reconstructed dipole field found using the Planck 217 GHz temperature map (green, solid). The power is nearly flat for $\ell<10^3$, consistent with the expected level of white reconstruction noise (black dashed) for $\ell \lesssim 10^3$ and signal plus noise (red dashed). We also reproduce the power spectrum of a simulated reconstruction based on a mock 217 GHz with a kSZ signal built from unWISE galaxy density (orange solid).}
    \label{fig:extended_217_pspec}
\end{figure}

To investigate the reconstruction power spectrum in greater detail, in Fig.~\ref{fig:extended_217_pspec} we show the power spectrum of the reconstruction over the full range of multipoles $1 \leq \ell \leq 4000$ for the reconstruction based on the 217 GHz channel. The reconstruction power spectrum is very nearly flat for $\ell < 10^{3}$, consistent with our expectation of scale-independent reconstruction noise. Finally, we compare with the reconstruction power spectrum from the simulations described in Appendix~\ref{appendix:pipeline_validation}. There is excellent agreement between the actual and simulated reconstructions, with both displaying a nearly flat power spectrum for $\ell < 10^{3}$, turning over at high $\ell$. We therefore conclude that the reconstruction based on Planck 217 GHz and unWISE blue is consistent with simple random Gaussian simulations at the level of the power spectrum.

\subsection{Statistically isotropic foregrounds: the reconstruction monopole}\label{sec:foregroundssystematics_mono}

The clearest impact of foregrounds is on the reconstruction monopole. As discussed in Sec.~\ref{sec:possible_systematics}, statistically {\em isotropic} cross-correlations between the temperature and galaxy map contribute to the reconstruction monopole. To estimate the contribution from foregrounds, we perform a reconstruction using CMB-subtracted maps, where the blackbody component is estimated using the SMICA component separation technique. To determine the contribution from CMB detector noise, we perform reconstructions using noise realizations from the FFP10 simulations. 

In Table~\ref{table:monodipo}, we compare the monopole for the reconstruction based on temperature maps, CMB-subtracted maps, and noise simulations. Table~\ref{table:monodipo} also includes the reconstruction noise for each data combination. The monopole (map average) expected from reconstruction noise is $\langle \hat{v} \rangle \simeq \sqrt{N/4\pi}$. This can be used to quantify the  significance of a measured monopole. 

We can draw a number of conclusions from the results presented in Table~\ref{table:monodipo}. First, we see that detector noise does not contribute to the monopole at a significant level. This is consistent with the expected absence of a correlation between CMB detector noise and the unWISE blue maps. At 100, 217 and 353 GHz we obtain a monopole from the temperature and CMB-subtracted maps that is similar in sign and magnitude, implying that the measured monopole can be largely accounted for by foregrounds (this trend continues at 545 and 857 GHz). At 143 GHz, the monopole is smallest, with similarly small values found for the CMB-subtracted and noise simulation reconstructions. In all cases except 143 GHz, the measured monopoles are significantly larger than expected from reconstruction noise, and therefore are statistically significant under the hypothesis that the maps contain only reconstruction noise. Performing the reconstruction without applying the unWISE systematics weights yields values for the monopole that vary at the percent-level from those recorded in Table~\ref{table:monodipo}. 

We can compare these results to our expectations for the contribution from extragalactic foregrounds. The dominant extragalactic foreground at 100 GHz is the thermal Sunyaev Zel'dovich (tSZ) effect (a decrement in temperature at this frequency), which contributes a positive reconstruction monopole from the negative cross-correlation with unWISE blue. We expect that the 143 GHz map has the least contamination from extragalactic foregrounds, and therefore the smallest reconstruction monopole. At 217 and 353 GHz, the dominant extragalactic foreground is the cosmic infrared background (CIB), which contributes a negative reconstruction monopole due to the positive correlation with unWISE blue. Further, we expect the contribution from CIB to increase with frequency. The results in Table~\ref{table:monodipo} are consistent with these predictions. 

Galactic thermal dust also contributes significantly at high frequencies, and could be correlated with unWISE blue number counts. If dust emission is correlated with extinction of unWISE sources (and therefore a deficit in number counts), we expect a positive reconstruction monopole. This is inconsistent with what is observed. We further explore the possibility of a reconstruction monopole arising from galactic dust by performing the reconstruction using the FFP10 353 GHz galactic thermal dust map. We find a reconstruction monopole of $-2.5 \times 10^{-4}$, two orders of magnitude smaller than the observed 353 GHz reconstruction monopole, and comparable in magnitude to the prediction for pure reconstruction noise. We therefore rule out galactic thermal dust as the source of the large reconstruction monopole at high frequencies. 

In summary, we find strong evidence for statistically isotropic correlations between unWISE blue and CMB extragalactic foregrounds in the reconstruction monopole.

\subsection{Measuring and correcting for systematics}\label{sec:foregroundssystematics_template}

As discussed in detail in Sec.~\ref{sec:possible_systematics}, an additive estimator bias results when a systematic modulates the isotropic cross-correlation between the CMB and galaxy maps. For example, variations in depth, stellar contamination, and other forms of 'calibration error' in the unWISE galaxy map cause a modulation of the number counts of the form:
\begin{equation}\label{eq:calibration_errors}
    N^g_{\rm obs} (\hat{n}) = (1+P(\hat{n}))  N^g (\hat{n})
\end{equation}
Ref.~\cite{2023arXiv230905659F} solved for a set of systematics weights $W_{\rm sys} (\hat{n})$ based on the sum of contributions proportional to Gaia stellar density and the limiting magnitude in the W2 (4.6 $\mu$m) WISE band, such that $N^g_{\rm obs, weighted} = W_{\rm sys} (\hat{n}) N^g_{\rm obs} (\hat{n})$ is a more faithful representation of the true number counts. If accurate, these weights are simply the inverse of the modulating systematics in Eq.~\eqref{eq:calibration_errors}. 

Here, we use a fully empirical method to determine and correct for the additive estimator bias due to foregrounds and survey systematics. The reconstruction performed using high frequency Planck maps at 353, 545, and 857 GHz is dominated by foregrounds and systematics on large angular scales. We focus on the 857 GHz map here, since it is most strongly dominated by CIB anisotropies that correlate with unWISE blue. Using the reconstruction map from 857 GHz and unWISE blue, we obtain an estimate of the survey systematics on large angular scales from Eq.~\eqref{eq:lss_addbias}:
\begin{equation}
\hat{P} (\hat{n}) = \frac{\hat{v}^{857} (\hat{n})}{\hat{v}_{00}^{857}} - 1
\end{equation}
This estimated systematics map is noise-dominated on small angular scales. We therefore smooth the unmasked regions of the map with a Gaussian kernel of FWHM $28^\circ$. The result is shown in the lower panel of Fig.~\ref{fig:unwise_mask} along with the $W_{\rm sys} (\hat{n})$ template for stellar density and W2 depth described above. The two maps are visually highly correlated, in particular the decrements at the north and south ecliptic pole from the unWISE scan strategy.

With the systematics map, we can either predict the additive estimator bias at another frequency labeled by $x$ as
\begin{equation}\label{eq:predict_sys_map} 
\hat{v}^{\rm sys; x} (\hat{n}) = \hat{v}_{00}^x \hat{P} (\hat{n}) 
\end{equation}
and subtract this template from the reconstruction or we can correct the unWISE blue maps by dividing number counts by $1+\hat{P}$. In Fig.~\ref{fig:weight_vs_unweight} we demonstrate that these two choices for correcting systematics produce equivalent results. In our analysis, we choose to correct the unWISE maps for systematics, and then reconstruct. From Fig.~\ref{fig:weight_vs_unweight} we see that correcting for systematics has a large impact at $\ell = 2$. This could be expected from the clearly quadrupolar shape of the systematics maps in Fig.~\ref{fig:unwise_mask}. 

\subsection{The one-point function}

\begin{figure*}
\centering
    \includegraphics[width=1.75\columnwidth]{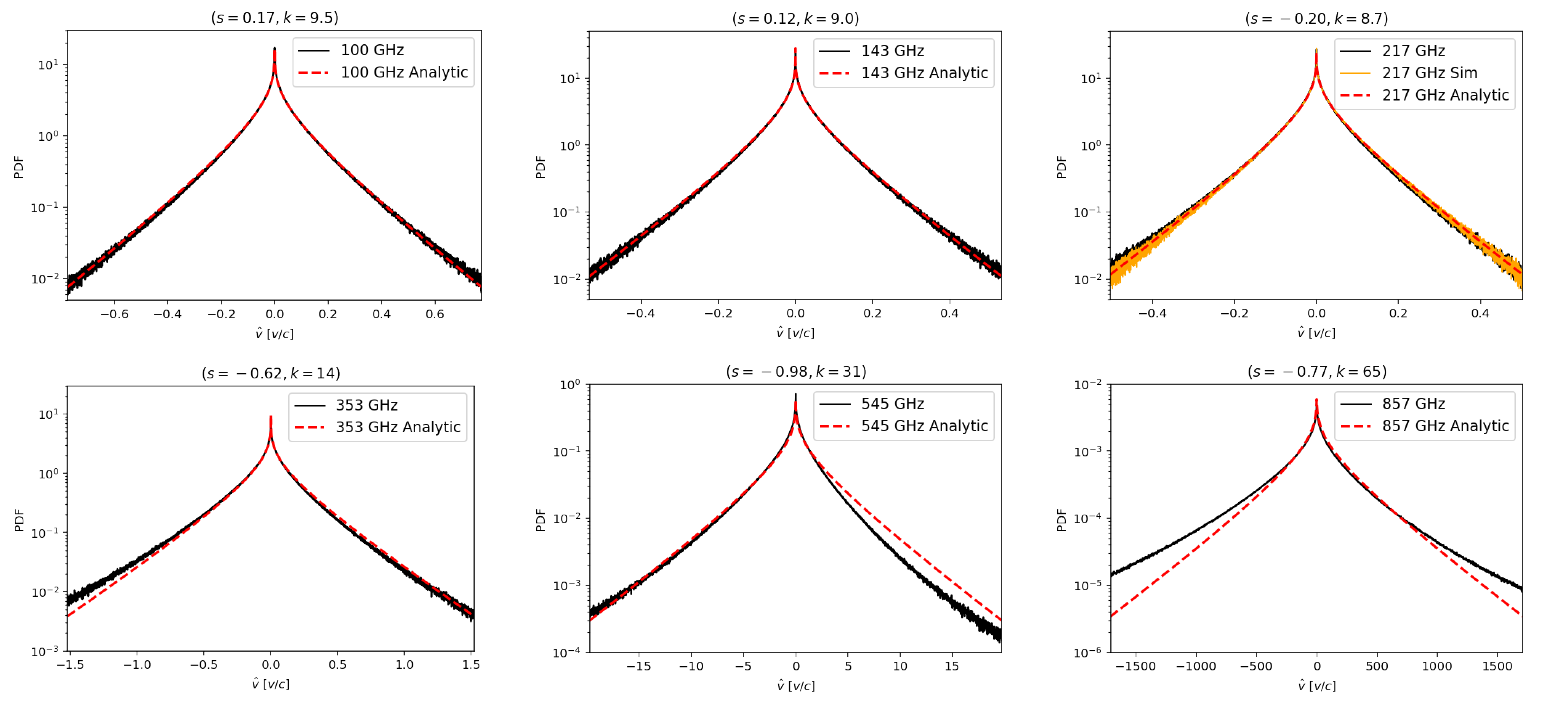}
    \caption{The one-point function of the reconstruction maps at each frequency (black solid) plotted against the expected normal product distribution Eq.~\eqref{eq:normalproductdist} (red dashed). For the 217 GHz reconstruction, we overplot the distribution for a simulated reconstruction map. At the top of the plot, we show the skewness $s$ and excess kurtosis $k$ for each measured distribution. The expected skewness of the normal product distribution is $s=0$, and the expected kurtosis is $k=6$.}
    \label{fig:1pt_freqrecons}
\end{figure*}

After correcting the reconstruction maps for systematics, we compute the one-point function of the masked reconstruction maps at each frequency. We compare the resulting distributions to the expected normal product distribution for reconstruction noise Eq.~\eqref{eq:normalproductdist} in Fig.~\ref{fig:1pt_freqrecons}. At the top of each plot, we show the skewness $s$ and  kurtosis $k$ of each map. The normal product distribution has $s=0$ and $k=6$. 

The 143 GHz reconstruction, which we argued above has the smallest contribution from foregrounds and systematics, has the smallest skewness, nearly the smallest kurtosis, and is by eye a good fit to the expected distribution. At increasing frequencies, the distributions visibly deviate from normal product, and develop significant negative skewness and a large kurtosis. This can be anticipated because as the input temperature maps become dominated by foregrounds, they become increasingly non-Gaussian, and the estimator can no longer be expected to follow a normal product distribution. We interpret the negative skewness as a correlation between large over-densities in unWISE and large emission in the temperature maps, consistent with the interpretation of the monopole. 

We also compare with the simulated 217 GHz reconstruction map (described above, see discussion around Fig.~\ref{fig:extended_217_pspec}). The simulated distribution is visually quite close to the actual. For the simulated map, we can compute the distribution on the full sky and the masked sky, and with the real and mock unWISE galaxies. We find negligible difference between the distributions on the full vs masked sky for mock unWISE. We find that the mock unWISE x mock CMB reconstruction has a larger skewness and kurtosis  than the real unWISE x mock CMB, going from $s = 5.4 \times 10^{-3}$ to $s = -1.2 \times 10^{-2}$ and $k=5.98$ (as expected for normal product) to $k=7.9$, respectively. This is an estimate of the impact on the one-point function from the intrinsic non-Gaussianity of the unWISE number counts map (inducing an increase in skewness of $\mathcal{O}(10^{-2})$ and kurtosis of $\mathcal{O}(1)$). These  shifts do not account for the deviations from the normal product distribution observed in the actual reconstructions, implying that correlated foregrounds are important.

\begin{figure*}
\centering
    \includegraphics[width=1.5\columnwidth]{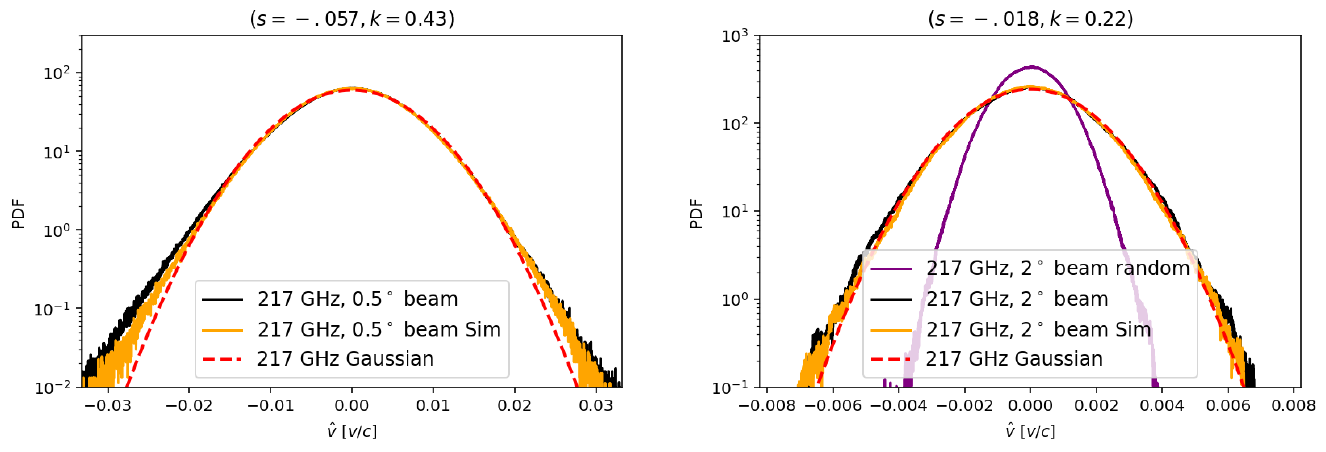}
    \caption{The one-point distribution for 217 GHz reconstructions smoothed with a $0.5^\circ$ (left, black solid) and $2^\circ$ (right, black solid) Gaussian beam. The predicted Gaussian distribution with variance Eq.\eqref{eq:pixelnoise} that is expected from the predicted scale-independent reconstruction noise is shown as the red-dashed curves. The distribution for a smoothed simulated reconstruction using real unWISE x mock 217 GHz is shown in orange, and the distribution for a map where pixel values are randomized before smoothing is shown in purple. The values of skewness $s$ and kurtosis $k$ at each level of smoothing is shown above the plots.}
    \label{fig:1pt_217_vssims}
\end{figure*}

Smoothing the maps, for sufficiently large beam size, we expect to recover a Gaussian one point function. In Fig.~\ref{fig:1pt_217_vssims} we plot the one point distribution for the 217 GHz reconstruction map smoothed with a $0.5^\circ$ and $2^\circ$ Gaussian beam (solid black). This is compared to a Gaussian distribution (red dashed) with a pixel variance given by the flat reconstruction noise convolved with a Gaussian beam:
\begin{eqnarray}\label{eq:pixelnoise}
N_i = \sum_{\ell} \frac{2\ell+1}{4\pi} N B_\ell^2
\end{eqnarray}
We also report the skewness $s$ and kurtosis $k$ at the top of each plot. As expected, the skewness and kurtosis decrease as the maps are smoothed on larger scales, agreeing well with the expected Gaussian distribution for smoothing scales $\sim 2^\circ$ and larger. We also compare with the smoothed simulated 217 GHz reconstructions. For smaller beam sizes, the excess skewness of the actual reconstructions is apparent. However, in general there is good agreement between the simulated and actual maps for smoothing scales $\sim 2^\circ$ and larger.

It is important to note that the large non-Gaussianity of the reconstruction noise implies that there are significant pixel-pixel correlations at native resolution. This is true even in the case where the reconstruction noise has a constant angular power spectrum. To demonstrate this, we randomize the pixels in the reconstruction map before smoothing (purple line). In this case, we obtain a Gaussian from the normal product distribution upon smoothing, but with a smaller variance. This is because randomizing the pixels destroys correlations, and averaging these statistically independent pixels within the beam yields a smaller variance.

In summary, these results demonstrate that the one-point distribution can be a useful diagnostic of the importance of CMB foreground residuals on the reconstruction maps. 

\subsection{Contributions to the power spectrum from foregrounds}\label{sec:foregroundssystematics}

After applying the systematics weights as described above, we assess the impact of foregrounds and detector noise on the full reconstruction power spectrum by comparing reconstructions based on individual frequency maps to CMB-subtracted and noise maps in Fig.~\ref{fig:freq_pspec} (the blue-shaded and red-shaded regions, respectively). Note that the CMB-subtracted maps contain both foreground residuals and instrumental noise. As discussed in detail in Sec.~\ref{sec:possible_systematics}, statistically {\rm anisotropic} cross-correlations between the temperature and galaxy maps are necessary to influence the estimator power spectrum beyond the monopole. Uncorrelated foregrounds contribute to the estimator variance only through their impact on the reconstruction noise, which is independent of $\ell$. In addition, as discussed in the previous subsection, foregrounds have an impact on the estimator one-point function.

At 100, 143, and 217 GHz, the CMB subtracted and noise reconstructions have a nearly flat power spectrum. The contributions are of nearly equal magnitude. Since the CMB-subtracted maps contain both foregrounds and noise, this suggests that the foreground contribution to the reconstruction is not important at these frequencies. Further, this contribution is no more than 30-50$\%$ of the total estimator variance, which we therefore conclude is dominated by the blackbody CMB. 

\begin{figure}
\centering
    \includegraphics[width=.8\columnwidth]{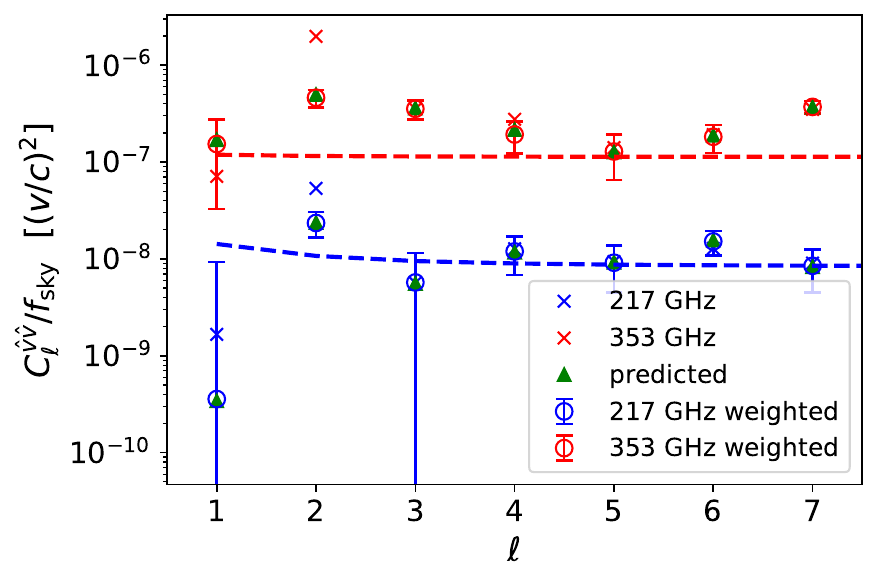}
    \caption{The reconstruction power spectrum produced using unWISE blue maps with (circles) and without (crosses) the systematics weights applied for 217 GHz (blue) and 353 GHz (red) maps. The error bars indicate the expected variance from reconstruction noise alone. The green triangles are the result when reconstructing without weights and adding the bias predicted by Eq.~\eqref{eq:predict_sys_map}. The dashed lines indicate the expected signal and reconstruction noise.}
    \label{fig:weight_vs_unweight}
\end{figure}

At 353 GHz, the CMB subtracted reconstruction accounts for all of the estimator variance; the contribution from noise is negligible. This implies that neither the blackbody CMB nor the noise contributes significantly to the estimator variance - the estimator variance is entirely due to foregrounds. In Fig.~\ref{fig:freq_pspec} there is additional power at low-$\ell$ beyond what is expected from the reconstruction noise. This excess is dramatically larger when the reconstruction is performed on the unWISE map without the systematics weights applied as shown in Fig.~\ref{fig:weight_vs_unweight}. It is possible that a more comprehensive analysis of systematics could further reduce the large-scale power in the reconstructions at high frequencies; we defer such an analysis to future work.

As one explicit check that the CIB is responsible for the low-$\ell$ signal in the high frequency reconstructions, we perform reconstructions using galactic foreground maps. The FFP10 suite of simulations also provides templates for various individual galactic foreground components. By performing the reconstruction on these templates, we determine the foreground with the highest estimator response at each frequency. For all FFP10 simulated foreground maps the most significant foreground at 100, 143, 217, and 353 GHz is galactic thermal dust, increasing sharply in importance with frequency. At 217 and 353 GHz no other simulated galactic foregrounds are within an order of magnitude of the thermal dust reconstructions. However, the magnitude of the response for the thermal dust templates is not significant compared to the measured spectra at corresponding frequencies. This is strong evidence that extragalactic foregrounds are the primary foreground contribution.

\begin{figure}
\centering
    \includegraphics[width=.9\columnwidth]{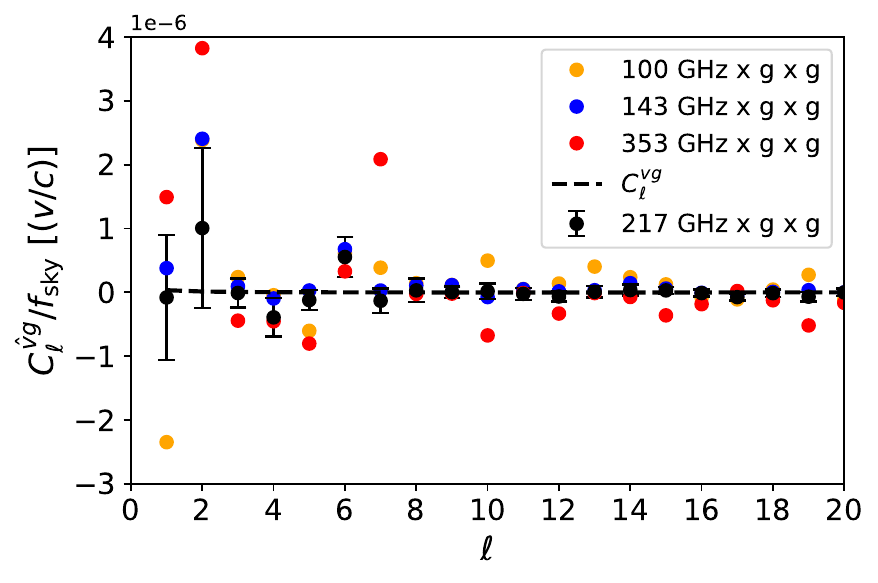}
    \caption{The cut-sky cross-power spectrum between unWISE blue and the reconstruction at various frequencies, normalized by $f_{\rm sky}$. The error bars on the 217 GHz data points represents the expected covariance in the absence of a correlation, using the empirical reconstruction and unWISE blue auto-spectra in Eq.~\eqref{eq:vgnoise}.} 
    \label{fig:frequencies_x_galaxies}
\end{figure}

Another handle on foregrounds and residual systematics is $C_\ell^{\hat{v} g}$. The cut-sky cross-correlation for reconstruction maps at each frequency is shown in Fig.~\ref{fig:frequencies_x_galaxies}. The expected level of correlation from the kSZ signal is shown as the black dashed curve (see Fig.~\ref{fig:gv} for a clearer picture of the theory signal). The error bars on the 217 GHz data points are the expected covariance from uncorrelated unWISE blue and the reconstruction~\footnote{Note that these error bars are inflated compared to the theory covariance in Fig.~\ref{fig:gv} due to the very large power in unWISE on large angular scales compared to the theory expectation.}. Deviations from the expected spectra are largest for 100 and 353 GHz, which as we discussed above, have significant reconstruction monopoles consistent with correlated extragalactic CMB foregrounds. Such correlations, in combination with systematics residuals, lead to a non-zero $C_\ell^{\hat{v} g}$.

\begin{figure}
\centering
    \includegraphics[width=.9\columnwidth]{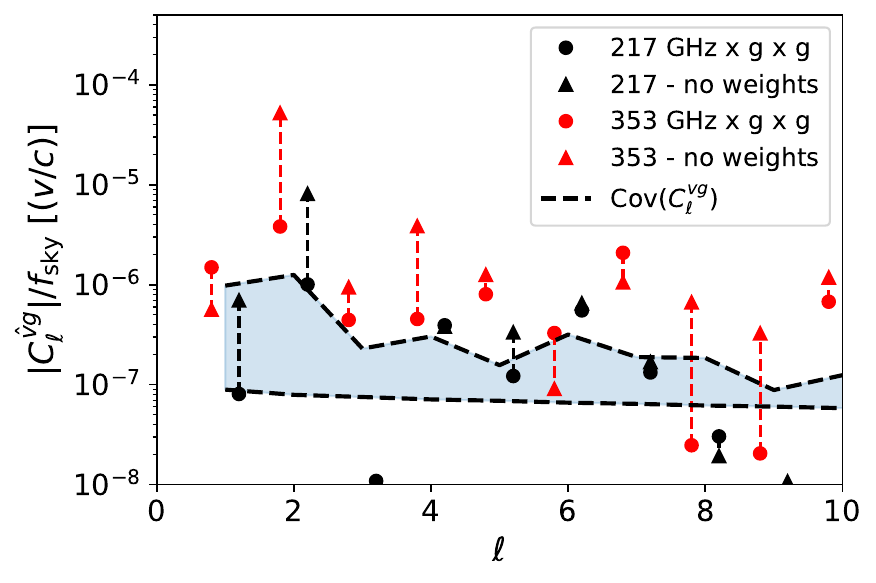}
    \caption{The magnitude of the cut-sky cross-power spectrum between unWISE blue and reconstruction with (circles) and without (triangles) applying the systematics weights. The boundaries of the shaded region are the expected covariance for only reconstruction noise using the empirical (top) or theory (bottom) unWISE power spectrum.} 
    \label{fig:frequencies_x_galaxies_syst}
\end{figure}

To illustrate the impact of the systematics weights, we compare the results for the cross-power with and without correcting for unWISE systematics in Fig.~\ref{fig:frequencies_x_galaxies_syst}. The impact is largest for 217 and 353 GHz, as expected since the relative impact of systematics is proportional to the reconstruction monopole. For these channels, the systematics weights reduce the amplitude of the cross-power by an order of magnitude at $\ell = 2$ (the dominant mode in the systematics map). Overall, after the systematics weights are applied, the cross-power is consistent with the expected covariance (see Eq.~\eqref{eq:vgnoise}) from the measured unWISE blue power spectrum (top boundary of the blue shaded region). However, as demonstrated in this plot, there is still significant more cross-power on large angular scales than is predicted by the expected unWISE clustering signal (lower boundary of the blue shaded region). Performing an identical analysis using the W2 and stellar density template weights produces a comparable reduction in cross-power to the empirically derived weights, as could be expected given the high degree of correlation between the templates on large angular scales. In either case, the excess of power beyond what is expected from the clustering signal alone suggests further improvements can be made in correcting for systematics. 

We conclude that foregrounds can effectively be mitigated by masking, removing the map monopole, correcting for galaxy survey systematics that lead to an additive estimator bias, and focusing on frequency channels that are not strongly dominated by extragalactic foregrounds. With these strategies, it is possible to achieve the statistical error bar on velocity reconstruction with CMB at the depth of Planck and large photometric samples such as unWISE. We stress that this is an extremely optimistic conclusion for the future of kSZ velocity reconstruction/kSZ tomography. An important future direction is to fully assess the importance of various extragalactic foreground components using simulations, we require mock datasets with properly correlated kSZ, galaxies, and foregrounds. This is beyond the scope of the current work.

\subsection{Reconstruction from component separated CMB maps}\label{sec:componentseparation}

We perform kSZ velocity reconstruction for a wide variety of component separated maps. This includes data products from Planck PR3: SMICA, SMICA no SZ (where the thermal Sunyaev Zel'dovich signal is explicitly deprojected), Commander, SEVEM, and NILC maps. We also analyze blackbody maps produced using the {\tt pyilc} code, including pyilc (no deprojection), pyilc no SZ (where the thermal Sunyaev Zel'dovich signal is explicitly deprojected), and pyilc no CIB (where the CIB signal modeled as a 2 parameter modified blackbody is explicitly deprojected; see Ref.~\cite{McCarthy:2023hpa}). 

We begin by computing the reconstruction noise and the monopole on the masked reconstruction maps; the values are reported in Table~\ref{table:monodipo}. The reconstruction noise for the component separated maps is $\sim 30\%$ lower than the 217 GHz channel (lowest noise of the individual frequency maps). As discussed in the previous subsection, the reconstruction monopole is a measure of the statistically isotropic cross-correlation between the CMB map and unWISE blue. The Commander map has the largest reconstruction monopole, matching the monopole at 217 GHz in magnitude and sign. The monopole for the remaining maps are similar in magnitude (of order $10^{-4} \ v/c$), with varying sign, and within several standard deviations of the expectation from white reconstruction noise. For both the SMICA and pyILC maps, deprojecting tSZ or the CIB results in a statistically significant change in the monopole. This is another indication that extragalactic foregrounds play an important role in the reconstruction monopole. The spread of monopoles suggests that foreground residuals are the dominant source of the observed reconstruction monopoles, with the Commander map having the largest residual. 

\begin{figure*}
\centering
    \includegraphics[width=.75 \textwidth]{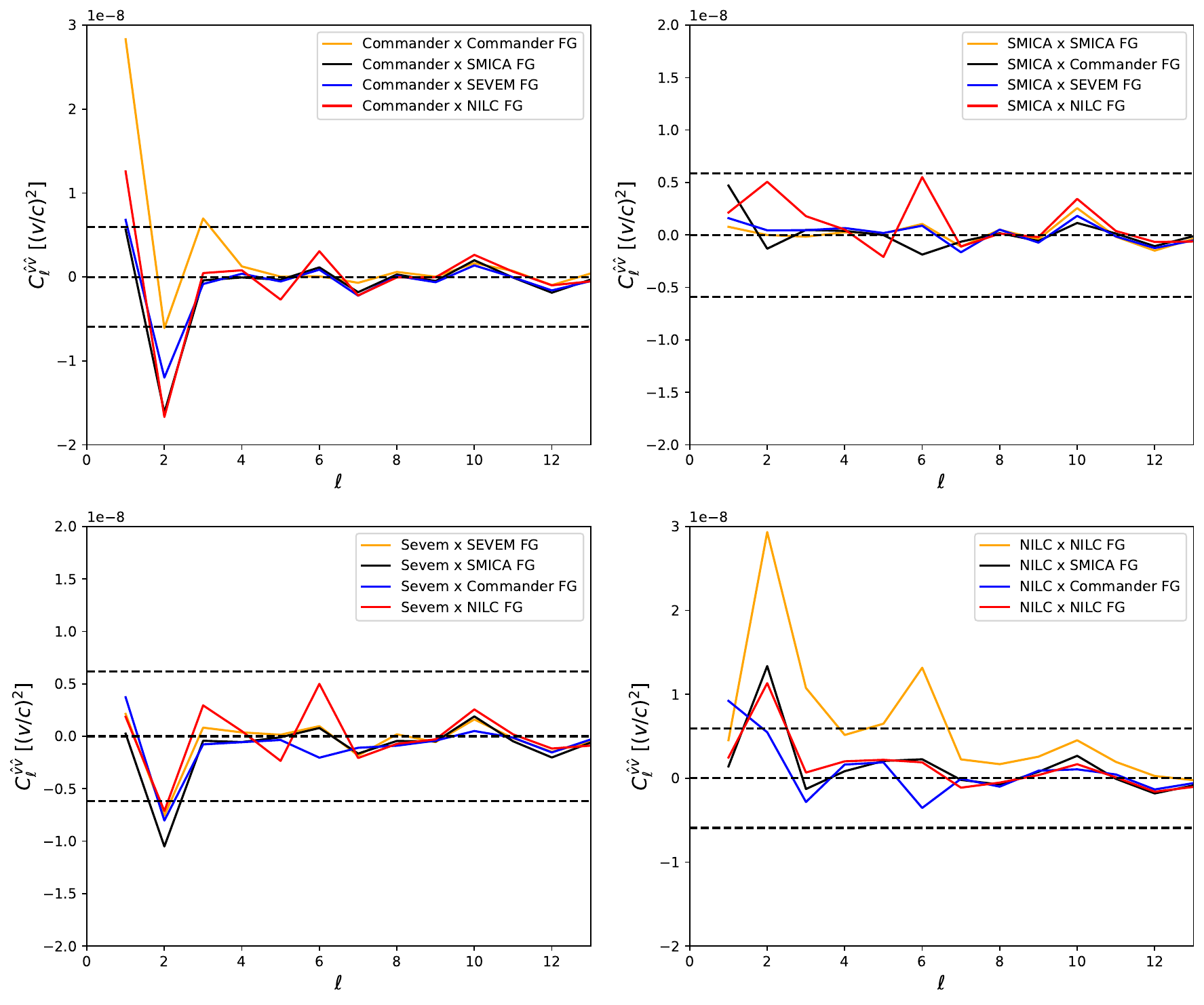}
    \caption{The cross-correlation between reconstructions produced using Planck component separated maps and reconstructions produced using various CMB-subtracted maps at 217 GHz. For reference, the dashed lines indicate the reconstruction noise in the component separated maps.}
    \label{fig:CMBxFG}
\end{figure*}

We perform the reconstruction on the 217 GHz CMB-subtracted maps with unWISE blue for each of the available component separation techniques. We then cross-correlate these foreground maps with the reconstructions for blackbody maps produced with each component separation technique. We would expect this to vanish in the absence of foreground residuals. The combinations are shown in Fig.~\ref{fig:CMBxFG}. At low-$\ell$, there is significant correlation with foreground maps for Commander, Sevem, and NILC. The SMICA-based reconstruction has the lowest correlation with foreground maps. Among the PR3 component separated maps, we therefore focus on SMICA. In addition, because we have found the CIB to be the dominant foreground in the reconstructions, we focus on the pyILC no CIB maps as an additional baseline map. Both the SMICA and pyilc maps also offer the advantage that data products with and without deprojected extragalactic components (tSZ and CIB) are available, allowing for additional foreground and systematics tests.

\begin{figure*}
\centering
    \includegraphics[width=.75 \textwidth]{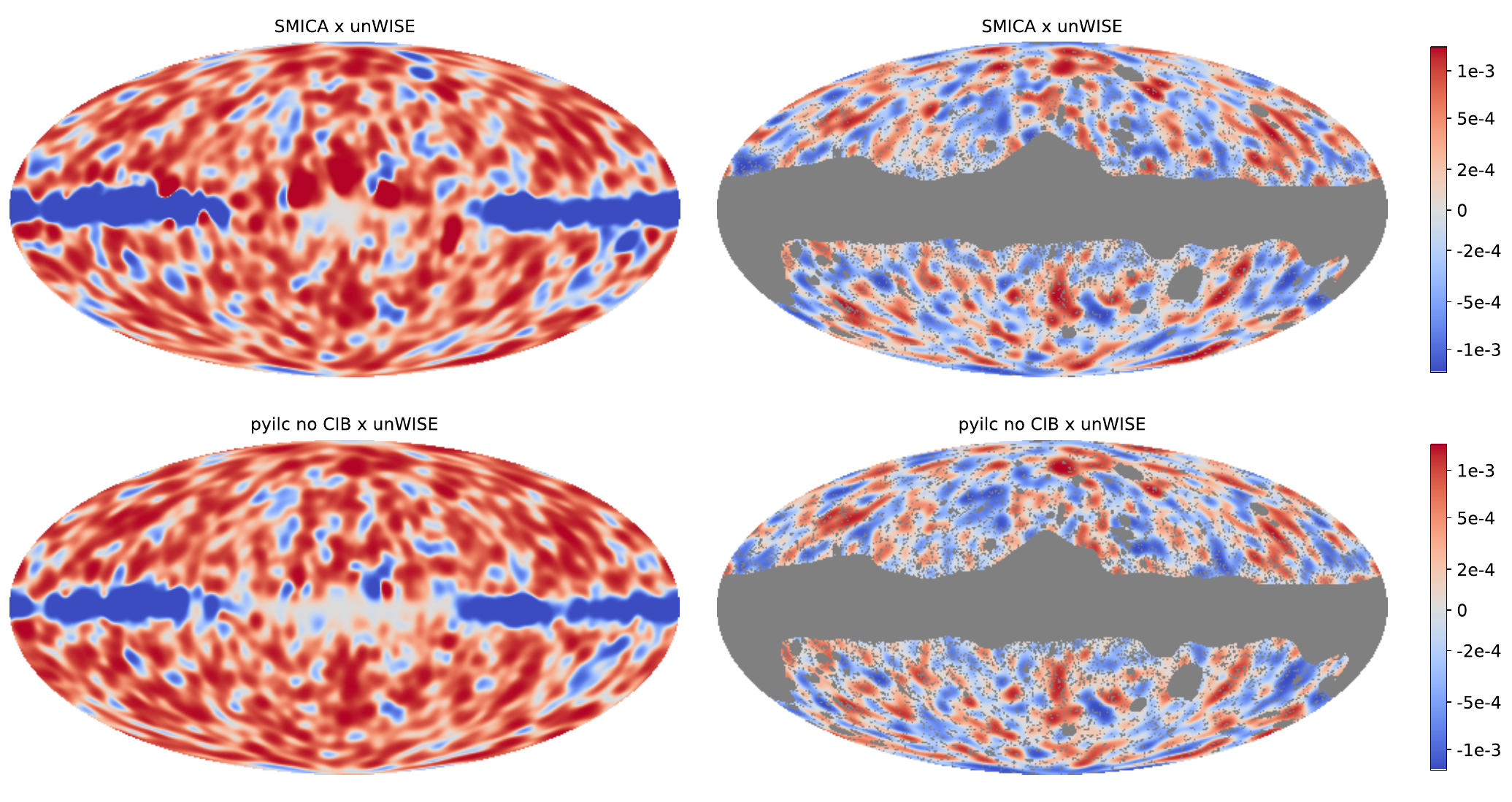}
    \caption{Dipole field reconstruction based on the Planck SMICA (top) or Planck pyilc no CIB (bottom) map with the unWISE blue sample in units of $v/c$ and smoothed with a $5.75^\circ$ Gaussian beam. Unmasked  (left) and masked maps (right) are plotted on a linear scale. As for the individual frequency maps in Fig.~\ref{fig:freq_maps}, large-amplitude features in the reconstruction are effectively removed by masking. A positive monopole of order $\sim 10^{-4}$ is visible in the unmasked maps, which is similar to the amplitude of fluctuations away from the galactic plane; the masked maps on the right is shown with the monopole removed.}
    \label{fig:CMBrec_maps}
\end{figure*}

We show the reconstruction maps from the SMICA and pyilc no CIB maps in Fig.~\ref{fig:CMBrec_maps}, smoothed with a $5.75^\circ$ Gaussian beam. The features of these reconstruction maps are visibly highly correlated at this smoothing scale. More quantitatively, the pixel-space correlation coefficient between the maps at this smoothing scale is $r = 0.91$; the correlation coefficient of the unsmoothed maps is $r = 0.44$. We conclude that there is significant correlation between the reconstruction noises for these component separated maps.

 \begin{figure}
\centering
    \includegraphics[width=.8\columnwidth]{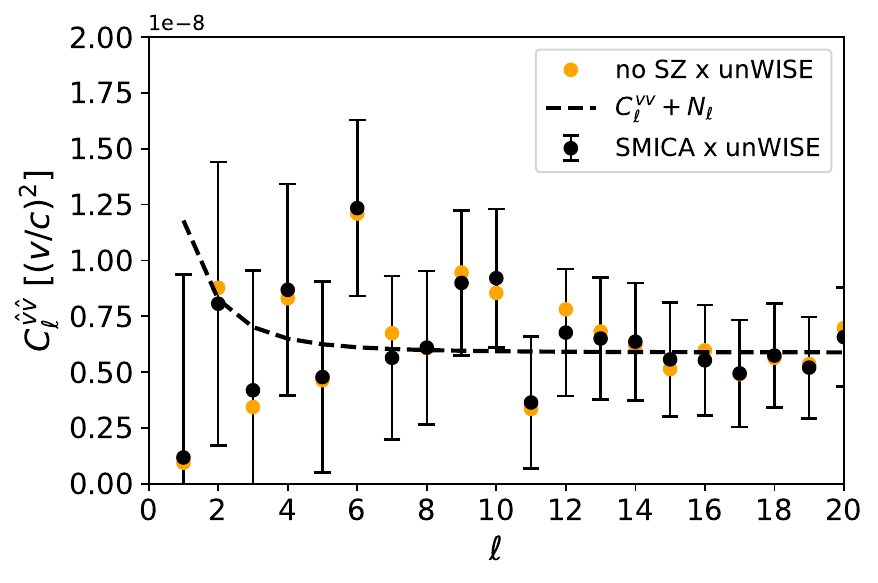}
    \caption{Power spectra for reconstructions based on SMICA (black) and the SMICA no SZ (orange) maps. The spectra were computed on the masked sky and re-scaled by $f_{\rm sky}^{-1}$ to approximate the power on the full sky. The error bars reflect the expected variance from reconstruction noise. We also show the expected full-sky signal plus noise spectrum (black dashed).}
    \label{fig:smica_pspec}
\end{figure}

In Fig.~\ref{fig:smica_pspec} we show the reconstruction spectra for SMICA, computed on the masked sky and re-scaled by $f_{\rm sky}^{-1}$ to estimate the full-sky power spectrum. The measured spectrum is consistent with the expected signal and reconstruction noise. Quantifying the goodness of fit to the fiducial model in the range $1<\ell<50$ we obtain $\chi^2 = 43.3$, corresponding to a PTE of 0.75. This choice of scales is somewhat arbitrary, but gives some sense of the goodness of fit on scales where the reconstruction noise is Gaussian. We also show the cut sky spectra for the reconstruction using the SMICA no SZ map in Fig.~\ref{fig:smica_pspec}, which illustrates the impact of deprojecting tSZ residuals from the component separated map. We find that this has a negligible effect on the reconstruction power spectrum. This is one measure of the robustness of the SMICA reconstruction. Unfortunately, there are no Planck SMICA products that deproject the CIB, which we have identified as the dominant foreground of concern.

\begin{figure}
\centering
    \includegraphics[width=.8\columnwidth]{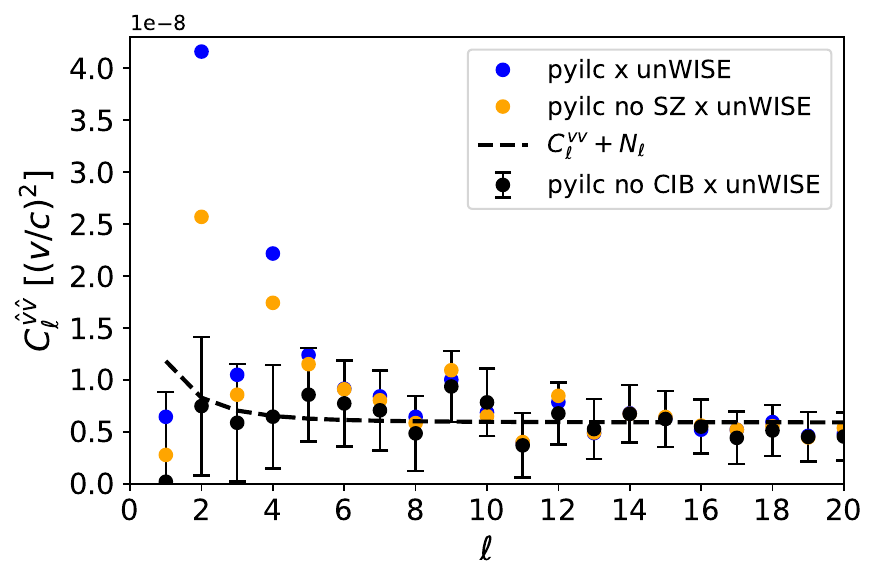}
    \caption{Power spectra for reconstructions based on the pyilc  (blue), pyilc no SZ (orange) and the pyilc no CIB (black) maps. The spectra were computed on the masked sky and re-scaled by $f_{\rm sky}^{-1}$ to approximate the power on the full sky. The error bars reflect the expected variance from reconstruction noise. We also show the expected full-sky signal plus noise spectrum (black dashed).}
    \label{fig:pyilc_pspec}
\end{figure}

In Fig.~\ref{fig:pyilc_pspec} we show the reconstruction spectra for the {\tt pyilc} maps (fiducial, as well as with the tSZ or CIB deprojected), computed on the masked sky and re-scaled by $f_{\rm sky}^{-1}$. In this case, there is quite a bit of variation at low-$\ell$ between the reconstruction spectra for the fiducial pyilc map and the maps deprojecting tSZ and CIB. The excess of power in the pyilc and pyilc no SZ maps (particularly at $\ell =2, 4$) compared to the pyilc no CIB map strongly indicates that the former maps have residual CIB contamination and that our empirical systematics weights have not completely nulled the statistically anisotropic CIB-unWISE correlation observed in the 217-857 GHz reconstructions. Based on this behavior, we conclude that the pyilc no CIB map provides a more reliable reconstruction than its kin. Quantifying the fit of the pyilc no CIB reconstruction to the predicted constant reconstruction noise plus the expected signal we obtain $\chi^2 = 26$ including data in the range $1<\ell<50$, corresponding to a PTE of 0.99. This is not as good as found for SMICA, but it demonstrates rough consistency.

We also examine the one-point statistics of the reconstructions and compare with the theoretically expected normal product distribution (Eq.~\eqref{eq:normalproductdist}). For all reconstructions besides Commander, we find $s \simeq 5 \times 10^{-2}$ and $k \simeq 8.5$. These values of skewness and kurtosis are consistent with our estimates above for the contribution from the intrinsic non-Gaussianity of the unWISE blue map. The Commander reconstruction has several hundred pixels with extreme values that drive the skewness and kurtosis to $s=1.3$ and $k=260$. Masking pixels with values greater than 15 standard deviations in magnitude brings the skewness and kurtosis in line with the other component separated maps. Coarse-graining to large angular scales, the distribution of pixel values in each case approaches a Gaussian with the expected variance based on Eq.~\eqref{eq:pixelnoise}. On large angular scales, it is therefore an excellent approximation to treat the reconstruction as a random Gaussian field that, in the absence of a signal, is fully characterized by the constant estimator variance.

\begin{figure*}
\centering
    \includegraphics[width=.8\columnwidth]{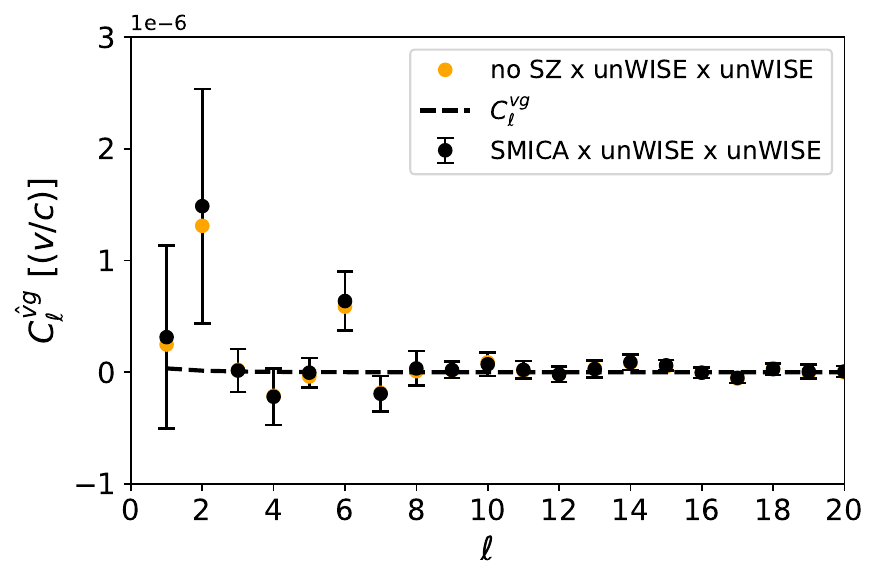}
    \includegraphics[width=.8\columnwidth]{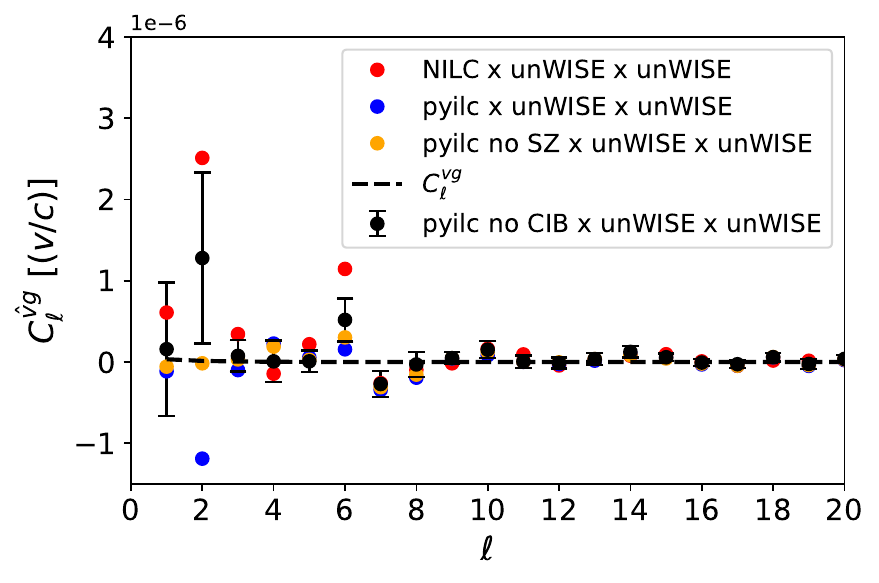}
    \caption{The cross-power between SMICA (left panel) and pyilc (right panel) reconstructed dipole fields and the unWISE galaxy density. In the left panel we show the SMICA no SZ reconstruction (orange points); in the right panel we show the reconstruction based on the Planck NILC (red points) and fiducial pyilc map (blue points) as well as the SZ deprojected (orange points) and CIB deprojected (black points) maps. The cut-sky spectra are scaled by a factor of $f_{\rm sky}^{-1}$ to approximate the full-sky cross-power. The error bars represent the expected variance from Eq.~\eqref{eq:vgnoise} using the empirical unWISE and reconstruction power spectra.}
    \label{fig:cross_compseparated}
\end{figure*}

\begin{figure}
\centering
    \includegraphics[width=.8\columnwidth]{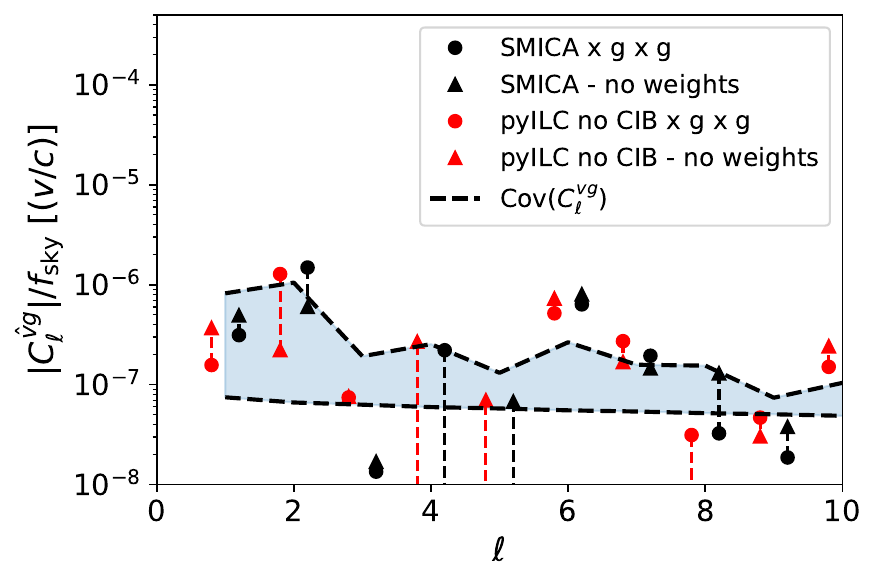}
    \caption{The magnitude of the cut-sky cross-power spectrum between unWISE blue and reconstruction with (circles) and without (triangles) applying the systematics weights for SMICA (black) and pyilc no CIB (red). The boundaries of the shaded region are the expected covariance for only reconstruction noise using the empirical (top) or theory (bottom) unWISE power spectrum.}
    \label{fig:cross_systematics}
\end{figure}

In Fig.~\ref{fig:cross_compseparated} we plot the cross-power between unWISE and reconstructions based on SMICA and SMICA no SZ (left panel) as well as Planck NILC, pyilc, pyilc no SZ, and pyilc no CIB (right panel). Spectra are computed on the masked sky and the maps are re-scaled by a factor of $f_{\rm sky}^{-1}$. There is little difference between the SMICA and SMICA no SZ results (left panel), and the cross-power at low-$\ell$ is roughly consistent with no signal in the cross-spectrum given the large expected covariance (from Eq.~\eqref{eq:vgnoise}). There is a far larger spread in results for the NILC-based maps (right panel) at low-$\ell$. These variations are still roughly within the range expected from the covariance. To investigate this further, we show the effect of including the unWISE systematics weights on the SMICA and pyilc no CIB reconstructions in Fig.~\ref{fig:cross_systematics}. Unlike for the individual frequency maps, there is not a systematic reduction in cross-power, and the cross-power remains within the expected covariance. We conclude that there is not evidence for residual correlated foregrounds and systematics in these component separated maps.

%% file: sections/posterior.tex
\section{Constraining the optical depth bias}
\label{sec:posterior}

\begin{figure}
\centering
    \includegraphics[width=.8\columnwidth]{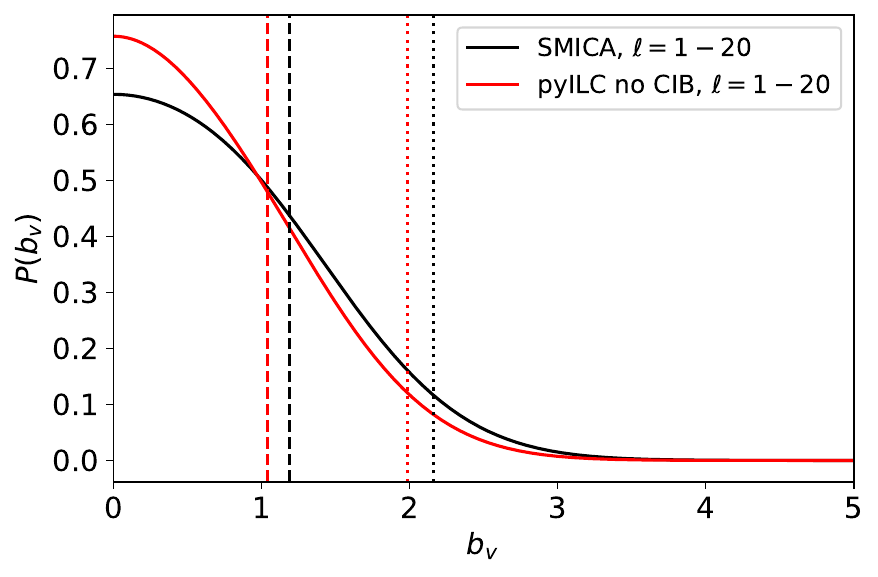}
    \caption{The posterior probability distribution over $b_v$ from the SMICA (black) and pyilc no CIB (red) dipole field reconstruction power spectra. The $68\%$ and $95\%$ confidence intervals are shown as dashed and dotted lines, respectively.}
    \label{fig:bv_posterior}
\end{figure}

We now attempt to constrain the optical depth bias, which controls the amplitude of the signal component of the reconstructed remote dipole field. As described in Sec.~\ref{sec:likelihoodandposterior}, we model the measured reconstruction power spectrum by Eq.~\eqref{eq:reconmodel}, which consists of the radial velocity autospectrum with fixed $\Lambda$CDM parameters scaled by the optical depth bias $b_v$ plus reconstruction noise set by the estimator prefactor values listed in Table~\ref{table:monodipo}. In Appendix~\ref{appendix:pdftest} we demonstrated that the posterior over $b_v$ can be obtained by evaluating the product of likelihood functions for the reconstruction power at each multipole over a grid of values for $b_v$ (Eq.~\eqref{eq:bvposterior}). We evaluate the posterior in Eq.~\eqref{eq:bvposterior} over a dense grid of values for $b_v$ in the range $0<b_v<10$ and compute the normalization by integrating over $b_v$. The input data are the reconstruction power spectra on the cut-sky. The result is shown in Fig.~\ref{fig:bv_posterior} for SMICA and pyilc no CIB reconstruction spectra. We choose $\ell_{\rm min} =1$ and $\ell_{\rm max} = 20$, corresponding to the range of multipoles shown in Figs.\ref{fig:smica_pspec} and~\ref{fig:pyilc_pspec}.  

The posteriors are both peaked at $b_v=0$, and we obtain an upper bound on $b_v$ by finding the values that encompass $68\%$ and $95\%$ of the posterior. For the SMICA reconstruction, we find $b_v < 1.04$ at $68\%$ and $b_v < 1.99$ at $95\%$ confidence. For the pyILC no CIB reconstruction, we find $b_v < 1.19$ at $68\%$ and $b_v < 2.16$ at $95\%$ confidence. These limits are consistent with the expected signal-to-noise of ${\rm SN} \sim 1$ computed in Eq.~\eqref{eq:Sndefinition} -- the constraint on signal amplitude scales inversely with signal-to-noise. 

We tested the robustness of our results to our choices for the minimum and maximum $\ell$ used in the posterior. Varying the maximum $\ell$, the posteriors remain peaked at $b_v=0$; comparing the constraint with only the dipole to the constraint keeping $\ell = 1-20$, the upper bounds shift by $\sim 10\%$. Because the signal falls steeply with $\ell$, changing the minimum $\ell$ has a far larger effect. Using $\ell = 2-20$ the posteriors shift and widen, yielding an upper limit of $b_v < 1.88$ at $68\%$ for SMICA and $b_v < 1.70$ at $68\%$ for pyILC no CIB. This is consistent with the forecast in Eq.~\eqref{eq:Sndefinition} that demonstrated the dipole carries roughly half the total signal-to-noise.

%% file: sections/conclusions.tex
\section{Conclusions}
\label{sec:conclude}

In this paper we have attempted to recover the remote dipole field using kSZ velocity reconstruction applied to CMB temperature data from Planck and galaxy density from the unWISE survey. We adapted the quadratic estimator formalism of Refs.~\cite{Terrana2017,Deutsch2018a,Smith2018,Cayuso2018,Cayuso2023} to large photometric redshift bins characterizing the unWISE blue sample. We characterized the expected signal, and the impact of possible systematics on the optical depth bias - a multiplicative bias on the estimator mean compared to the expected underlying signal. We applied our reconstruction pipeline to single-frequency Planck temperature maps as well as CMB maps produced using a variety component separation techniques. We identified the dominant foregrounds and systematics, and demonstrated that masking, monopole subtraction, and correcting large-scale galaxy survey systematics are effective strategies to minimize their impact. We found that the SMICA component separated map and the CIB-deprojected map computed using the {\tt pyilc} code have the smallest foreground residuals. These reconstructions were used to constrain the amplitude of the velocity bias to $b_v < 1.04$ at $68\%$ confidence (SMICA), where the fiducial value within $\Lambda$CDM with our modelling assumptions is $b_v=1.0$. This constraint is consistent with our forecasted signal-to-noise of $\mathcal{O}(1)$ for kSZ velocity reconstruction using Planck and the unWISE blue sample.

An important component of our analysis was to characterize the impact of foregrounds and systematics on the reconstruction. For reconstructions from individual frequency maps, we find a large, frequency-dependent reconstruction monopole in unmasked sky regions. A significant monopole is absent from reconstructions based on component-separated CMB maps. This is consistent with a statistically isotropic cross-correlation between non-blackbody CMB foregrounds and the unWISE galaxy density. Beyond the monopole, there is no statistically significant evidence for additive estimator biases arising from statistically anisotropic cross-correlations between CMB foregrounds and unWISE galaxy density. Both the monopole and the additive bias increase with frequency, which in combination with several other tests, strongly suggest that the CIB is the foreground that is responsible. Comparing reconstructions produced using various component separation techniques, we find that SMCIA and the CIB deprojected maps produced using {\tt pyilc} contained the lowest amplitude foreground residuals. We recommend using these products for the highest fidelity kSZ velocity reconstruction with Planck data.

Even in the absence of a detection, this data is useful for a wide variety of cosmological constraints. The measured reconstruction monopole is sensitive to bulk radial motion, as expected in e.g. void models. The observed reconstruction monopole can thus be seen as a constraint on homogeneity. The reconstruction dipole is the volume-average of the locally observed CMB dipole seen throughout the unWISE survey volume, and is sensitive to large bulk-flows and isocurvature modes. The observed reconstruction dipole can therefore be seen as a measurement of the difference between the rest frame of large-scale structure and the rest frame of the CMB. The cross-correlation between the reconstruction and a galaxy survey can be used to make inferences on large angular scales that are impossibly buried below systematics  in the galaxy autopower spectrum. This can place constraints on primordial non-Gaussianity and isocurvature. We explore the constraints on these scenarios in a companion paper~\cite{cosmopaper}.

There are several avenues for improving the current analysis. Moving from inverse-variance filtering to Wiener filtering the CMB on the cut sky would simplify the analysis and interpretation of reconstructions. Another improvement is to utilize better models of the galaxy-optical depth connection (with data-derived constraints on model parameters), and go beyond the linear filtering used to obtain the inferred optical depth map from the galaxy density. We have used only the unWISE blue sample, but applying our pipeline to the green and red samples is straightforward. One could then perform actual `tomography' by using the different redshift distributions to constrain the dipole field at different redshift. Finally, the techniques outlined here can be applied to other existing CMB and galaxy survey datasets.  The total signal-to-noise with various data combinations, e.g. ACT and DES, is comparable to the current analysis when trade-offs like sky coverage, CMB detector noise, and galaxy density are taken into account. However, these data combinations allow a more complete exploration of possible systematic effects.

Overall, our results strongly support the future program of kSZ velocity reconstruction/kSZ tomography. We have demonstrated that it is straightforward to mitigate the effects of systematics and foreground in the quadratic estimator formalism with Planck and unWISE-quality data. Looking ahead to data from Simons Observatory and large surveys such as Euclid, LSST and SPHEREx, we can expect dramatic improvements. Using spectroscopic surveys such as DESI and in the future MegaMapper, even more dramatic gains in information are possible. In preparation for these new datasets, it will be crucial to refine the analysis pipeline presented here using simulations and also to compare and integrate with complementary techniques and other kSZ estimators (see e.g. the overview in Ref.~\cite{Smith2018}). We can look forward to a bright future for exploring the most fundamental questions in cosmology using kSZ tomography.

\acknowledgments
We thank Boris Bolliet, Simone Ferraro, Gil Holder, Selim Hotinli, Moritz M\"unchmeyer, Emmanuel Schaan, and Kendrick Smith for helpful input at various stages of this project. In particular we thank Alex Krolewski for providing various unWISE data products and guiding us in their use. We thank Fiona McCarthy for access to the pyILC maps. We thank Gil Holder for suggesting that we examine the reconstruction monopole. MCJ is supported by the National Science and Engineering Research Council through a Discovery grant. Research at Perimeter Institute is supported in part by the Government of Canada through the Department of Innovation, Science and Economic Development Canada and by the Province of Ontario through the Ministry of Research, Innovation and Science. Some of the results in this paper have been derived using the HEALPix package~\cite{2005ApJ...622..759G} and the {\tt hmvec} code.

%% file: sections/bias_appendix.tex
\section{Estimating the magnitude of the optical depth bais}
\label{appendix:optical depth}

In this appendix we explore the plausible range over which the optical depth bias is expected to vary. We begin by deriving a formal expression for the optical depth bias. Taking the ensemble-average of the harmonic space quadratic estimator in Eq.~\eqref{eq:estimator} we have
\begin{eqnarray}
    \langle \hat{v}_{\ell m}\rangle = - N_\ell  \sum_{\ell_1 m_1;\ell_2 m_2} &\left(-1\right)^m&
    \begin{pmatrix}
        \ell_1 & \ell_2 & \ell \\
        m_1 & m_2 & -m
    \end{pmatrix}
    G_{\ell_1 \ell_2 \ell} \nonumber \\
    &\times&
    \langle \Theta_{\ell_1 m_1} \delta_{\ell_2 m_2} \rangle
\end{eqnarray}
Substituting with Eq.~\eqref{eq:map_crosspower} and utilizing the properties of the Wigner 3j symbols we have
\begin{eqnarray}
    \langle \hat{v}_{\ell m}\rangle^{\rm t} = \frac{N_\ell}{2\ell+1} && \sum_{\ell_1;\ell_2}  G_{\ell_1 \ell_2 \ell} [f_{\ell_1 \ell_2 \ell}]^{\rm t} \nonumber \\
   &\times& \int d\chi \ \frac{[C_{\ell=\bar{\ell}}^{\dot{\tau} g}\left(\chi\right)]^{\rm t}}{[\bar{C}_{\ell=\bar{\ell}}^{\tau g}]^{\rm t}} v_{\ell m} (\chi)
\end{eqnarray}
where we have explicitly labeled quantities that depend on the true underlying quantities by the `t' superscript. Re-arranging this slightly, we have
\begin{equation}
    \langle \hat{v}_{\ell m}\rangle^{\rm t} = \int d\chi \ b_v (\chi,\ell) W_v (\chi) v_{\ell m} (\chi)
\end{equation}
where $b_v$ is the optical depth bias, which in general depends on $\chi$ and on scale $\ell$, and which is defined as
\begin{eqnarray}\label{eq:bias_eqn}
    b_v (\chi,\ell) &\equiv& \frac{\sum_{\ell_1;\ell_2}  G_{\ell_1 \ell_2 \ell} f_{\ell_1 \ell_2 \ell} [\bar{C}_{\ell_2}^{\tau g}]^{\rm t}/\bar{C}_{\ell_2}^{\tau g} }{\sum_{\ell_1;\ell_2}  G_{\ell_1 \ell_2 \ell} f_{\ell_1 \ell_2 \ell}} \nonumber \\
    &\times& \frac{[C_{\ell=\bar{\ell}}^{\dot{\tau} g}\left(\chi\right)]^{\rm t}}{C_{\ell=\bar{\ell}}^{\dot{\tau} g}\left(\chi\right)} \frac{\bar{C}_{\ell=\bar{\ell}}^{\tau g}}{[\bar{C}_{\ell=\bar{\ell}}^{\tau g}]^{\rm t}} \\
    &\simeq& \frac{ \sum_{\ell_1} \frac{2 \ell_1 + 1}{4 \pi} \frac{\bar{C}_{\ell_1}^{\tau g}  [\bar{C}_{\ell_1}^{\tau g}]^{\rm t}}{C_{\ell_1}^{TT} C_{\ell_1}^{gg}}  }{  \sum_{\ell_2} \frac{2 \ell_2 + 1}{4 \pi} \frac{(\bar{C}_{\ell_2}^{\tau g})^2  }{C_{\ell_2}^{TT} C_{\ell_2}^{gg}}}
    \frac{[C_{\ell=\bar{\ell}}^{\dot{\tau} g}\left(\chi\right)]^{\rm t}}{C_{\ell=\bar{\ell}}^{\dot{\tau} g}\left(\chi\right)} \frac{\bar{C}_{\ell=\bar{\ell}}^{\tau g}}{[\bar{C}_{\ell=\bar{\ell}}^{\tau g}]^{\rm t}} \nonumber
\end{eqnarray}
The approximation in the last line is valid in the limit where $\ell \ll \ell_1, \ell_2$, which removes the scale-dependence of $b_v$. 

\begin{figure}
    \centering
    \includegraphics[width=1.\columnwidth]{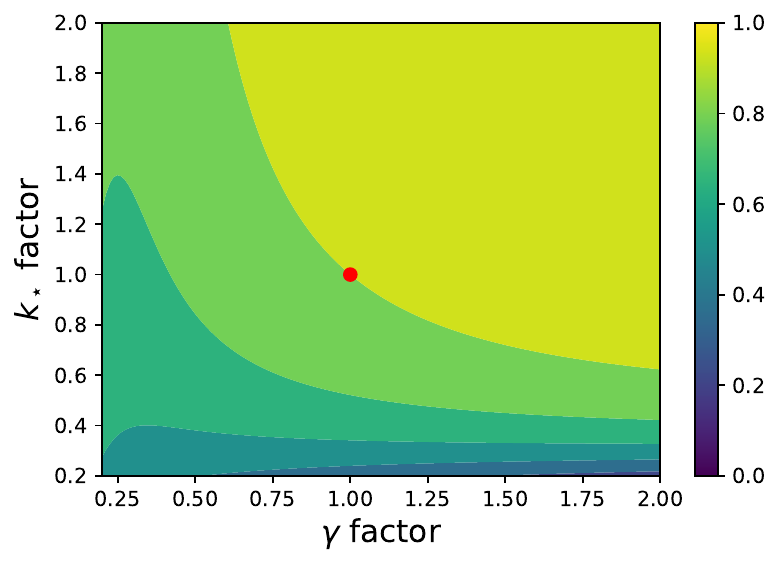}
    \caption{The weighted average of $b_v$ defined in Eq.~\eqref{eq:bvweightedavg} where $[b_e(\chi, \ell)]^{\rm t}$ is evaluated with $k_\star(z)$ and $\gamma(z)$ multiplied by a constant factor (values on the x- and y-axis). The contours shown are for $b_v= \{0.6,0.7,0.8,0.9,1.0 \}$ from dark to light. The red dot indicates the fiducial model.}
    \label{fig:bvsensitivity}
\end{figure}

\begin{figure}
    \centering
    \includegraphics[width=1.\columnwidth]{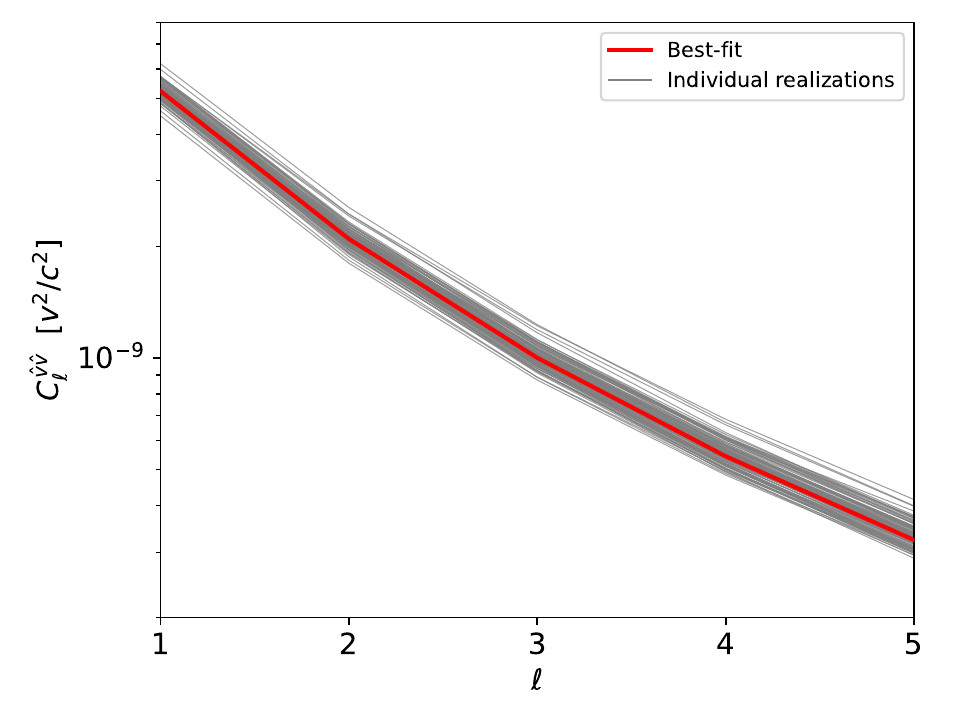}
    \caption{The estimator autopower spectrum evaluated with individual realizations of the galaxy window function (grey) plotted against the fidudial spectrum (red). The optical depth bias can be estimated as the square root of the ratio of the grey and red curves.}
    \label{fig:wg_bv_variations}
\end{figure}

If the true cross-spectra are well-described by our model with different parameter choices, we can use Eq.~\ref{eq:bias_eqn} to estimate the range of values we might plausibly expect the optical depth to take. Specializing to spectra described by Eq.~\eqref{eq:ctaugmodel}, we can write Eq.~\ref{eq:bias_eqn} as
\begin{eqnarray}
\label{eq:bvexpression}
    b_v (\chi) &\simeq& \frac{[\bar{n}_{e,0} ]^{\rm t}}{\bar{n}_{e,0}} \frac{b_e(\bar{\chi},\bar{\ell})}{[b_e(\bar{\chi},\bar{\ell})]^{\rm t}} \frac{ \sum_{\ell_1} \frac{2 \ell_1 + 1}{4 \pi} \frac{(\bar{C}_{\ell_2}^{\tau g})^2}{C_{\ell_1}^{TT} C_{\ell_1}^{gg}} \frac{[b_e(\bar{\chi},\ell_1)]^{\rm t}}{b_e(\bar{\chi},\ell_1)} }{  \sum_{\ell_2} \frac{2 \ell_2 + 1}{4 \pi} \frac{(\bar{C}_{\ell_2}^{\tau g})^2  }{C_{\ell_2}^{TT} C_{\ell_2}^{gg}}} \nonumber \\
    &\times& 
    \frac{[b_e(\chi,\bar{\ell})]^{\rm t}}{b_e(\chi,\bar{\ell})}
    \frac{[W_\mathrm{g}(\chi)]^{\rm t}}{W_\mathrm{g}(\chi)}
\end{eqnarray}
The factors on the first row are constant, while the factors on the second row depend on  $\chi$. For the model variations considered below, it is a reasonable approximation to replace $b_v(\chi)$ by a weighted average over the fiducial $W_v(\chi)$ and neglect the $\chi$-dependence. We define
\begin{eqnarray}\label{eq:bvweightedavg}
    b_v \equiv \int d\chi \ W_v(\chi) b_v(\chi)
\end{eqnarray}
It is straightforward to determine the impact of the various factors contributing to $b_v$, and to assess their expected relative importance. As a baseline, the results below assume Planck 217 GHz noise levels and unWISE blue number counts.
\begin{itemize}
    \item $\frac{[\bar{n}_{e,0} ]^{\rm t}}{\bar{n}_{e,0}}$: The mean electron density today is defined in Eq.~\eqref{eq:ne0}. Since the baryon abundance is well-measured, the most uncertain parameters in this expression are the gas fraction $f_{\rm gas}$ and the fraction of electrons that are ionized $X$. In our fiducial model we chose $f_{\rm gas} = 0.9$, which is plausible based on observational baryon inventories e.g.~\cite{2004ApJ...616..643F,2012ApJ...759...23S,PhysRevD.103.063513}. We also chose $X=1$, corresponding to the scenario where helium is completely ionized in the present Universe. Allowing for the gas fraction to be as low as $\sim 80\%$ and allowing for the scenario where helium is not completely ionized, this factor alone could produce an optical depth bias as small as $0.8$.  
    \item Factors involving $b_e$: The bias factor $b_e$ defined in Eq.~\eqref{eq:matter_electron_bias} depends on three redshift-dependent functions: $b_\star$, $k_\star$, and $\gamma$, which control the overall-amplitude, the scale at which feedback processes suppress structure, and how abrupt this suppression occurs, respectively. Among the model parameters, we expect $b_v$ to be most sensitive to $k_\star$ and $\gamma$ since $b_\star$ is a multiplicative constant that does not vary strongly over the redshifts spanned by unWISE. To provide a simple estimate of the sensitivity of $b_v$, we retain the fiducial $b_\star(z)$ and multiply the redshift-dependent parameters by a constant $f$ in the range $0.2 \leq f \leq 2$ such that $k_\star(z) \rightarrow f k_\star(z)$ and $\gamma \rightarrow f \gamma(z)$. The result is shown in Fig.~\ref{fig:bvsensitivity}. Qualitatively, $b_v> 1$ when there is less suppression of structure in the electron distribution than the fiducial model over the relevant scales/redshifts and $b_v < 1$ when there is more. If we consider these model variations as representative of the expected mis-match between our fiducial model and the `truth', then a plausible range we might expect due to modelling the electron-galaxy connection is $0.6 \lesssim b_v \lesssim 1.1$. 
    \item $\frac{[W_\mathrm{g}(\chi)]^{\rm t}}{W_\mathrm{g}(\chi)}$: As discussed in Sec.~\ref{ssec:spec_models}, there is significant uncertainty in the redshift distribution characterizing the unWISE blue sample. Here, we model $[W_g(\chi)]^{\rm t}$ by the individual realizations of the galaxy window function plotted in Fig.~\ref{fig:dndz}. Keeping the other factors that contribute to $b_v(\chi)$ in Eq.~\eqref{eq:bvexpression} fixed, note that $b_v(\chi) W_v(\chi)$ is simply given by $W_v(\chi)$ as computed with the realizations of the galaxy window function. Because the realizations vary strongly in redshift, we estimate $b_v$ by comparing $C_\ell^{\hat{v} \hat{v}}$ computed from the fiducial window function and $[C_\ell^{\hat{v} \hat{v}}]^{\rm t}$ computed from the individual galaxy window function realizations. These spectra are shown in Fig.~\ref{fig:wg_bv_variations}. At low-$\ell$, it can be seen that the shape of the power spectrum is retained, and we can estimate $b_v \simeq ( [C_\ell^{\hat{v} \hat{v}}]^{\rm t} / C_\ell^{\hat{v} \hat{v}} )^{1/2}$. This yields a range $0.9 \lesssim b_v \lesssim 1.1$.
\end{itemize}

From the investigation above, it appears difficult to increase the optical depth bias beyond $b_v \sim 1.1$. For example, when baryons perfectly trace dark matter on all scales (e.g. there is no feedback) we obtain $b_v \simeq 1.1$. The strongest influence in the context of our model is $b_e$, which can yield an optical depth bias as small as $b_v \sim 0.5$ when there is significant extra suppression in small-scale electron power. We therefore take our conservative, bottom-line plausible range of optical depth bias to be $0.5 < b_v < 1.1$.

\begin{figure}
    \centering
    \includegraphics[width=1.\columnwidth]{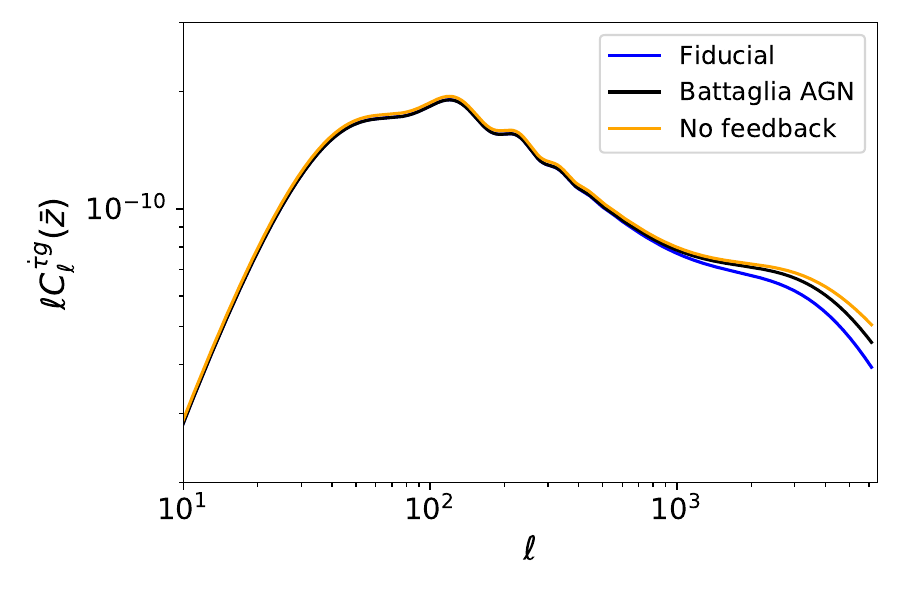}
    \caption{The galaxy-optical depth cross-power spectrum evaluated at the median redshift ($\bar{z} = 0.68$) for the fiducial model parameters chosen in this paper (Fiducial, blue) compared with the case where there is no baryonic feedback and electrons cluster like dark matter (No feedback, orange) as well as a halo model commonly adopted in the literature (Battaglia AGN, black).}
    \label{fig:Ctaug_comparisons}
\end{figure}

We note that our fiducial model for $b_e$ differs from a commonly adopted choice in other studies based on simulation-derived gas profiles in the context of the halo model. This model is described in Ref.~\cite{Smith2018}, and a computation of the electron-galaxy cross-power spectrum for this model is implemented in the {\tt hmvec} code using the 'Battaglia AGN' gas profiles. We approximate $[b_e(\chi,\ell)]^{\rm t} \simeq \sqrt{P_{ee}(\chi,\ell)/P_{mm}(\chi,\ell)}$. In Fig.~\ref{fig:Ctaug_comparisons} we compare the galaxy-optical depth power spectrum evaluated at $\bar{z}=0.68$ for our fiducial model (blue) against the Battaglia AGN halo model (black) and the case where electrons cluster like dark matter (orange) - the extreme scenario where there is no baryonic feedback. Since it has the largest power suppression, our fiducial model has stronger feedback than predicted by the Battaglia AGN gas profiles over the wavelengths and redshifts relevant to the unWISE blue, Planck data combination. The impact on $b_v$ is quite small, and we find that $b_v \simeq 1.05$ using our fiducual spectrum and assuming the Battaglia AGN model is the truth. This variation is smaller than many of the other uncertainties described above. 

We note that the numerical value of $b_v$ is entirely dependent on the combination of data considered. Therefore, caution should be exercised when comparing results obtained using different data combinations. For example, in the limit of no CMB noise (including scales $\ell<6000$), we estimate $b_v \simeq 1.15$ if the truth is Battaglia AGN, and $b_v \simeq 1.25$ if the truth is no feedback. If instead we had assumed Battaglia AGN as our baseline, and our fiducial model as the truth, then we would have assigned $b_v = 0.95$ for Planck and $b_v = 0.87$ in the noiseless limit. 

%% file: sections/simulation_appendix.tex
\section{Pipeline Validation and Tests}
\label{appendix:pipeline_validation}

In this appendix, we perform a set of validation tests of our kSZ velocity reconstruction pipeline. To facilitate some of these tests, we create a mock kSZ signal such that the cross-correlation between the CMB temperature and the galaxy field is precisely as assumed in Eq.~\eqref{eq:map_crosspower}. Specifically, we approximate the kSZ signal as 
\begin{eqnarray}\label{eq:mock_ksz}
    \Theta^{\rm kSZ} (\hat{n}) \simeq - \tau(\hat{n}) \int d\chi \ W_v(\chi) v(\hat{n},\chi)
\end{eqnarray}
where the projected optical depth and galaxy density are related in harmonic space by a linear filter
\begin{eqnarray}\label{eq:app_filter_gtau}
    \tau(\hat{n}) = \sum_{\ell m} \delta^g_{\ell m} \frac{\bar{C}^{\tau g}_\ell}{C_\ell^{gg}} Y_{\ell m}(\hat{n})
\end{eqnarray}
We apply this filtering operation to the actual unWISE blue map to create $\tau(\hat{n})$ assuming the fiducial spectra $\bar{C}^{\tau g}_\ell$ and $C_\ell^{gg}$ used in the estimator. We also create a full-sky Gaussian mock galaxy density map from the unWISE blue spectrum, as well as the corresponding optical depth map. We then draw random Gaussian realizations of $\int d\chi \ W_v(\chi) v(\hat{n},\chi)$ assuming the fiducial $\Lambda$CDM cosmological parameters, and multiply by $\tau(\hat{n})$ to create the mock signal Eq.~\eqref{eq:mock_ksz}. We create a mock random Gaussian CMB map $\Theta^{\rm CMB}(\hat{n})$ from the estimated 217 GHz power spectrum, which includes the primary CMB as well as noise and foregrounds at this frequency. We then create pairs of CMB maps: $\Theta (\hat{n}) = \Theta^{\rm CMB} (\hat{n}) + \Theta^{\rm kSZ} (\hat{n})$ and $\Theta^{\rm nokSZ} (\hat{n}) = \Theta^{\rm CMB}(\hat{n})$, where both share the same realization $\Theta^{\rm CMB} (\hat{n})$.

\subsection{Confirming the estimator provides an unbiased reconstruction}

As our first validation step, we apply the quadratic estimator pipeline on the full sky to the mock $\Theta$ and $\Theta^{\rm nokSZ}$ maps with the mock unWISE blue map to produce two reconstructions $\hat{v}$ and $\hat{v}^{\rm nokSZ}$. We repeat this for  maps with a mock kSZ signal from the actual unWISE blue map, masking the result with the fiducial reconstruction mask. In both cases, the pairs of reconstructions will have very nearly the same reconstruction noise (since the kSZ contribution to the reconstruction noise is small), and therefore if the estimator is faithfully reconstructing the velocity field, we expect that $\hat{v} (\hat{n}) - \hat{v}^{\rm nokSZ} (\hat{n}) = \langle \hat{v} (\hat{n}) \rangle \simeq v (\hat{n})$. 

\begin{figure*}
\centering
    \includegraphics[width=.48 \textwidth]{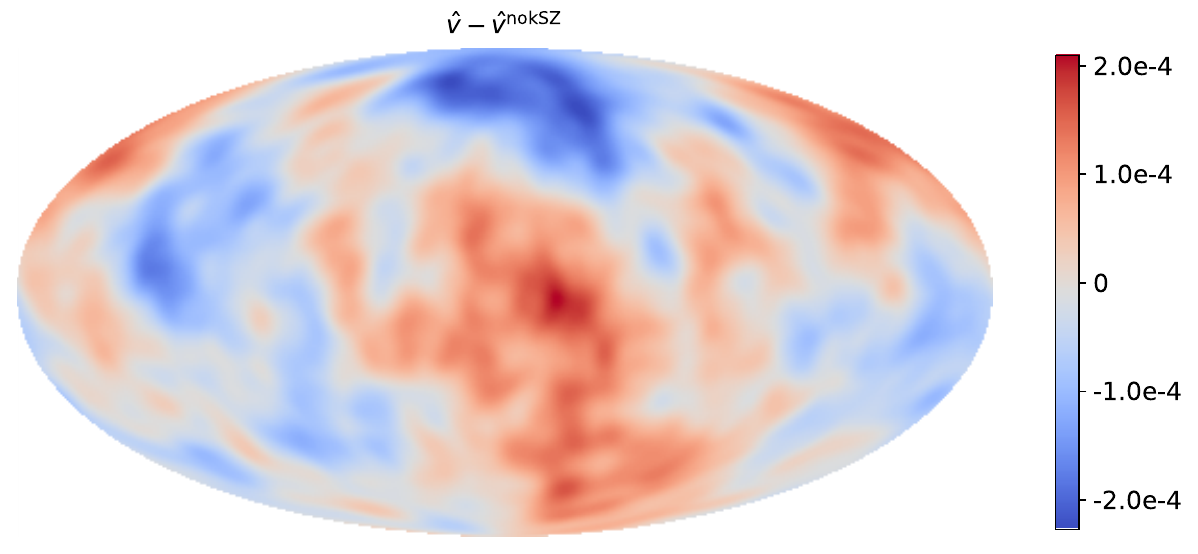}
    \includegraphics[width=.48 \textwidth]{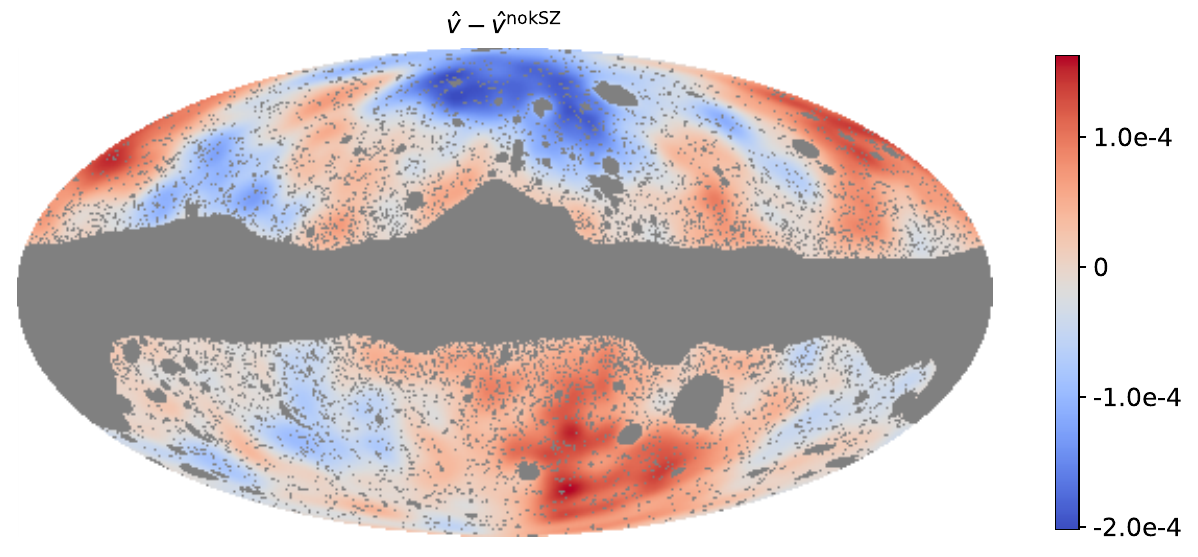}
    \includegraphics[width=.48 \textwidth]{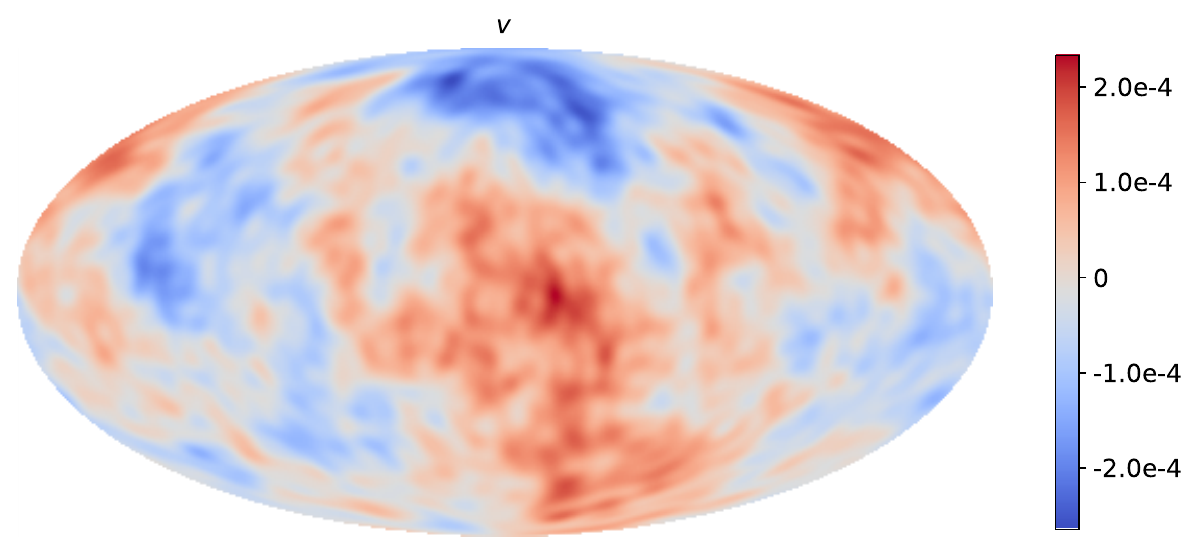}
    \includegraphics[width=.48 \textwidth]{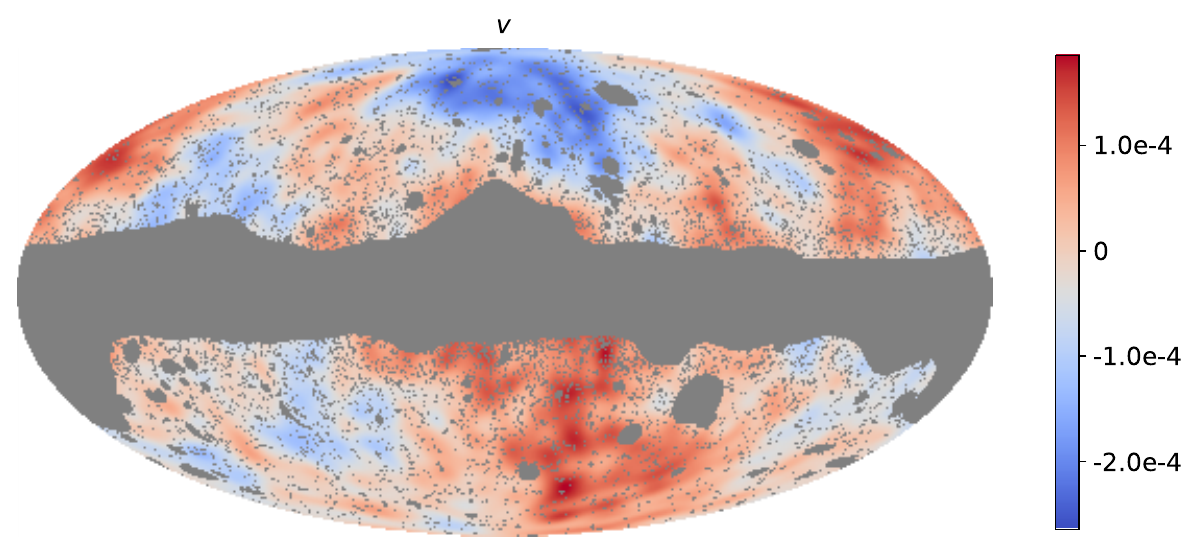}
    \caption{In the left column, we compare $\hat{v} (\hat{n}) - \hat{v}^{\rm nokSZ} (\hat{n})$ (top) with the true remote dipole signal (bottom) for mock unWISE galaxies and the corresponding optical depth template. In the right column, we compare $\hat{v} (\hat{n}) - \hat{v}^{\rm nokSZ} (\hat{n})$ (top) with the true remote dipole signal (bottom) for actual unWISE galaxies and the corresponding optical depth template. Reconstruction maps have been smoothed with a $5.75^\circ$ Gaussian beam. In both cases, the agreement is excellent.}
    \label{fig:simrecmaps_signal}
\end{figure*}

\begin{figure*}
    \centering
    \includegraphics[width=1.\columnwidth]{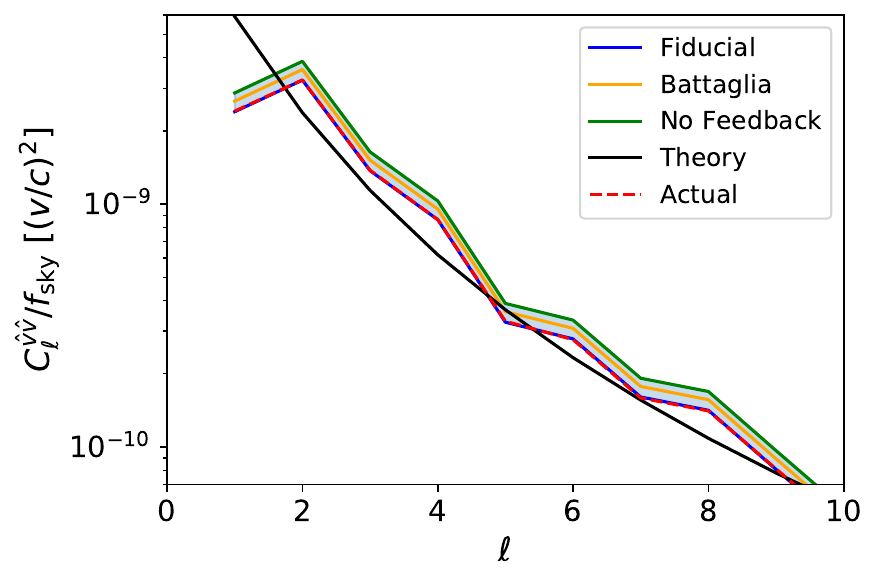}
    \includegraphics[width=1.\columnwidth]{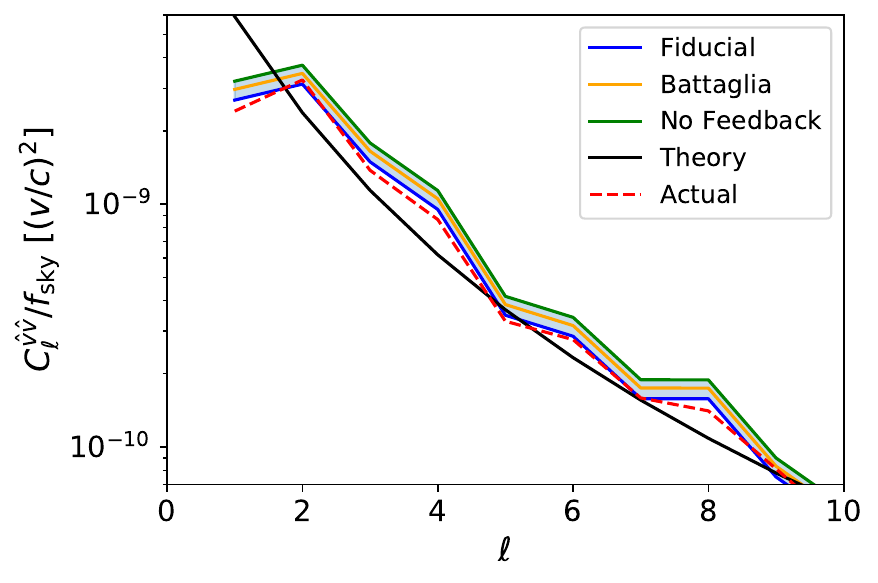}
    \caption{Cut-sky spectra of the reconstructions in Fig.~\ref{fig:simrecmaps_signal} based on mock unWISE (left) with mock 217 GHz CMB and actual unWISE wih mock 217 GHz CMB (right). The reconstruction difference spectrum for a kSZ signal created using the fiducial electron model (blue solid) can be compared with the actual remote dipole spectrum (red dashed), and against the cases where the true electron model is Battaglia AGN (orange solid) or no feedback (green solid).}
    \label{fig:sim_diffspec}
\end{figure*}

In Fig.~\ref{fig:simrecmaps_signal} we compare the reconstruction difference maps with the injected signal for the mock unWISE (left column) and actual unWISE (right column) optical depth templates. In both cases, the agreement is excellent. In Fig.~\ref{fig:sim_diffspec} we compare the cut-sky power spectra for difference reconstructions made using mock unWISE (left; blue solid) and actual unWISE (right; blue solid) to the actual underlying remote dipole field (red dashed). For mock unWISE, the agreement is nearly exact. We expect this, since all of the assumptions underlying the quadratic estimator are exactly satisfied. For actual unWISE, the agreement is still good, but there are visible differences at the $\sim 5 \%$-level. We speculate that this small difference represents the impact of going from a random Gaussian galaxy density to the true non-Gaussian statistics of the  unWISE map. This small difference is irrelevant in the small signal-to-noise regime accessible with Planck, and we conclude that the estimator produces an unbiased reconstruction of the remote dipole field when the fiducial galaxy-optical depth model matches the truth.

To investigate the optical depth bias, we create mock kSZ maps using a modified filter in Eq.~\eqref{eq:app_filter_gtau} where $\bar{C}_{\ell}^{\tau g}$ is determined by the Battaglia AGN and no-feedback cases discussed in Appendix~\ref{appendix:optical depth}. Keeping the fiducial model in the estimator, the reconstruction difference spectra are shown in Fig.~\ref{fig:sim_diffspec}. Here, we see the reconstruction is biased. As anticipated in Appendix~\ref{appendix:optical depth}, the bias is scale-independent at low-$\ell$. Furthermore, the numerical value of the bias is almost precisely equal to the estimates made there: $b_v = 1.052$ for Battaglia AGN and $b_v = 1.094$ for no feedback. Within the assumptions made in deriving the estimator, and the construction of our mock signal, we confirm the description of the optical depth bias in Appendix~\ref{appendix:optical depth}.

In Fig.~\ref{fig:sim_specwnoise}, we compare the full spectra of the reconstructions based on actual unWISE galaxy density where the mock 217 GHz CMB does (green solid) or doesn't (orange solid) contain the kSZ signal. This is compared against the expected reconstruction noise (the estimator pre-factor, black dashed) as well as the difference spectra shown previously in Fig.~\ref{fig:sim_diffspec} (same color scheme). Focusing on the spectrum at $\ell < 500$, we see that the scale-independent estimator prefactor defined in Eq.~\eqref{eq:const_est_n} is an excellent approximation to the estimator variance on large angular scales. This can be viewed as a self-consistency check on the approximate position-space estimator. The deviations at high-$\ell$ are irrelevant for any of the measurements presented in this paper, but could be accounted for by using the full expression for the harmonic space estimator's variance Eq.~\eqref{eq:variance_general}. Turning to the low-$\ell$ part of the signal, there is a non-negligible difference in the reconstruction spectra at $\ell < 4$. However, the addition in power from the signal at low multipoles is quite small, and would be difficult to identify.


\begin{figure}
    \centering  \includegraphics[width=1.\columnwidth]{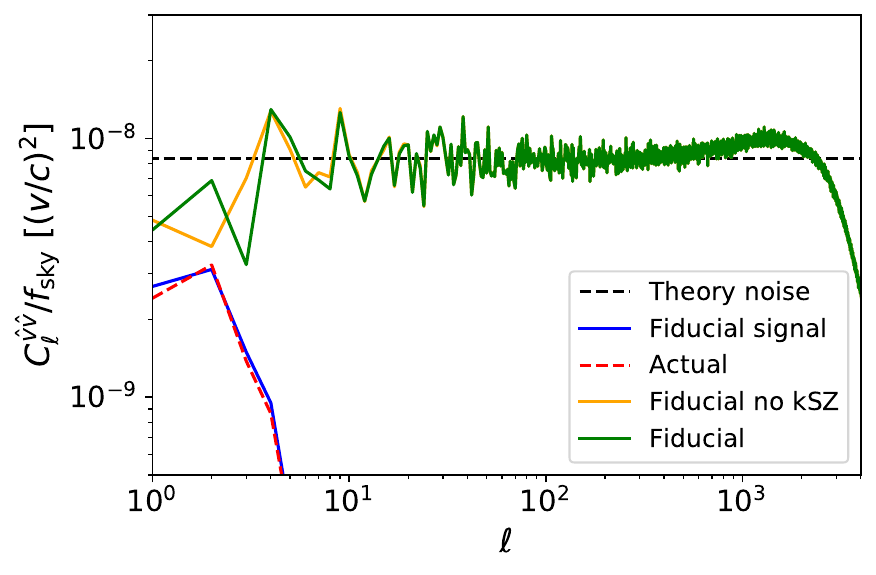}
    \caption{Full reconstruction angular power spectrum of maps with (green solid) and without (orange solid) a kSZ signal against the estimator pre-factor (Theory noise, black dashed) and the reconstruction difference map (blue) as well as the spectrum of the true  remote dipole field (red dashed). }
    \label{fig:sim_specwnoise}
\end{figure}

\subsection{Sensitivity to estimator weights}

The filtered fields defined in Eq.~\eqref{eq:map_filters} and the normalization defined in Eq.~\eqref{eq:const_est_n} depend on estimates of the full-sky power spectra $C_\ell^{TT}$, $C_\ell^{gg}$, and $\bar{C}_\ell^{\tau g}$. As discussed in Appendix~\ref{appendix:optical depth}, mis-estimating $\bar{C}_\ell^{\tau g}$ results in a biased reconstruction. A poor estimate of $C_\ell^{TT}$ or $C_\ell^{gg}$ from data on the maksed sky also results in a disagreement between the estimator noise variance (e.g. reconstruction noise) and the estimator pre-factor $N$. 

 In our fiducial pipeline, we use the Master technique to deconvolve the mask and arrive at an estimate of the full-sky power spectrum. In this case, we found good agreement between the estimator variance and the pre-factor. We can compare this against the case where power spectra are estimated by dividing the cut-sky power spectra (computed after removing the monopole) by the fraction of masked pixels. Here we find a $\sim 5-10\%$ difference between the estimator variance and pre-factor. In contrast, deconvolving the mask to compute the unWISE power spectrum had a far smaller effect, and we find nearly identical results to simply dividing the cut-sky spectrum by the unmasked sky fraction. We therefore utilize the Master technique as implemented in {\tt pymaster} to compute the temperature power spectra.

\section{The masked spectra}
\label{appendix:pdftest}

\begin{figure*}
\centering
    \includegraphics[width=.4
    \textwidth]{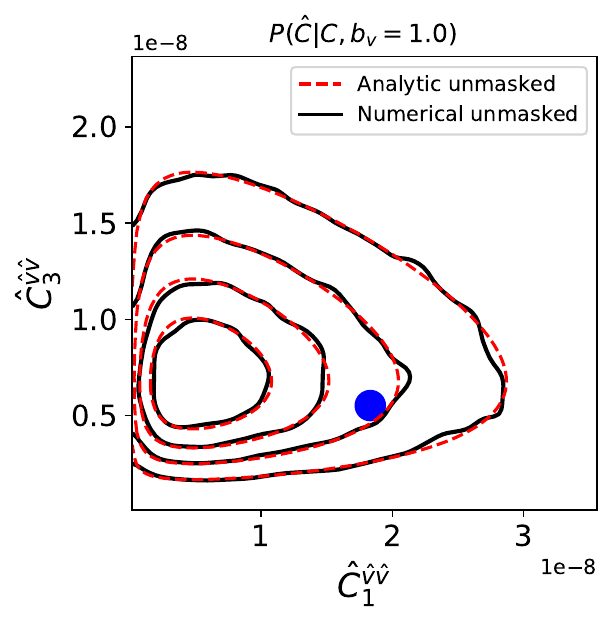}
    \includegraphics[width=.4 \textwidth]{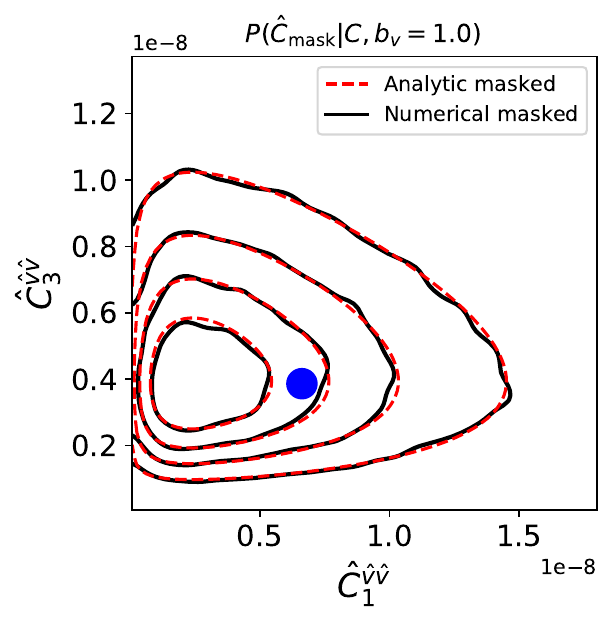}
    \includegraphics[width=.4 \textwidth]{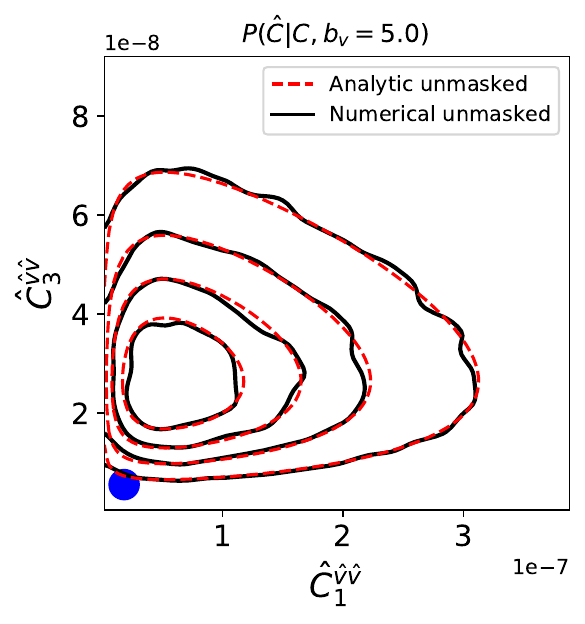}
    \includegraphics[width=.4 \textwidth]{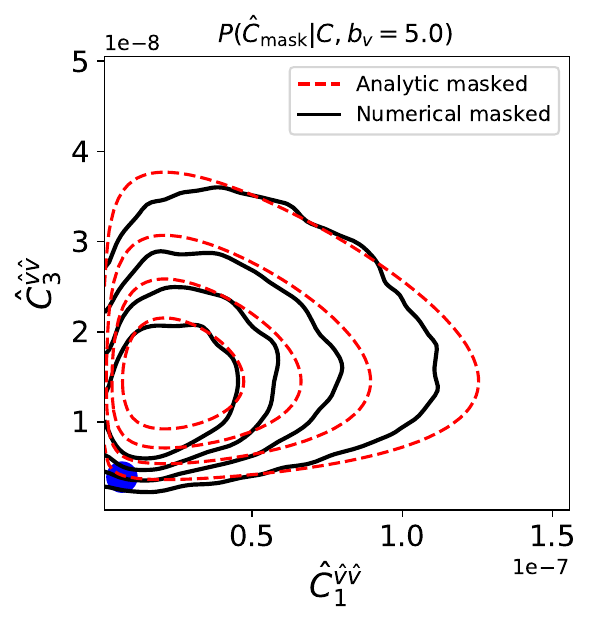}
    \caption{The posterior over $\hat{C}^{\hat{v} \hat{v}}_\ell$ in the $\ell=1,3$ plane without (left) and with (right) the unWISE mask for $b_v=1.$ (top) and $b_v=5.$ (bottom). We compare the analytic posterior Eq.~\eqref{eq:cvvlikelihood} (red solid) to a numerical posterior generated from random Gaussian samples. For reference, the blue dot represents the values of the full-sky and masked-sky spectra for the simulated Planck 217 GHz x unWISE reconstruction shown in Fig.~\eqref{fig:sim_specwnoise}).    }
    \label{fig:likelihood_comparisons}
\end{figure*}

In this appendix we validate our use of Eq.~\eqref{eq:bvposterior} to compute the posterior over $b_v$ using spectra computed on the masked sky. We first create $10^5$ random Gaussian simulations of the remote dipole field signal and $10^5$ realizations of the reconstruction noise expected for Planck 217 GHz and unWISE on the full sky, assuming the fiducial cosmological parameters. For each realization, we create a map of the signal plus noise on a grid of values of $b_v = [0,5]$. We then compute the power spectrum of each map both with and without a mask. We use these samples to create a numerical multi-variate probability distribution for the full-sky and cut-sky spectra across $1 \leq \ell \leq 10$, at each value of $b_v$. This numerical distribution preserves any coupling between the spectra at different values of $\ell$. We also compute the assumed distribution Eq.~\eqref{eq:bvposterior} based on the product of the distributions Eq.~\eqref{eq:cvvlikelihood} at each $\ell$ (evaluated using cut-sky spectra).

In Fig.~\ref{fig:likelihood_comparisons} we plot a slice through the numerical (black solid) and analytic (red dashed) likelihood for $\hat{C}^{\hat{v} \hat{v}}_\ell$ with $b_v = 1.0$ (top row) and $b_v = 5.0$ (bottom row) in the $\ell=1,3$ plane for full-sky (left) and masked (right) spectra. For small $b_v$, where the signal is dominated by reconstruction noise, there is no coupling between different $\ell$ and the analytic and numerical likelihoods agree well. For larger $b_v$, where the signal becomes significantly more important than the reconstruction noise, the mask-induced mode coupling is apparent in the covariance between $\hat{C}^{\hat{v} \hat{v}}_1$ and $\hat{C}^{\hat{v} \hat{v}}_3$ for the numerical likelihood in the bottom-right panel. For reference, we also plot (blue point) the measured full-sky and masked spectra from the simulation realization used in Sec.~\eqref{appendix:pipeline_validation}. Note that this lies near the peak of the likelihood for $b_v=1$ and near the tails for $b_v=5$.

\begin{figure*}
    \centering  \includegraphics[width=.65\columnwidth]{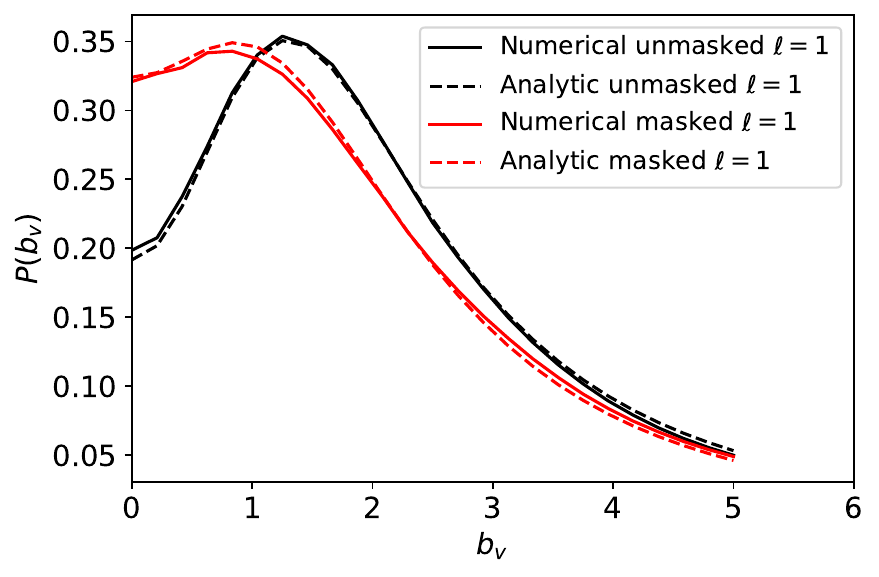}
    \includegraphics[width=.65\columnwidth]{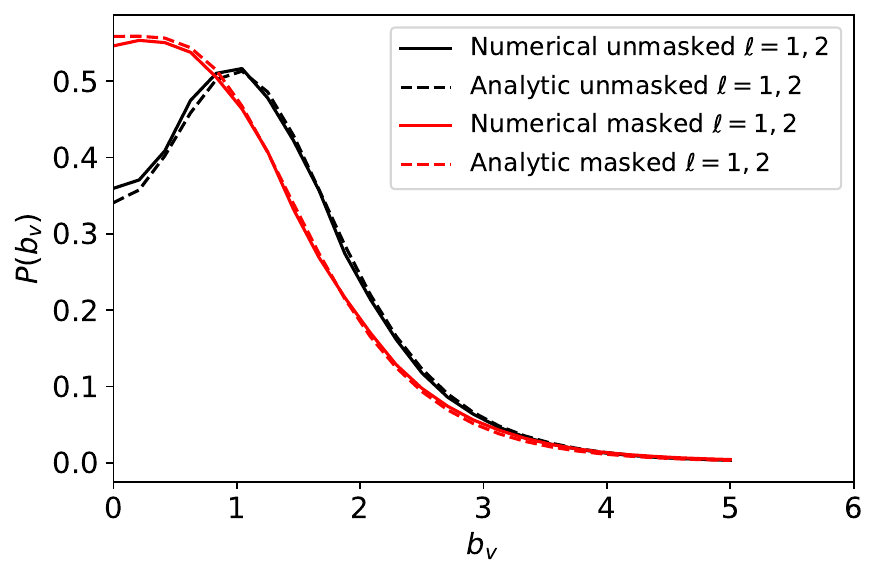}
    \includegraphics[width=.65\columnwidth]{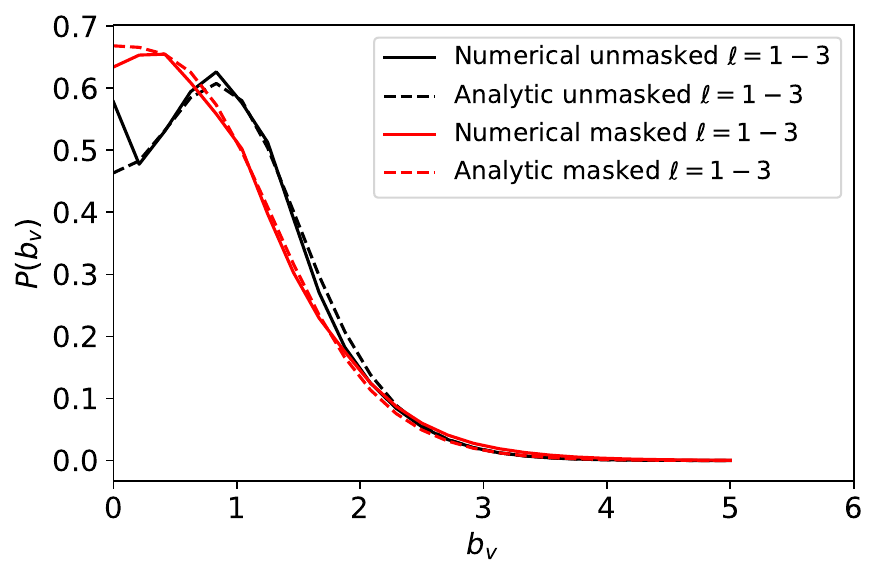}
    \caption{The posterior over $b_v$ including $\ell =1$ (left), $\ell=1,2$ (middle), and $\ell=1,2,3$ (right). }
    \label{fig:bvpost_sims}
\end{figure*}

In Fig.~\ref{fig:bvpost_sims} we show the posterior for $b_v$ that results from the analytic and numerical likelihoods over the full sky and masked spectra given the measured spectra from the simulated 217 GHz x unWISE realization. The agreement between the two is excellent on both the full and masked sky. The mask-induced mode coupling for large $b_v$ in Fig.~\eqref{fig:likelihood_comparisons} is simply too far in the tails to make a significant difference in the final posterior. We include three panels in Fig.~\ref{fig:bvpost_sims} where the product over likelihoods Eq.~\eqref{eq:bvposterior} is truncated at $\ell=1$ (left; this constraint includes only information from the dipole), $\ell=2$ (middle), and $\ell = 3$ (right). The posterior tightens as more information is included, but because the signal spectrum falls steeply with $\ell$, there is not a significant change in the posterior for $\ell_{\rm max} > 3$. Comparing the posteriors from full-sky to masked sky spectra, we see that the posterior shifts towards zero and broadens as a result of masking.

\begin{figure}
    \centering  \includegraphics[width=.8\columnwidth]{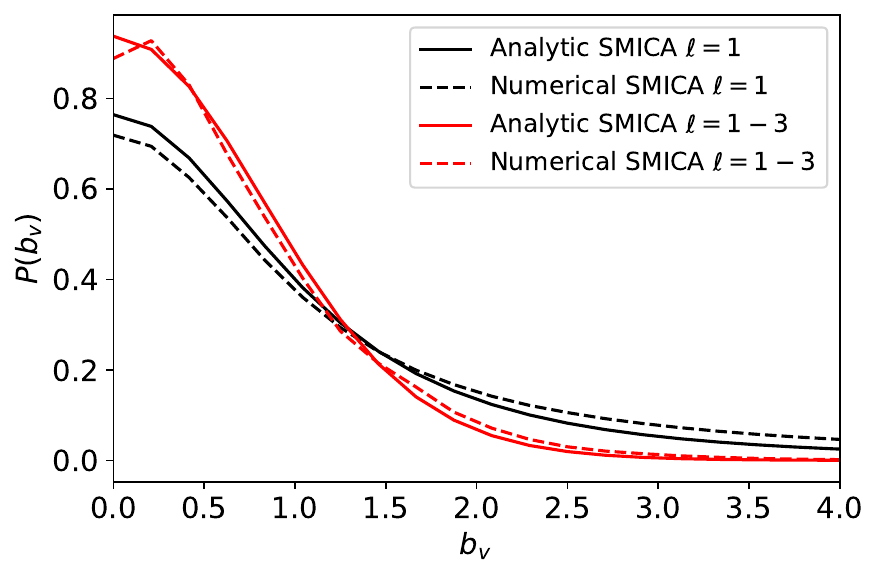}
    \caption{The posterior over $b_v$ from the SMICA unWISE reconstruction including $\ell =1$ (black) and $\ell=1,2,3$ (red) using the analytic (solid) or numerical (dashed) likelihoods. }
    \label{fig:bvpost_sims2}
\end{figure}

To determine if these results transfer to our analysis of actual maps, we compute the numerical and analytic likelihoods appropriate for the SMICA and unWISE reconstructions on the masked sky. We compare the resulting posteriors over $b_v$ in Fig.~\ref{fig:bvpost_sims2} for the case where the likelihood has only the $\ell=1$ spectrum and the case where the likelihood has the $\ell = 1, 2, 3$ spectra. We see again that there is good agreement between the analytic and numerical posteriors for this data combination.

In summary, we conclude that it is appropriate to use an analytic likelihood that neglects mode coupling to compute the posterior over $b_v$.

%% file: planck_unwise_refs.bib
@article{Planck:2020olo,
    author = "Akrami, Y. and others",
    collaboration = "Planck",
    title = "{$Planck$ intermediate results. LVII. Joint Planck LFI and HFI data processing}",
    eprint = "2007.04997",
    archivePrefix = "arXiv",
    primaryClass = "astro-ph.CO",
    doi = "10.1051/0004-6361/202038073",
    journal = "Astron. Astrophys.",
    volume = "643",
    pages = "A42",
    year = "2020"
}

@article{McCarthy:2023hpa,
    author = "McCarthy, Fiona and Hill, J. Colin",
    title = "{Component-separated, CIB-cleaned thermal Sunyaev-Zel{\textquoteright}dovich maps from Planck PR4 data with a flexible public needlet ILC pipeline}",
    eprint = "2307.01043",
    archivePrefix = "arXiv",
    primaryClass = "astro-ph.CO",
    doi = "10.1103/PhysRevD.109.023528",
    journal = "Phys. Rev. D",
    volume = "109",
    number = "2",
    pages = "023528",
    year = "2024"
}

@article{Lague:2024czc,
    author = {Lagu{\"e}, Alex and Madhavacheril, Mathew S. and Smith, Kendrick M. and Ferraro, Simone and Schaan, Emmanuel},
    title = "{Constraints on Local Primordial Non-Gaussianity with 3D Velocity Reconstruction from the Kinetic Sunyaev-Zeldovich Effect}",
    eprint = "2411.08240",
    archivePrefix = "arXiv",
    primaryClass = "astro-ph.CO",
    doi = "10.1103/PhysRevLett.134.151003",
    journal = "Phys. Rev. Lett.",
    volume = "134",
    number = "15",
    pages = "151003",
    year = "2025"
}

@article{McCarthy:2024nik,
    author = "McCarthy, Fiona and others",
    title = "{The Atacama Cosmology Telescope: Large-scale velocity reconstruction with the kinematic Sunyaev-Zel'dovich effect and DESI LRGs}",
    eprint = "2410.06229",
    archivePrefix = "arXiv",
    primaryClass = "astro-ph.CO",
    doi = "10.1088/1475-7516/2025/05/057",
    journal = "JCAP",
    volume = "05",
    pages = "057",
    year = "2025"
}

@misc{Lai:2025qdw,
      title="{KSZ Velocity Reconstruction with ACT and DESI-LS using a Tomographic QML Power Spectrum Estimator}", 
      author={Anderson C. M. Lai and Yurii Kvasiuk and Moritz Münchmeyer},
      year={2025},
      eprint={2506.21684},
      archivePrefix={arXiv},
      primaryClass={astro-ph.CO},
      url={https://arxiv.org/abs/2506.21684}, 
}

@misc{Krywonos:2024mpb,
      title="{Constraints on cosmology beyond LCDM with kinetic Sunyaev Zel'dovich velocity reconstruction}", 
      author={Jordan Krywonos and Selim C. Hotinli and Matthew C. Johnson},
      year={2024},
      eprint={2408.05264},
      archivePrefix={arXiv},
      primaryClass={astro-ph.CO},
      url={https://arxiv.org/abs/2408.05264}, 
}

@article{Shaw:2011sy,
    author = "Shaw, Laurie D. and Rudd, Douglas H. and Nagai, Daisuke",
    title = "{Deconstructing the kinetic SZ Power Spectrum}",
    eprint = "1109.0553",
    archivePrefix = "arXiv",
    primaryClass = "astro-ph.CO",
    doi = "10.1088/0004-637X/756/1/15",
    journal = "Astrophys. J.",
    volume = "756",
    pages = "15",
    year = "2012"
}

@article{Springel:2017tpz,
    author = "Springel, Volker and others",
    title = "{First results from the IllustrisTNG simulations: matter and galaxy clustering}",
    eprint = "1707.03397",
    archivePrefix = "arXiv",
    primaryClass = "astro-ph.GA",
    journal = "Mon. Not. Roy. Astron. Soc.",
    volume = "475",
    number = "1",
    pages = "676--698",
    year = "2018"
}

@article{Smith:2016lnt,
    author = "Smith, Kendrick M. and Ferraro, Simone",
    title = "{Detecting Patchy Reionization in the Cosmic Microwave Background}",
    eprint = "1607.01769",
    archivePrefix = "arXiv",
    primaryClass = "astro-ph.CO",
    doi = "10.1103/PhysRevLett.119.021301",
    journal = "Phys. Rev. Lett.",
    volume = "119",
    number = "2",
    pages = "021301",
    year = "2017"
}

@misc{Hotinli:2025tul,
      title={Velocity Reconstruction from KSZ: Measuring $f_{NL}$ with ACT and DESILS}, 
      author={Selim C. Hotinli and Kendrick M. Smith and Simone Ferraro},
      year={2025},
      eprint={2506.21657},
      archivePrefix={arXiv},
      primaryClass={astro-ph.CO},
      url={https://arxiv.org/abs/2506.21657}, 
}

@article{Louis_2017,
   title={The Atacama Cosmology Telescope: two-season ACTPol spectra and parameters},
   volume={2017},
   ISSN={1475-7516},
   url={http://dx.doi.org/10.1088/1475-7516/2017/06/031},
   DOI={10.1088/1475-7516/2017/06/031},
   number={06},
   journal={Journal of Cosmology and Astroparticle Physics},
   publisher={IOP Publishing},
   author={Louis, Thibaut and others},
   year={2017},
   month=jun, pages={031–031} }

@article{Krolewski2021,
	author = {Krolewski, Alex and Ferraro, Simone and White, Martin},
	doi = {10.1088/1475-7516/2021/12/028},
	issn = {1475-7516},
	journal = {Journal of Cosmology and Astroparticle Physics},
	month = dec,
	number = {12},
	pages = {028},
	publisher = {IOP Publishing},
	title = {Cosmological constraints from unWISE and Planck CMB lensing tomography},
	url = {http://dx.doi.org/10.1088/1475-7516/2021/12/028},
	volume = {2021},
	year = {2021}}

@article{Laigle:2016jxn,
	archiveprefix = {arXiv},
	author = {Laigle, C. and others},
	doi = {10.3847/0067-0049/224/2/24},
	eprint = {1604.02350},
	journal = {Astrophys. J. Suppl.},
	number = {2},
	pages = {24},
	primaryclass = {astro-ph.GA},
	reportnumber = {APJS, 224, 24-(2016)},
	title = {{The COSMOS2015 Catalog: Exploring the 1\ensuremath{<} z\ensuremath{<} 6 Universe with half a million galaxies}},
	volume = {224},
	year = {2016}}

@article{Yan:2023okq,
	archiveprefix = {arXiv},
	author = {Yan, Ziang and Maniyar, Abhishek S. and van Waerbeke, Ludovic},
	eprint = {2310.10848},
	month = {10},
	primaryclass = {astro-ph.CO},
	title = {{The star formation, dust, and abundance of galaxies with unWISE-CIB cross-correlations}},
    journal = "",
	year = {2023}}

@article{2023arXiv230905659F,
    author = "Farren, Gerrit S. and others",
    collaboration = "ACT",
    title = "{The Atacama Cosmology Telescope: Cosmology from cross-correlations of unWISE galaxies and ACT DR6 CMB lensing}",
    eprint = "2309.05659",
    archivePrefix = "arXiv",
    journal = "",
    primaryClass = "astro-ph.CO",
    month = "9",
    year = "2023"
}

@article{2010AJ....140.1868W,
	adsurl = {https://ui.adsabs.harvard.edu/abs/2010AJ....140.1868W},
	archiveprefix = {arXiv},
	author = {{Wright}, Edward L. and others},
	doi = {10.1088/0004-6256/140/6/1868},
	eprint = {1008.0031},
	journal = {The Astronomical Journal},
	month = dec,
	number = {6},
	pages = {1868-1881},
	primaryclass = {astro-ph.IM},
	title = {{The Wide-field Infrared Survey Explorer (WISE): Mission Description and Initial On-orbit Performance}},
	volume = {140},
	year = 2010}

@article{Meisner_2017a,
	author = {Meisner, A. M. and Lang, D. and Schlegel, D. J.},
	doi = {10.3847/1538-3881/aa894e},
	issn = {1538-3881},
	journal = {The Astronomical Journal},
	month = sep,
	number = {4},
	pages = {161},
	publisher = {American Astronomical Society},
	title = {Deep Full-sky Coadds from Three Years of WISE and NEOWISE Observations},
	url = {http://dx.doi.org/10.3847/1538-3881/aa894e},
	volume = {154},
	year = {2017}}

@article{Meisner_2017,
	author = {Meisner, Aaron M. and Lang, Dustin and Schlegel, David J.},
	doi = {10.3847/1538-3881/153/1/38},
	issn = {1538-3881},
	journal = {The Astronomical Journal},
	month = jan,
	number = {1},
	pages = {38},
	publisher = {American Astronomical Society},
	title = {FULL-DEPTH COADDS OF THE WISE AND FIRST-YEAR NEOWISE-REACTIVATION IMAGES},
	url = {http://dx.doi.org/10.3847/1538-3881/153/1/38},
	volume = {153},
	year = {2017}}

@article{Lang_2014,
	author = {Lang, Dustin},
	doi = {10.1088/0004-6256/147/5/108},
	issn = {1538-3881},
	journal = {The Astronomical Journal},
	month = apr,
	number = {5},
	pages = {108},
	publisher = {American Astronomical Society},
	title = {unWISE: UNBLURRED COADDS OF THE WISE IMAGING},
	url = {http://dx.doi.org/10.1088/0004-6256/147/5/108},
	volume = {147},
	year = {2014}}

@article{2004ApJ...616..643F,
	adsurl = {https://ui.adsabs.harvard.edu/abs/2004ApJ...616..643F},
	archiveprefix = {arXiv},
	author = {{Fukugita}, Masataka and {Peebles}, P.~J.~E.},
	doi = {10.1086/425155},
	eprint = {astro-ph/0406095},
	journal = {\apj},
	month = dec,
	number = {2},
	pages = {643-668},
	primaryclass = {astro-ph},
	title = {{The Cosmic Energy Inventory}},
	volume = {616},
	year = 2004}

@article{PhysRevD.103.063513,
	author = {Schaan, Emmanuel and others},
	collaboration = {Atacama Cosmology Telescope Collaboration},
	doi = {10.1103/PhysRevD.103.063513},
	issue = {6},
	journal = {Phys. Rev. D},
	month = {Mar},
	numpages = {26},
	pages = {063513},
	publisher = {American Physical Society},
	title = {Atacama Cosmology Telescope: Combined kinematic and thermal Sunyaev-Zel'dovich measurements from BOSS CMASS and LOWZ halos},
	url = {https://link.aps.org/doi/10.1103/PhysRevD.103.063513},
	volume = {103},
	year = {2021}}

@article{2012ApJ...759...23S,
	adsurl = {https://ui.adsabs.harvard.edu/abs/2012ApJ...759...23S},
	archiveprefix = {arXiv},
	author = {{Shull}, J. Michael and {Smith}, Britton D. and {Danforth}, Charles W.},
	doi = {10.1088/0004-637X/759/1/23},
	eid = {23},
	eprint = {1112.2706},
	journal = {\apj},
	month = nov,
	number = {1},
	pages = {23},
	primaryclass = {astro-ph.CO},
	title = {{The Baryon Census in a Multiphase Intergalactic Medium: 30\% of the Baryons May Still be Missing}},
	volume = {759},
	year = 2012}

@article{Bolton_2016,
	author = {Bolton, James S. and Puchwein, Ewald and Sijacki, Debora and Haehnelt, Martin G. and Kim, Tae-Sun and Meiksin, Avery and Regan, John A. and Viel, Matteo},
	doi = {10.1093/mnras/stw2397},
	issn = {1365-2966},
	journal = {Monthly Notices of the Royal Astronomical Society},
	month = sep,
	number = {1},
	pages = {897--914},
	publisher = {Oxford University Press (OUP)},
	title = {The Sherwood simulation suite: overview and data comparisons with the Lyman alpha forest at redshifts 2 < z < 5},
	url = {http://dx.doi.org/10.1093/mnras/stw2397},
	volume = {464},
	year = {2016}}

@article{Plante_2018,
	author = {La Plante, Paul and Trac, Hy and Croft, Rupert and Cen, Renyue},
	doi = {10.3847/1538-4357/aae693},
	issn = {1538-4357},
	journal = {The Astrophysical Journal},
	month = nov,
	number = {2},
	pages = {106},
	publisher = {American Astronomical Society},
	title = {Helium Reionization Simulations. III. The Helium Ly$\alpha$ Forest},
	url = {http://dx.doi.org/10.3847/1538-4357/aae693},
	volume = {868},
	year = {2018}}

@article{La_Plante_2017,
	author = {La Plante, Paul and Trac, Hy and Croft, Rupert and Cen, Renyue},
	doi = {10.3847/1538-4357/aa7136},
	issn = {1538-4357},
	journal = {The Astrophysical Journal},
	month = may,
	number = {2},
	pages = {87},
	publisher = {American Astronomical Society},
	title = {Helium Reionization Simulations. II. Signatures of Quasar Activity on the IGM},
	url = {http://dx.doi.org/10.3847/1538-4357/aa7136},
	volume = {841},
	year = {2017}}

@article{Upton_Sanderbeck_2016,
	author = {Upton Sanderbeck, Phoebe R. and D'Aloisio, Anson and McQuinn, Matthew J.},
	doi = {10.1093/mnras/stw1117},
	issn = {1365-2966},
	journal = {Monthly Notices of the Royal Astronomical Society},
	month = may,
	number = {2},
	pages = {1885--1897},
	publisher = {Oxford University Press (OUP)},
	title = {Models of the thermal evolution of the intergalactic medium after reionization},
	url = {http://dx.doi.org/10.1093/mnras/stw1117},
	volume = {460},
	year = {2016}}

@article{Boera_2015,
	author = {Boera, Elisa and Murphy, Michael T. and Becker, George D. and Bolton, James S.},
	doi = {10.1093/mnrasl/slv172},
	issn = {1745-3933},
	journal = {Monthly Notices of the Royal Astronomical Society: Letters},
	month = dec,
	number = {1},
	pages = {L79--L83},
	publisher = {Oxford University Press (OUP)},
	title = {Constraining the temperature--density relation of the intergalactic medium with the Lyman $\alpha$ and $\beta$ forests},
	url = {http://dx.doi.org/10.1093/mnrasl/slv172},
	volume = {456},
	year = {2015}}

@article{Viel_2013,
	author = {Viel, Matteo and Becker, George D. and Bolton, James S. and Haehnelt, Martin G.},
	doi = {10.1103/physrevd.88.043502},
	issn = {1550-2368},
	journal = {Physical Review D},
	month = aug,
	number = {4},
	publisher = {American Physical Society (APS)},
	title = {Warm dark matter as a solution to the small scale crisis: New constraints from high redshift Lyman-alpha forest data},
	url = {http://dx.doi.org/10.1103/PhysRevD.88.043502},
	volume = {88},
	year = {2013}}

@article{Calura_2012,
	author = {Calura, F. and Tescari, E. and D'Odorico, V. and Viel, M. and Cristiani, S. and Kim, T.-S. and Bolton, J. S.},
	doi = {10.1111/j.1365-2966.2012.20811.x},
	issn = {0035-8711},
	journal = {Monthly Notices of the Royal Astronomical Society},
	month = apr,
	number = {4},
	pages = {3019--3036},
	publisher = {Oxford University Press (OUP)},
	title = {The Lyman $\alpha$ forest flux probability distribution at $z>3$},
	url = {http://dx.doi.org/10.1111/j.1365-2966.2012.20811.x},
	volume = {422},
	year = {2012}}

@article{Alizadeh_2012,
	author = {Alizadeh, Esfandiar and Hirata, Christopher M.},
	issn = {1550-2368},
	journal = {Physical Review D},
	month = jun,
	number = {12},
	publisher = {American Physical Society (APS)},
	title = {How to detect gravitational waves through the cross correlation of the galaxy distribution with the CMB polarization},
	volume = {85},
	year = {2012}}

@article{Madhavacheril:2019buy,
	archiveprefix = {arXiv},
	author = {Madhavacheril, Mathew S. and Battaglia, Nicholas and Smith, Kendrick M. and Sievers, Jonathan L.},
	doi = {10.1103/PhysRevD.100.103532},
	eprint = {1901.02418},
	journal = {Phys. Rev. D},
	number = {10},
	pages = {103532},
	primaryclass = {astro-ph.CO},
	title = {{Cosmology with the kinematic Sunyaev-Zeldovich effect: Breaking the optical depth degeneracy with fast radio bursts}},
	volume = {100},
	year = {2019}}

@article{Kumar:2022bly,
	archiveprefix = {arXiv},
	author = {Kumar, Neha Anil and Hotinli, Selim C. and Kamionkowski, Marc},
	doi = {10.1103/PhysRevD.107.043504},
	eprint = {2208.02829},
	journal = {Phys. Rev. D},
	number = {4},
	pages = {043504},
	primaryclass = {astro-ph.CO},
	title = {{Uncorrelated compensated isocurvature perturbations from kinetic Sunyaev-Zeldovich tomography}},
	volume = {107},
	year = {2023}}

@article{AnilKumar:2022flx,
	archiveprefix = {arXiv},
	author = {Anil Kumar, Neha and Sato-Polito, Gabriela and Kamionkowski, Marc and Hotinli, Selim C.},
	doi = {10.1103/PhysRevD.106.063533},
	eprint = {2205.03423},
	journal = {Phys. Rev. D},
	number = {6},
	pages = {063533},
	primaryclass = {astro-ph.CO},
	title = {{Primordial trispectrum from kinetic Sunyaev-Zel\textquoteright{}dovich tomography}},
	volume = {106},
	year = {2022}}

@article{Carlstrom_2011,
	author = {Carlstrom, J. E. and others},
	doi = {10.1086/659879},
	issn = {1538-3873},
	journal = {Publications of the Astronomical Society of the Pacific},
	month = may,
	number = {903},
	pages = {568--581},
	publisher = {IOP Publishing},
	title = {The 10 Meter South Pole Telescope},
	url = {http://dx.doi.org/10.1086/659879},
	volume = {123},
	year = {2011}}

@article{Farren:2023yna,
	archiveprefix = {arXiv},
	author = {Farren, Gerrit S. and Sherwin, Blake D. and Bolliet, Boris and Namikawa, Toshiya and Ferraro, Simone and Krolewski, Alex},
	eprint = {2311.04213},
	month = {11},
    journal = "",
	primaryclass = {astro-ph.CO},
	title = {{Detection of the CMB lensing -- galaxy bispectrum}},
	year = {2023}}

@article{Kusiak:2022xkt,
	archiveprefix = {arXiv},
	author = {Kusiak, Aleksandra and Bolliet, Boris and Krolewski, Alex and Hill, J. Colin},
	doi = {10.1103/PhysRevD.106.123517},
	eprint = {2203.12583},
	journal = {Phys. Rev. D},
	number = {12},
	pages = {123517},
	primaryclass = {astro-ph.CO},
	title = {{Constraining the galaxy-halo connection of infrared-selected unWISE galaxies with galaxy clustering and galaxy-CMB lensing power spectra}},
	volume = {106},
	year = {2022}}

@article{Krolewski:2021znk,
	archiveprefix = {arXiv},
	author = {Krolewski, Alex and Ferraro, Simone},
	doi = {10.1088/1475-7516/2022/04/033},
	eprint = {2110.13959},
	journal = {JCAP},
	number = {04},
	pages = {033},
	primaryclass = {astro-ph.CO},
	title = {{The Integrated Sachs Wolfe effect: unWISE and Planck constraints on dynamical dark energy}},
	volume = {04},
	year = {2022}}

@unpublished{cosmopaper,
	author = {Bloch, Richard and Hotinli, Selim C and Krywonos, Jordan and Johnson, Matthew C},
	note = {To appear},
	title = {Cosmological constraints from kSZ tomography}}

@article{Ferraro:2016ymw,
	archiveprefix = {arXiv},
	author = {Ferraro, Simone and Hill, J. Colin and Battaglia, Nick and Liu, Jia and Spergel, David N.},
	doi = {10.1103/PhysRevD.94.123526},
	eprint = {1605.02722},
	journal = {Phys. Rev. D},
	number = {12},
	pages = {123526},
	primaryclass = {astro-ph.CO},
	title = {{Kinematic Sunyaev-Zel\textquoteright{}dovich effect with projected fields. II. Prospects, challenges, and comparison with simulations}},
	volume = {94},
	year = {2016}}

@article{Ferraro:2014msa,
	archiveprefix = {arXiv},
	author = {Ferraro, Simone and Sherwin, Blake D. and Spergel, David N.},
	doi = {10.1103/PhysRevD.91.083533},
	eprint = {1401.1193},
	journal = {Phys. Rev. D},
	number = {8},
	pages = {083533},
	primaryclass = {astro-ph.CO},
	title = {{WISE measurement of the integrated Sachs-Wolfe effect}},
	volume = {91},
	year = {2015}}

@article{Marques:2019aug,
	archiveprefix = {arXiv},
	author = {Marques, Gabriela A. and Bernui, Armando},
	doi = {10.1088/1475-7516/2020/05/052},
	eprint = {1908.04854},
	journal = {JCAP},
	pages = {052},
	primaryclass = {astro-ph.CO},
	title = {{Tomographic analyses of the CMB lensing and galaxy clustering to probe the linear structure growth}},
	volume = {05},
	year = {2020}}

@article{Hill:2016dta,
	archiveprefix = {arXiv},
	author = {Hill, J. Colin and Ferraro, Simone and Battaglia, Nick and Liu, Jia and Spergel, David N.},
	doi = {10.1103/PhysRevLett.117.051301},
	eprint = {1603.01608},
	journal = {Phys. Rev. Lett.},
	number = {5},
	pages = {051301},
	primaryclass = {astro-ph.CO},
	title = {{Kinematic Sunyaev-Zel\textquoteright{}dovich Effect with Projected Fields: A Novel Probe of the Baryon Distribution with Planck, WMAP, and WISE Data}},
	volume = {117},
	year = {2016}}

@article{2005ApJ...622..759G,
	adsurl = {https://ui.adsabs.harvard.edu/abs/2005ApJ...622..759G},
	archiveprefix = {arXiv},
	author = {{G{\'o}rski}, K.~M. and {Hivon}, E. and {Banday}, A.~J. and {Wandelt}, B.~D. and {Hansen}, F.~K. and {Reinecke}, M. and {Bartelmann}, M.},
	doi = {10.1086/427976},
	eprint = {astro-ph/0409513},
	journal = {\apj},
	month = apr,
	number = {2},
	pages = {759-771},
	primaryclass = {astro-ph},
	title = {{HEALPix: A Framework for High-Resolution Discretization and Fast Analysis of Data Distributed on the Sphere}},
	volume = {622},
	year = 2005}

@article{Schlafly_2019,
	author = {Schlafly, Edward F. and Meisner, Aaron M. and Green, Gregory M.},
	doi = {10.3847/1538-4365/aafbea},
	issn = {1538-4365},
	journal = {The Astrophysical Journal Supplement Series},
	month = feb,
	number = {2},
	pages = {30},
	publisher = {American Astronomical Society},
	title = {The unWISE Catalog: Two Billion Infrared Sources from Five Years of WISE Imaging},
	url = {http://dx.doi.org/10.3847/1538-4365/aafbea},
	volume = {240},
	year = {2019}}

@article{Planck:2015txa,
	archiveprefix = {arXiv},
	author = {Ade, P. A. R. and others},
	collaboration = {Planck},
	doi = {10.1051/0004-6361/201527103},
	eprint = {1509.06348},
	journal = {Astron. Astrophys.},
	pages = {A12},
	primaryclass = {astro-ph.CO},
	title = {{Planck 2015 results. XII. Full Focal Plane simulations}},
	volume = {594},
	year = {2016}}

@article{Planck:2018yye,
	archiveprefix = {arXiv},
	author = {Akrami, Y. and others},
	collaboration = {Planck},
	doi = {10.1051/0004-6361/201833881},
	eprint = {1807.06208},
	journal = {Astron. Astrophys.},
	pages = {A4},
	primaryclass = {astro-ph.CO},
	title = {{Planck 2018 results. IV. Diffuse component separation}},
	volume = {641},
	year = {2020}}

@article{Planck:2020qil,
	archiveprefix = {arXiv},
	author = {Akrami, Y. and others},
	collaboration = {Planck},
	doi = {10.1051/0004-6361/202038053},
	eprint = {2003.12646},
	journal = {Astron. Astrophys.},
	pages = {A100},
	primaryclass = {astro-ph.CO},
	title = {{Planck intermediate results. LVI. Detection of the CMB dipole through modulation of the thermal Sunyaev-Zeldovich effect: Eppur si muove II}},
	volume = {644},
	year = {2020}}

@article{Battaglia:2016xbi,
	archiveprefix = {arXiv},
	author = {Battaglia, Nicholas},
	doi = {10.1088/1475-7516/2016/08/058},
	eprint = {1607.02442},
	journal = {JCAP},
	pages = {058},
	primaryclass = {astro-ph.CO},
	title = {{The Tau of Galaxy Clusters}},
	volume = {08},
	year = {2016}}

@article{Contreras:2019bxy,
	archiveprefix = {arXiv},
	author = {Contreras, Dagoberto and Johnson, Matthew C. and Mertens, James B.},
	doi = {10.1088/1475-7516/2019/10/024},
	eprint = {1904.10033},
	journal = {JCAP},
	pages = {024},
	primaryclass = {astro-ph.CO},
	title = {{Towards detection of relativistic effects in galaxy number counts using kSZ Tomography}},
	volume = {10},
	year = {2019}}

@article{Cayuso:2019hen,
	archiveprefix = {arXiv},
	author = {Cayuso, Juan I. and Johnson, Matthew C.},
	doi = {10.1103/PhysRevD.101.123508},
	eprint = {1904.10981},
	journal = {Phys. Rev. D},
	number = {12},
	pages = {123508},
	primaryclass = {astro-ph.CO},
	title = {{Towards testing CMB anomalies using the kinetic and polarized Sunyaev-Zel\textquoteright{}dovich effects}},
	volume = {101},
	year = {2020}}

@article{Kusiak:2021hai,
	archiveprefix = {arXiv},
	author = {Kusiak, Aleksandra and Bolliet, Boris and Ferraro, Simone and Hill, J. Colin and Krolewski, Alex},
	doi = {10.1103/PhysRevD.104.043518},
	eprint = {2102.01068},
	journal = {Phys. Rev. D},
	number = {4},
	pages = {043518},
	primaryclass = {astro-ph.CO},
	title = {{Constraining the baryon abundance with the kinematic Sunyaev-Zel\textquoteright{}dovich effect: Projected-field detection using Planck, WMAP, and unWISE}},
	volume = {104},
	year = {2021}}

@article{DESI:2016fyo,
	archiveprefix = {arXiv},
	author = {Aghamousa, Amir and others},
	collaboration = {DESI},
	eprint = {1611.00036},
	month = {10},
	primaryclass = {astro-ph.IM},
	reportnumber = {FERMILAB-PUB-16-517-AE},
    journal = "",
	title = {{The DESI Experiment Part I: Science,Targeting, and Survey Design}},
	year = {2016}}

@article{Contreras:2022zdz,
	archiveprefix = {arXiv},
	author = {Contreras, Dagoberto and McCarthy, Fiona and Johnson, Matthew C.},
	doi = {10.1103/PhysRevD.107.023521},
	eprint = {2205.15779},
	journal = {Phys. Rev. D},
	number = {2},
	pages = {023521},
	primaryclass = {astro-ph.CO},
	title = {{Maximum likelihood kinetic Sunyaev-Zel\textquoteright{}dovich velocity reconstruction}},
	volume = {107},
	year = {2023}}

@article{Kvasiuk:2023nje,
    author = {Kvasiuk, Yurii and M\"unchmeyer, Moritz},
    title = "{Autodifferentiable likelihood pipeline for the cross-correlation of CMB and large-scale structure due to the kinetic Sunyaev-Zeldovich effect}",
    eprint = "2305.08903",
    archivePrefix = "arXiv",
    primaryClass = "astro-ph.CO",
    doi = "10.1103/PhysRevD.109.083515",
    journal = "Phys. Rev. D",
    volume = "109",
    number = "8",
    pages = "083515",
    year = "2024"
}

@article{Giri:2020pkk,
	archiveprefix = {arXiv},
	author = {Giri, Utkarsh and Smith, Kendrick M.},
	doi = {10.1088/1475-7516/2022/09/028},
	eprint = {2010.07193},
	journal = {JCAP},
	pages = {028},
	primaryclass = {astro-ph.CO},
	title = {{Exploring KSZ velocity reconstruction with N-body simulations and the halo~model}},
	volume = {09},
	year = {2022}}

@article{Bernardis2017,
	author = {Bernardis, F. De and others},
	doi = {10.1088/1475-7516/2017/03/008},
	issn = {1475-7516},
	journal = {Journal of Cosmology and Astroparticle Physics},
	month = {mar},
	number = {03},
	pages = {008--008},
	title = {{Detection of the pairwise kinematic Sunyaev-Zel'dovich effect with BOSS DR11 and the Atacama Cosmology Telescope}},
	url = {https://iopscience-iop-org.ezproxy.library.yorku.ca/article/10.1088/1475-7516/2017/03/008/pdf http://stacks.iop.org/1475-7516/2017/i=03/a=008?key=crossref.78b444f57ff6c7877a813335c464e085},
	volume = {2017},
	year = {2017}}

@article{Chen2022,
	archiveprefix = {arXiv},
	arxivid = {2109.04092},
	author = {Chen, Ziyang and Zhang, Pengjie and Yang, Xiaohu and Zheng, Yi},
	doi = {10.1093/mnras/stab3604},
	eprint = {2109.04092},
	issn = {0035-8711},
	journal = {Monthly Notices of the Royal Astronomical Society},
	month = {jan},
	number = {4},
	pages = {5916--5928},
	title = {{Detection of pairwise kSZ effect with DESI galaxy clusters and Planck}},
	url = {https://academic.oup.com/mnras/article/510/4/5916/6461109},
	volume = {510},
	year = {2022}}

@article{Hand2012,
	author = {Hand, Nick and others},
	doi = {10.1103/PhysRevLett.109.041101},
	issn = {0031-9007},
	journal = {Physical Review Letters},
	month = {jul},
	number = {4},
	pages = {041101},
	title = {{Evidence of Galaxy Cluster Motions with the Kinematic Sunyaev-Zel'dovich Effect}},
	url = {https://journals-aps-org.ezproxy.library.yorku.ca/prl/pdf/10.1103/PhysRevLett.109.041101 https://link.aps.org/doi/10.1103/PhysRevLett.109.041101},
	volume = {109},
	year = {2012}}

@article{Munchmeyer2018,
	archiveprefix = {arXiv},
	arxivid = {1810.13424},
	author = {M{\"{u}}nchmeyer, Moritz and Madhavacheril, Mathew S and Ferraro, Simone and Johnson, Matthew C and Smith, Kendrick M},
	doi = {10.1103/PhysRevD.100.083508},
	eprint = {1810.13424},
	issn = {2470-0010},
	journal = {Physical Review D},
	month = {oct},
	number = {8},
	pages = {083508},
	title = {{Constraining local non-Gaussianities with kinetic Sunyaev-Zel'dovich tomography}},
	url = {https://arxiv.org/pdf/1810.13424.pdf http://arxiv.org/abs/1810.13424 https://link.aps.org/doi/10.1103/PhysRevD.100.083508},
	volume = {100},
	year = {2019}}

@article{Pan2019,
	archiveprefix = {arXiv},
	arxivid = {1906.04208},
	author = {Pan, Zhen and Johnson, Matthew C.},
	doi = {10.1103/PhysRevD.100.083522},
	eprint = {1906.04208},
	issn = {2470-0010},
	journal = {Physical Review D},
	month = {oct},
	number = {8},
	pages = {083522},
	title = {{Forecasted constraints on modified gravity from Sunyaev-Zel'dovich tomography}},
	url = {https://link.aps.org/doi/10.1103/PhysRevD.100.083522},
	volume = {100},
	year = {2019}}

@article{Terrana2017,
	archiveprefix = {arXiv},
	arxivid = {1610.06919v2},
	author = {Terrana, Alexandra and Harris, Mary-Jean and Johnson, Matthew C},
	doi = {10.1088/1475-7516/2017/02/040},
	eprint = {1610.06919v2},
	issn = {1475-7516},
	journal = {Journal of Cosmology and Astroparticle Physics},
	month = {feb},
	number = {02},
	pages = {040--040},
	title = {{Analyzing the cosmic variance limit of remote dipole measurements of the cosmic microwave background using the large-scale kinetic Sunyaev Zel'dovich effect}},
	url = {https://arxiv.org/pdf/1610.06919.pdf http://stacks.iop.org/1475-7516/2017/i=02/a=040?key=crossref.1ba3b1ca0174bb00e5dcbb97d328a226},
	volume = {2017},
	year = {2017}}

@article{Deutsch2018a,
	author = {Deutsch, Anne-Sylvie and Dimastrogiovanni, Emanuela and Johnson, Matthew C. and M{\"{u}}nchmeyer, Moritz and Terrana, Alexandra},
	doi = {10.1103/PhysRevD.98.123501},
	issn = {2470-0010},
	journal = {Physical Review D},
	month = {dec},
	number = {12},
	pages = {123501},
	publisher = {American Physical Society},
	title = {{Reconstruction of the remote dipole and quadrupole fields from the kinetic Sunyaev Zel'dovich and polarized Sunyaev Zel'dovich effects}},
	url = {https://doi.org/10.1103/PhysRevD.98.123501 https://link.aps.org/doi/10.1103/PhysRevD.98.123501},
	volume = {98},
	year = {2018}}

@article{Cayuso2018,
	archiveprefix = {arXiv},
	arxivid = {1806.01290},
	author = {Cayuso, Juan I. and Johnson, Matthew C. and Mertens, James B.},
	doi = {10.1103/PhysRevD.98.063502},
	eprint = {1806.01290},
	issn = {2470-0010},
	journal = {Physical Review D},
	month = {sep},
	number = {6},
	pages = {063502},
	title = {{Simulated reconstruction of the remote dipole field using the kinetic Sunyaev Zel'dovich effect}},
	url = {http://arxiv.org/abs/1806.01290 http://dx.doi.org/10.1103/PhysRevD.98.063502 https://link.aps.org/doi/10.1103/PhysRevD.98.063502},
	volume = {98},
	year = {2018}}

@article{Tegmark2003,
	author = {Tegmark, Max and {De Oliveira-Costa}, Ang{\'{e}}lica and Hamilton, Andrew J.S.},
	doi = {10.1103/PhysRevD.68.123523},
	issn = {05562821},
	journal = {Physical Review D},
	month = {dec},
	number = {12},
	pages = {123523},
	title = {{High resolution foreground cleaned CMB map from WMAP}},
	url = {http://www.eso.org/ https://link.aps.org/doi/10.1103/PhysRevD.68.123523},
	volume = {68},
	year = {2003}}

@article{Akrami2020,
	author = {{Planck Collaboration}},
	issn = {0004-6361},
	journal = {Astronomy {\&} Astrophysics},
	month = {sep},
	pages = {A4},
	title = {{Planck 2018 results. IV. Diffuse component separation}},
	url = {https://ui.adsabs.harvard.edu/abs/2020A{\%}26A...641A...1P/abstract https://www.aanda.org/10.1051/0004-6361/201833881},
	volume = {641},
	year = {2020}}

@article{Ade2019,
	archiveprefix = {arXiv},
	arxivid = {1808.07445v2},
	author = {{The Simons Observatory Collaboration}},
	doi = {10.1088/1475-7516/2019/02/056},
	eprint = {1808.07445v2},
	issn = {1475-7516},
	journal = {Journal of Cosmology and Astroparticle Physics},
	month = {feb},
	number = {02},
	pages = {056--056},
	title = {{The Simons Observatory: science goals and forecasts}},
	url = {https://simonsobservatory.org/publications. https://iopscience.iop.org/article/10.1088/1475-7516/2019/02/056},
	volume = {2019},
	year = {2019}}

@article{Sunyaev1980,
	author = {Sunyaev, R A and Zeldovich, Ya B},
	doi = {10.1093/mnras/190.3.413},
	issn = {0035-8711},
	journal = {Monthly Notices of the Royal Astronomical Society},
	month = {mar},
	number = {3},
	pages = {413--420},
	title = {{The velocity of clusters of galaxies relative to the microwave background. The possibility of its measurement}},
	url = {http://articles.adsabs.harvard.edu/cgi-bin/nph-iarticle{\_}query?1980MNRAS.190..413S{\&}data{\_}type=PDF{\_}HIGH{\&}whole{\_}paper=YES{\&}type=PRINTER{\&}filetype=.pdf https://academic.oup.com/mnras/article-lookup/doi/10.1093/mnras/190.3.413},
	volume = {190},
	year = {1980}}

@article{Krolewski2020,
	archiveprefix = {arXiv},
	arxivid = {1909.07412v2},
	author = {Krolewski, Alex and Ferraro, Simone and Schlafly, Edward F and White, Martin},
	doi = {10.1088/1475-7516/2020/05/047},
	eprint = {1909.07412v2},
	issn = {1475-7516},
	journal = {Journal of Cosmology and Astroparticle Physics},
	month = {may},
	number = {05},
	pages = {047--047},
	title = {{unWISE tomography of Planck CMB lensing}},
	url = {https://iopscience.iop.org/article/10.1088/1475-7516/2020/05/047},
	volume = {2020},
	year = {2020}}

@article{Alonso_2019,
   title={A unified pseudo-Cl framework},
   volume={484},
   ISSN={1365-2966},
   number={3},
   journal={Monthly Notices of the Royal Astronomical Society},
   publisher={Oxford University Press (OUP)},
   author={Alonso, David and Sanchez, Javier and Slosar, Anže},
   year={2019},
   month=jan, pages={4127–4151} }

@article{Smith2018,
	archiveprefix = {arXiv},
	arxivid = {1810.13423},
	author = {Smith, Kendrick M and Madhavacheril, Mathew S and M{\"{u}}nchmeyer, Moritz and Ferraro, Simone and Giri, Utkarsh and Johnson, Matthew C},
	eprint = {1810.13423},
	month = {oct},
    journal = "",
	title = {{KSZ tomography and the bispectrum}},
	url = {http://arxiv.org/abs/1810.13423},
	year = {2018}}

@article{Hotinli2019,
	archiveprefix = {arXiv},
	arxivid = {1908.08953},
	author = {Hotinli, Selim C. and Mertens, James B. and Johnson, Matthew C. and Kamionkowski, Marc},
	doi = {10.1103/PhysRevD.100.103528},
	eprint = {1908.08953},
	issn = {2470-0010},
	journal = {Physical Review D},
	month = {nov},
	number = {10},
	pages = {103528},
	title = {{Probing correlated compensated isocurvature perturbations using scale-dependent galaxy bias}},
	url = {https://link.aps.org/doi/10.1103/PhysRevD.100.103528},
	volume = {100},
	year = {2019}}

@article{Takahashi2020,
	archiveprefix = {arXiv},
	arxivid = {2010.01560},
	author = {Takahashi, Ryuichi and Ioka, Kunihito and Mori, Asuka and Funahashi, Koki},
	doi = {10.1093/mnras/stab170},
	eprint = {2010.01560},
	issn = {13652966},
	journal = {Monthly Notices of the Royal Astronomical Society},
	month = {oct},
	number = {2},
	pages = {2615--2629},
	title = {{Statistical modelling of the cosmological dispersion measure}},
	url = {http://arxiv.org/abs/2010.01560 http://dx.doi.org/10.1093/mnras/stab170},
	volume = {502},
	year = {2020}}

@article{Cayuso2023,
	archiveprefix = {arXiv},
	arxivid = {2111.11526},
	author = {Cayuso, Juan and Bloch, Richard and Hotinli, Selim C. and Johnson, Matthew C and McCarthy, Fiona},
	doi = {10.1088/1475-7516/2023/02/051},
	eprint = {2111.11526},
	issn = {1475-7516},
	journal = {Journal of Cosmology and Astroparticle Physics},
	month = {feb},
	number = {02},
	pages = {051},
	title = {{Velocity reconstruction with the cosmic microwave background and galaxy surveys}},
	url = {https://iopscience.iop.org/article/10.1088/1475-7516/2023/02/051},
	volume = {2023},
	year = {2023}}

@article{LSSTScienceCollaboration2009,
	archiveprefix = {arXiv},
	arxivid = {0912.0201},
	author = {{LSST Science and LSST Project Collaborations}},
	eprint = {0912.0201},
	month = {dec},
    journal = "",
	title = {{LSST Science Book, Version 2.0}},
	url = {http://arxiv.org/abs/0912.0201},
	year = {2009}}

@article{PlanckCollaboration2020,
	author = {{Planck Collaboration}},
	doi = {10.1051/0004-6361/201833880},
	issn = {0004-6361},
	journal = {Astronomy {\&} Astrophysics},
	month = {sep},
	pages = {A1},
	title = {{Planck 2018 results. I. Overview and the cosmological legacy of Planck}},
	url = {https://www.aanda.org/10.1051/0004-6361/201833880},
	volume = {641},
	year = {2020}}

@article{Mainzer2014,
	archiveprefix = {arXiv},
	arxivid = {1406.6025},
	author = {Mainzer, A. and others},
	doi = {10.1088/0004-637X/792/1/30},
	eprint = {1406.6025},
	issn = {1538-4357},
	journal = {The Astrophysical Journal},
	month = {aug},
	number = {1},
	pages = {30},
	title = {{Initial Performance of the NEOWISE Reactivation Mission}},
	url = {https://iopscience.iop.org/article/10.1088/0004-637X/792/1/30},
	volume = {792},
	year = {2014}}

@article{GaiaCollaboration2018,
	archiveprefix = {arXiv},
	arxivid = {1804.09365},
	author = {{Gaia Collaboration}},
	doi = {10.1051/0004-6361/201833051},
	eprint = {1804.09365},
	issn = {0004-6361},
	journal = {Astronomy {\&} Astrophysics},
	month = {aug},
	pages = {A1},
	title = {{Gaia Data Release 2}},
	url = {http://arxiv.org/abs/1804.09365{\%}0Ahttp://dx.doi.org/10.1051/0004-6361/201833051 https://www.aanda.org/10.1051/0004-6361/201833051},
	volume = {616},
	year = {2018}}

@article{2012ApJ...755...70R,
	adsurl = {https://ui.adsabs.harvard.edu/abs/2012ApJ...755...70R},
	archiveprefix = {arXiv},
	author = {{Reichardt}, C.~L. and others},
	doi = {10.1088/0004-637X/755/1/70},
	eid = {70},
	eprint = {1111.0932},
	journal = {\apj},
	month = aug,
	number = {1},
	pages = {70},
	primaryclass = {astro-ph.CO},
	title = {{A Measurement of Secondary Cosmic Microwave Background Anisotropies with Two Years of South Pole Telescope Observations}},
	volume = {755},
	year = 2012}

@article{Caliskan:2023yov,
	archiveprefix = {arXiv},
	author = {\c{C}al\i{}\c{s}kan, Mesut and Anil Kumar, Neha and Hotinli, Selim C. and Kamionkowski, Marc},
	eprint = {2312.00118},
	month = {11},
	primaryclass = {astro-ph.CO},
	title = {{Reconstructing patchy helium reionization using the cosmic microwave background and large-scale structure}},
    journal = "",
	year = {2023}}

@article{Hotinli:2022jnt,
	archiveprefix = {arXiv},
	author = {Hotinli, Selim C.},
	doi = {10.1103/PhysRevD.108.043528},
	eprint = {2212.08004},
	journal = {Phys. Rev. D},
	number = {4},
	pages = {043528},
	primaryclass = {astro-ph.CO},
	title = {{Cosmological probes of helium reionization}},
	volume = {108},
	year = {2023}}

@article{Hotinli:2022jna,
	archiveprefix = {arXiv},
	author = {Hotinli, Selim C. and Ferraro, Simone and Holder, Gilbert P. and Johnson, Matthew C. and Kamionkowski, Marc and La Plante, Paul},
	doi = {10.1103/PhysRevD.107.103517},
	eprint = {2207.07660},
	journal = {Phys. Rev. D},
	number = {10},
	pages = {103517},
	primaryclass = {astro-ph.CO},
	title = {{Probing helium reionization with kinetic Sunyaev-Zel\textquoteright{}dovich tomography}},
	volume = {107},
	year = {2023}}

@article{Zhang:2015uta,
    author = "Zhang, Pengjie and Johnson, Matthew C.",
    title = "{Testing eternal inflation with the kinetic Sunyaev Zel'dovich effect}",
    eprint = "1501.00511",
    archivePrefix = "arXiv",
    primaryClass = "astro-ph.CO",
    doi = "10.1088/1475-7516/2015/06/046",
    journal = "JCAP",
    volume = "06",
    pages = "046",
    year = "2015"
}

@article{Zhang:2010fc,
    author = "Zhang, Pengjie",
    title = "{The dark flow induced small scale kinetic Sunyaev Zel'dovich effect}",
    eprint = "1004.0990",
    archivePrefix = "arXiv",
    primaryClass = "astro-ph.CO",
    doi = "10.1111/j.1745-3933.2010.00899.x",
    journal = "Mon. Not. Roy. Astron. Soc.",
    volume = "407",
    pages = "L36",
    year = "2010"
}

@article{Zhang:2010fa,
    author = "Zhang, Pengjie and Stebbins, Albert",
    title = "{Confirmation of the Copernican Principle at Gpc Radial Scale and above from the Kinetic Sunyaev Zel'dovich Effect Power Spectrum}",
    eprint = "1009.3967",
    archivePrefix = "arXiv",
    primaryClass = "astro-ph.CO",
    reportNumber = "FERMILAB-PUB-10-373-A",
    doi = "10.1103/PhysRevLett.107.041301",
    journal = "Phys. Rev. Lett.",
    volume = "107",
    pages = "041301",
    year = "2011"
}

@article{Pen:2012hn,
    author = "Pen, Ue-Li and Zhang, Pengjie",
    title = "{Observational Consequences of Dark Energy Decay}",
    eprint = "1202.0107",
    archivePrefix = "arXiv",
    primaryClass = "astro-ph.CO",
    doi = "10.1103/PhysRevD.89.063009",
    journal = "Phys. Rev. D",
    volume = "89",
    number = "6",
    pages = "063009",
    year = "2014"
}

@article{Shao:2010md,
    author = "Shao, Jiawei and Zhang, Pengjie and Lin, Weipeng and Jing, Yipeng and Pan, Jun",
    title = "{The kinetic SZ tomography with spectroscopic redshift surveys}",
    eprint = "1004.1301",
    archivePrefix = "arXiv",
    primaryClass = "astro-ph.CO",
    doi = "10.1111/j.1365-2966.2011.18166.x",
    journal = "Mon. Not. Roy. Astron. Soc.",
    volume = "413",
    pages = "628--642",
    year = "2011"
}

@article{Erickcek:2008jp,
    author = "Erickcek, Adrienne L. and Carroll, Sean M. and Kamionkowski, Marc",
    title = "{Superhorizon Perturbations and the Cosmic Microwave Background}",
    eprint = "0808.1570",
    archivePrefix = "arXiv",
    primaryClass = "astro-ph",
    doi = "10.1103/PhysRevD.78.083012",
    journal = "Phys. Rev. D",
    volume = "78",
    pages = "083012",
    year = "2008"
}
